\documentclass[aps, prl, twocolumn,amsmath,amssymb,floatfix, reprint]{revtex4-2}
\usepackage{graphicx}
\usepackage{physics}
\usepackage{times}
\usepackage{hyperref} 
\hypersetup{colorlinks, allcolors=blue}

\begin{document}

\title{Topological Superconductivity Mediated by Skyrmionic Magnons}

\author{Kristian M{\ae}land} 
\affiliation{\mbox{Center for Quantum Spintronics, Department of Physics, Norwegian University of Science and Technology, NO-7491 Trondheim, Norway}} 
\author{Asle Sudb{\o}}
\email[Corresponding author: ]{asle.sudbo@ntnu.no}
\affiliation{\mbox{Center for Quantum Spintronics, Department of Physics, Norwegian University of Science and Technology, NO-7491 Trondheim, Norway}}

\begin{abstract}
Topological superconductors are associated with the appearance of Majorana bound states, with promising applications in topologically protected quantum computing. 
In this Letter, we study a system where a skyrmion crystal is interfaced with a normal metal. Through interfacial exchange coupling, spin fluctuations in the skyrmion crystal mediate an effective electron-electron interaction in the normal metal. 
We study superconductivity within a weak-coupling approach and solve gap equations both close to the critical temperature and at zero temperature.
Special features in the effective electron-electron interaction due to the noncolinearity of the magnetic ground state yield topological superconductivity at the interface.
\end{abstract}

\maketitle

\paragraph*{Introduction.}
Quantum computation aims to solve problems of potentially great societal impact considerably faster than conventional computers. To achieve this, the intention is to utilize the quantum mechanical notions of superposition and entanglement.
Quantum decoherence due to small perturbations is a major challenge for proposed realizations, introducing the need for error correction \cite{ladd2010quantumcomp}.
In topological quantum computers, the idea is to use braiding of anyons as logic gates. Computations based on this are topologically protected against small perturbations \cite{TopoQuantumCompRevModPhys}. In this setting, topological superconductivity (TSC) offers the prospect of realizing Majorana bound states (MBSs), which are non-Abelian anyons. Braiding of MBSs is one of the most prominent propositions for topologically protected quantum computations \cite{Bernevig2013, TopoQuantumCompRevModPhys, menard2019isolated, MicrosoftTGP, TopoSCrevSato, TopoSCandSkRev}.

There has been considerable interest in creating TSC at the interface between chiral magnets and conventional superconductors (SCs) \cite{SkTopoSCNagaosa, SkTopoSCMajoranaChen, SkTopoSCMajoranaLossFM, SkMajoranaRex, SkTopoSCMajoranaLossAFM, SkTopoSCMajoranaDagotto, SkTopoSCgarnier, SkTopoSCMajoranaMascot, ExpSkHeterostructure, TopoSCandSkRev}. The case of noncoplanar, skyrmion ground states in the chiral magnets has been given much attention, since MBSs can be localized at the centers of skyrmions \cite{SkTopoSCMajoranaLossFM} or bound states of skyrmions and vortices \cite{SkMajoranaRex}. 
Bound states of skyrmions and superconducting vortices have been observed in Ref.~\cite{ExpSkHeterostructure}. Signatures of MBSs have been observed in a SC monolayer proximitized to magnetic islands \cite{menard2019isolated}, and at the ends of one-dimensional nanowire SCs \cite{MicrosoftTGP}.
Binding Majoranas to skyrmions is particularly interesting, since skyrmions can be moved by electric currents \cite{nagaosaRev, KlauiRev2016,SkTopoSCMajoranaLossFM, SkMajoranaRex}. To reach the topologically nontrivial regime, the theoretical proposals typically study a strong interaction between the spins in the magnet and the electrons in the SC \cite{SkTopoSCNagaosa, SkTopoSCMajoranaChen, SkTopoSCMajoranaLossFM, SkMajoranaRex, SkTopoSCMajoranaLossAFM, TopoSCandSkRev}.

Magnon-mediated superconductivity in heterostructures of colinear magnets and conductors has received a great deal of attention, often considering much weaker coupling across the interface \cite{KargarianFMTI, HugdalTIFMAFM, EirikTIAFM, ArneFMNM,ArneAFMNM_Umklapp,EirikNMAFM, EirikEliashberg}. All of these studies find topologically trivial SCs. Superconductivity mediated by spin fluctuations has been observed in a bilayer of bismuth and nickel \cite{ExpMagnonInducedHeterostruct}. 

In this Letter, we study magnon-mediated superconductivity in a normal metal (NM) due to spin fluctuations in skyrmion crystals (SkXs). Their noncolinearity leads to fundamentally new effects in the effective electron-electron interactions that give rise to TSC. 
We study superconductivity with a weak-coupling approach and solve both the linearized gap equation and the zero temperature gap equation. A bulk topological invariant is calculated to determine which parts of the superconducting phase diagram are topologically nontrivial.

\paragraph*{Model.}

\begin{figure}
    \centering
    \includegraphics[width=0.95\linewidth]{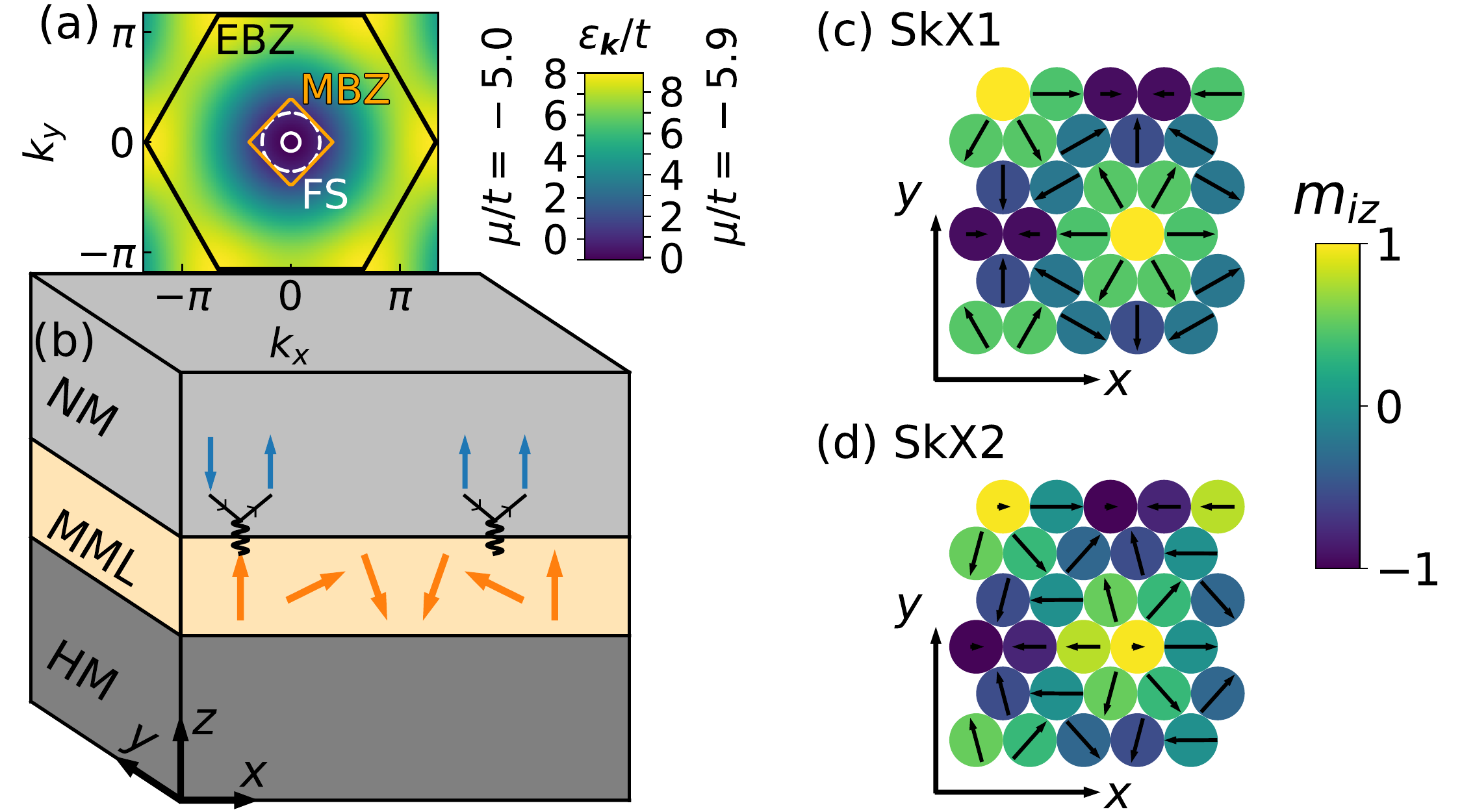}
    \caption{(a) Plot of the electron energy $\epsilon_{\boldsymbol{k}}$ showing the electron first Brillouin zone (EBZ) in black, the magnetic first Brillouin zone (MBZ) in orange and the Fermi surface (FS) in white. Two choices of $\mu$ are shown, where the solid FS corresponds to $\mu/t = -5.9$ and the dashed FS corresponds to $\mu/t = -5.0$. (b) An illustration of the system under consideration. The itinerant electrons (blue arrows) in a normal metal (NM) are coupled to the spins (orange arrows) in a magnetic monolayer (MML). The MML is deposited on a heavy metal (HM) such that skyrmion crystal (SkX) ground states (GSs) are preferred. The (c) SkX1 and (d) SkX2 GSs are shown with periodic boundary conditions. Colors give the $z$ component of the spins, $m_{iz}$, and arrows show their inplane component.}
    \label{fig:system}
\end{figure}

The system is shown in Fig.~\ref{fig:system}(b). The insulating magnetic monolayer (MML) is modeled by a nearest-neighbor ferromagnetic exchange interaction of strength $J$, a Dzyaloshinskii-Moriya interaction (DMI) of strength $D$, a four-spin interaction of strength $U$, and an easy-axis anisotropy of strength $K$ \cite{QSkOP, QSkQTPT, Suppl}. Motivated by Refs.~\cite{TuneK,TuneK_PRB}, we consider $K$ to be a tunable parameter. Throughout, we set $\hbar = a = 1$, where $a$ is the lattice constant.
The role of the heavy metal (HM) is to provide the spin-orbit coupling necessary for DMI. In addition, hydridization \cite{HeinzeSkX} can lead to an unusually small nearest-neighbor exchange interaction so that the four-spin interaction is not negligible. Other than that, the HM has no effect on our model, which focuses on the two-dimensional (2D) interface between the NM and the MML.
DMI prefers noncolinear magnetically ordered ground states (GSs). Among those, the four-spin interaction prefers noncoplanar, dense SkXs \cite{HeinzeSkX}. The two classical GSs in the MML are shown in Fig.~\ref{fig:system}(c) and (d). SkX1 is the GS for $K < K_t$ and SkX2 is the GS for $K>K_t$, where $K_t/J \in (0.518, 0.519)$ as previously reported in Refs.~\cite{QSkOP, QSkQTPT}. 
Note that the centers of the skyrmions in SkX2 are shifted compared to SkX1, giving SkX2 a lower symmetry \cite{QSkOP}.
These SkXs both feature 15 magnon bands, $\omega_{\boldsymbol{q}n}$, which we take as inputs in this Letter, $H_{\text{MML}} = \sum_{\boldsymbol{q}\in \text{MBZ},n} \omega_{\boldsymbol{q}n} b_{\boldsymbol{q}n}^\dagger b_{\boldsymbol{q}n}$. The quasimomentum $\boldsymbol{q}$ is restricted to the magnetic first Brillouin zone (MBZ) corresponding to the centered rectangular lattice set up by the SkX GSs \cite{QSkOP, QSkQTPT}. 

The NM at the interface is modeled by a hopping term with energy $t$ and a controllable chemical potential, $\mu$, which is diagonalized by a Fourier transform (FT) $c_{i\sigma} = \frac{1}{\sqrt{N}} \sum_{\boldsymbol{k}\in \text{EBZ}} c_{\boldsymbol{k}\sigma} e^{i\boldsymbol{k}\cdot\boldsymbol{r}_i}$. The quasimomentum $\boldsymbol{k}$ is restricted to the electron first Brillouin zone (EBZ) corresponding to the triangular lattice, and $N$ is the total number of lattice sites at the interface. $c_{i\sigma}$ annihilates an electron with spin $\sigma$ and site index $i$, located at $\boldsymbol{r}_i$. This yields $H_{\text{NM}} = \sum_{\boldsymbol{k}\in \text{EBZ}, \sigma} \epsilon_{\boldsymbol{k}} c_{\boldsymbol{k}\sigma}^\dagger c_{\boldsymbol{k}\sigma}$, with $\epsilon_{\boldsymbol{k}} =  -\mu - 2t[\cos k_x +2 \cos(k_x/2)\cos (\sqrt{3}k_y/2)]$ shown in Fig.~\ref{fig:system}(a).

The interaction between electrons in the NM and localized spins in the MML is modeled as an interfacial exchange interaction \cite{KargarianFMTI, HugdalTIFMAFM,  ArneFMNM,ArneAFMNM_Umklapp,EirikNMAFM, EirikTIAFM, EirikEliashberg, SkTopoSCNagaosa, SkTopoSCMajoranaChen, SkTopoSCMajoranaLossFM, SkMajoranaRex, SkTopoSCMajoranaLossAFM, SkTopoSCMajoranaDagotto, SkTopoSCgarnier, SkTopoSCMajoranaMascot, ExpMagnonInducedHeterostruct, ExpSkHeterostructure, ExpInterfaceExchange},
$
    H_{\text{em}} = -2\Bar{J}\sum_i \boldsymbol{c}_i^\dagger \boldsymbol{\sigma} \boldsymbol{c}_i \cdot \boldsymbol{S}_i,
$
where $\boldsymbol{c}_i = (c_{i\uparrow}, c_{i\downarrow})^T$, $\boldsymbol{\sigma}$ is a vector of the Pauli matrices, and $\boldsymbol{S}_i$ is the spin operator at site $i$. We treat this term as a perturbation to the NM, and focus on the magnon-mediated effective electron-electron interaction. Assuming $\Bar{J} \ll t$ we keep the $z$ axis as quantization axis for the electron spins in the NM. The spins in the MML are each quantized along the direction of the spin in the classical GS giving 15 separate quantization axes. Performing such rotations of the spins in the MML and applying the Holstein-Primakoff transformation, yields \cite{Suppl}
\begin{align}
\label{eq:Hem_i}
    H_{\text{em}} =& -\Bar{J}\sqrt{2S} \sum_{i\sigma} [e^{-i\sigma\phi_i}(\cos\theta_i-n_\sigma)a_i c_{i\sigma}^\dagger c_{i,-\sigma} + \text{H.c.}] \nonumber \\
    &+\Bar{J}\sqrt{2S}\sum_{i\sigma}(n_\sigma \sin\theta_i a_i c_{i\sigma}^\dagger c_{i\sigma} + \text{H.c.}).
\end{align}
Here, H.c.~denotes Hermitian conjugate, $n_\uparrow = 1$, $n_\downarrow = -1$, $a_i$ annihilates a magnon at site $i$, $S$ is the spin quantum number in the MML, and $\theta_i, \phi_i$ are the polar and azimuthal angles specifying the direction of the classical spin at site $i$. 
We have ignored terms that contain only two electron operators. Such renormalizations of the electron spectrum are higher order in perturbation theory than our weak-coupling treatment of the effective electron-electron interaction.  
Self-energy effects due to electron-magnon coupling could renormalize the electron spectrum. Given the existence of a magnon gap \cite{QSkOP, QSkQTPT}, such effects are negligible close to the Fermi surface (FS) \cite{Self-energy}.

Compared to earlier studies of superconductivity induced by colinear spin structures, Eq.~\eqref{eq:Hem_i} features a fundamental difference. Namely, given that $\theta_i \neq \{0,\pi\}$, a magnon can be involved in both spin flip processes as well as processes where the $z$-component of the electron spin is not changed. We illustrate this in Fig.~\ref{fig:system}(b). When the spin in the magnet points in the $z$ direction, the electron spin will always get a spin flip. When an itinerant electron interacts with a spin in the magnet with a nonzero inplane component, the electron spin need not change.

\paragraph*{Effective interaction.}
As shown in Fig.~\ref{fig:system}(a), the MBZ is far smaller than the EBZ. Thus, Umklapp processes must be included when applying FTs to Eq.~\eqref{eq:Hem_i}. 
The FT of the electron operators is modified to $c_{i\sigma} = \frac{1}{\sqrt{N}} \sum_{\boldsymbol{k}\in \text{MBZ}} \sum_\nu c_{\boldsymbol{k}+\boldsymbol{Q}_\nu, \sigma} e^{i(\boldsymbol{k}+\boldsymbol{Q}_\nu) \cdot\boldsymbol{r}_i}$, where
$\boldsymbol{Q}_\nu$ is a set of 15 reciprocal lattice vectors specified in Ref.~\cite{Suppl}. 
If site $i$ is located on sublattice $r$ we have $a_{i} = \frac{1}{\sqrt{N'}} \sum_{\boldsymbol{q}\in \text{MBZ}} a_{\boldsymbol{q}}^{(r)} e^{i\boldsymbol{q}\cdot\boldsymbol{r}_i}$, where $N'$ is the number of magnetic unit cells. 
The magnon operators $a_{\boldsymbol{q}}^{(r)}$ are transformed to their diagonal basis $b_{\boldsymbol{q}n}$ through a paraunitary matrix $T_{\boldsymbol{q}}$ \cite{COLPA, QSkOP, QSkQTPT}.
We obtain an effective electron-electron interaction mediated by the magnons in the SkXs by applying a Schrieffer-Wolff transformation \cite{SchriefferWolff}. 
Assuming oppositely directed momenta, we obtain \cite{Suppl}
\begin{equation}
    H_{\text{ee}} = \frac{1}{2}\sum_{\boldsymbol{k}\boldsymbol{k}'}^{\text{EBZ}} \sum_{\substack{\sigma_1 \sigma_2 \\\sigma_3 \sigma_4}} \Bar{V}_{\boldsymbol{k}\boldsymbol{k}'}^{\sigma_1 \sigma_2 \sigma_3 \sigma_4} c_{\boldsymbol{k}'\sigma_1}^\dagger c_{-\boldsymbol{k}'\sigma_2}^\dagger c_{-\boldsymbol{k}\sigma_3}c_{\boldsymbol{k}\sigma_4},
\end{equation}
where $\boldsymbol{k}' = \boldsymbol{k}+\boldsymbol{q} +\boldsymbol{Q}_\nu$.  
The coupling functions $\Bar{V}_{\boldsymbol{k}\boldsymbol{k}'}^{\sigma_1 \sigma_2 \sigma_3 \sigma_4}$ are of order $\Bar{J}^2/t$ and contain linear combinations of magnon transformation coefficients divided by $\epsilon_{\boldsymbol{k}} -\epsilon_{\boldsymbol{k}'} \pm \omega_{\pm\boldsymbol{q},n}$. Detailed expressions are given in Ref.~\cite{Suppl}. 
Due to the noncolinear spin structure, they are in general nonzero for all combinations of spin-indices. This is a crucial difference from colinear spin structures, where only a subset are nonzero \cite{ArneFMNM,ArneAFMNM_Umklapp,EirikNMAFM, EirikTIAFM, EirikEliashberg}. This endows the superconducting order parameter with richer spin-texture than in the colinear case. 
While we treat Umklapp effects in the effective interaction, their effect on the NM energy dispersion are ignored. Hence, our results are valid when the FS is smaller than the MBZ, i.e.~low filling $\mu/t \leq -5.0$, where such effects do not influence occupied states. 

\paragraph*{Superconductivity.}
We follow the generalized BCS theory outlined in Ref.~\cite{Sigrist} for unconventional SCs. Since all $\Bar{V}_{\boldsymbol{k}\boldsymbol{k}'}^{\sigma_1 \sigma_2 \sigma_3 \sigma_4} $ are nonzero in general, we can have coexistence of singlet SC gap, $\Delta_{\boldsymbol{k}\uparrow\downarrow}^{O(s)} = (\Delta_{\boldsymbol{k}\uparrow\downarrow}-\Delta_{\boldsymbol{k}\downarrow\uparrow})/2$, and all triplet gaps $\Delta_{\boldsymbol{k}\uparrow\downarrow}^{E(s)} = (\Delta_{\boldsymbol{k}\uparrow\downarrow}+\Delta_{\boldsymbol{k}\downarrow\uparrow})/2$, $\Delta_{\boldsymbol{k}\uparrow\uparrow}$, and $\Delta_{\boldsymbol{k}\downarrow\downarrow}$. This yields two distinct bands in the SC, 
$E_{\boldsymbol{k}\pm} = (\epsilon_{\boldsymbol{k}}^2 + \Tr\hat{\Delta}_{\boldsymbol{k}} \hat{\Delta}_{\boldsymbol{k}}^\dagger/2 \pm \sqrt{A_{\boldsymbol{k}}}/2)^{1/2},$ with $(\hat{\Delta}_{\boldsymbol{k}})_{\sigma\sigma'} = \Delta_{\boldsymbol{k}\sigma\sigma'}$ and, in our case, $A_{\boldsymbol{k}}/16 = (\Delta_{\boldsymbol{k}\uparrow\downarrow}^{E(s)})^2 (\Delta_{\boldsymbol{k}\uparrow\downarrow}^{O(s)})^2 -\Delta_{\boldsymbol{k}\uparrow\uparrow}\Delta_{\boldsymbol{k}\downarrow\downarrow}(\Delta_{\boldsymbol{k}\uparrow\downarrow}^{O(s)})^2$. A more general expression for $A_{\boldsymbol{k}}$ is given in Ref.~\cite{Suppl}.

The gap equation 
is \cite{Suppl}
\begin{equation}
    \boldsymbol{\Delta_k} = -\sum_{\boldsymbol{k}'} \mathcal{V}_{\boldsymbol{k}'\boldsymbol{k}} \sum_\eta \pqty{\frac{\boldsymbol{\Delta}_{\boldsymbol{k}'}}{2}+\eta \boldsymbol{B}_{\boldsymbol{k}'} }\chi_{\boldsymbol{k}'\eta}.
\end{equation}
Here, $\boldsymbol{\Delta}_{\boldsymbol{k}} = (\Delta_{\boldsymbol{k}\uparrow\downarrow}^{O(s)}, \Delta_{\boldsymbol{k}\uparrow\uparrow}, \Delta_{\boldsymbol{k}\downarrow\downarrow}, \Delta_{\boldsymbol{k}\uparrow\downarrow}^{E(s)})^T$, $\boldsymbol{B}_{\boldsymbol{k}} = (B_{\boldsymbol{k}\uparrow\downarrow}^{O(s)}, B_{\boldsymbol{k}\uparrow\uparrow}, B_{\boldsymbol{k}\downarrow\downarrow}, B_{\boldsymbol{k}\uparrow\downarrow}^{E(s)})^T$, $B_{\boldsymbol{k}\sigma_1\sigma_2}^\dagger = \frac{1}{4\sqrt{A_{\boldsymbol{k}}}}\pdv{A_{\boldsymbol{k}}}{\Delta_{\boldsymbol{k}\sigma_1\sigma_2}}$, and $\chi_{\boldsymbol{k}\eta} = \tanh(\beta E_{\boldsymbol{k}\eta}/2)/2E_{\boldsymbol{k}\eta}$. $\mathcal{V}_{\boldsymbol{k}'\boldsymbol{k}}$ is a matrix containing the 16 coupling functions \cite{Suppl}.

\begin{figure}
    \centering
    \includegraphics[width=0.9\linewidth]{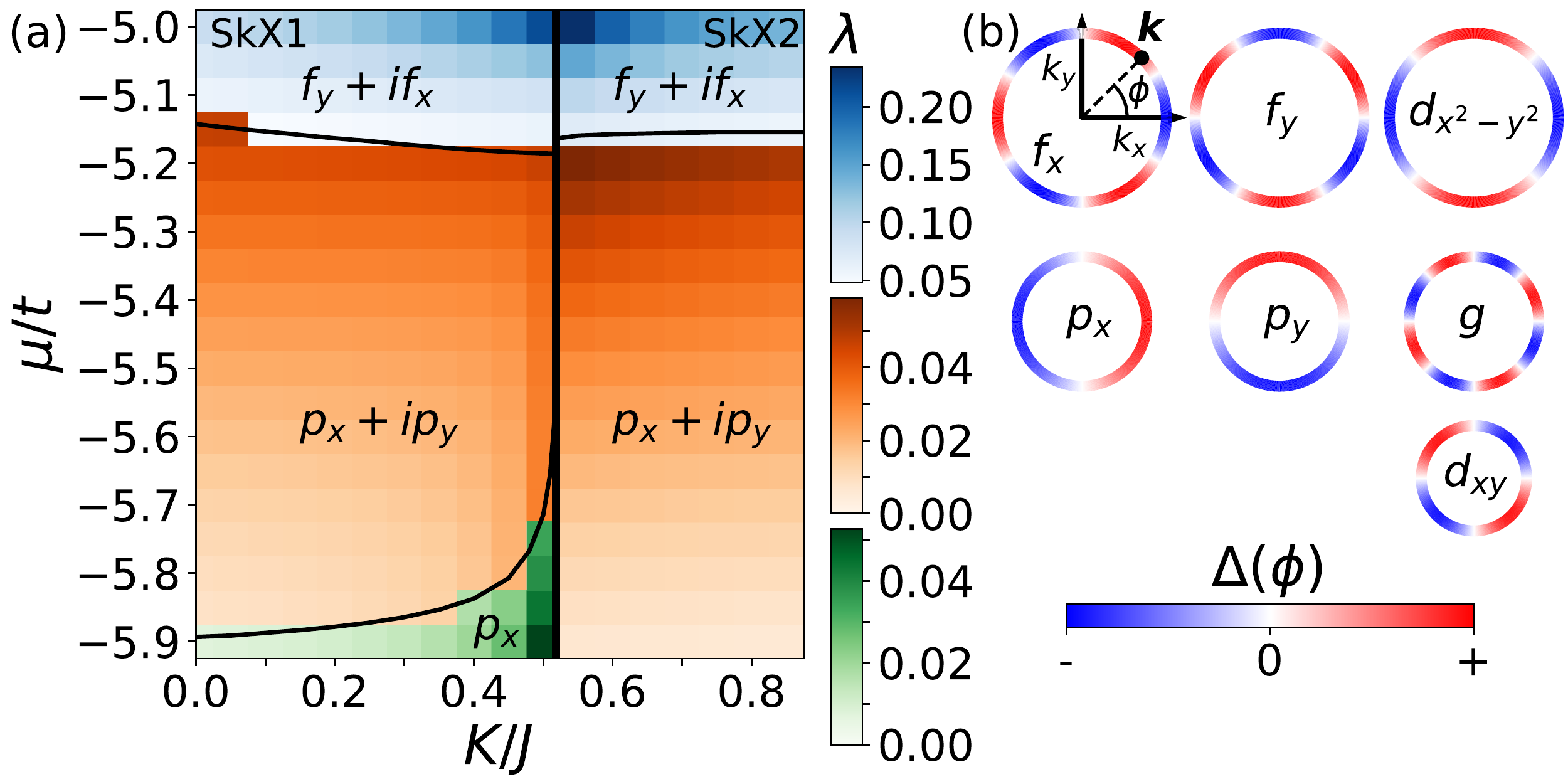}
    \caption{(a) Phase diagram and dimensionless coupling $\lambda$ in the superconducting state close to $T_{\text{c}}$. The vertical, black line shows the transition between SkX1 and SkX2 in the MML. The other black lines show the locations of the phase transitions with better resolution. In the green region, the unpolarized triplet gap $\Delta_{\boldsymbol{k}\uparrow\downarrow}^{E(s)}$ shows $p_x$-wave symmetry, while the other gaps are zero. In the orange and blue regions, the stated symmetry refers to the $\Delta_{\boldsymbol{k}\downarrow\downarrow}$ gap. The relevant gap symmetries are shown in (b). The parameters are $t/J = 1000$, $\Bar{J}/J = 50$, $D/J = 2.16$, $U/J = 0.35$, and $S=1$.}
    \label{fig:PDlam}
\end{figure}

For temperatures close to the critical temperature $T_{\text{c}}$ we
linearize the gap equation. Since $\chi_{\boldsymbol{k}'\eta}$ is peaked at the FS for small $T_{\text{c}}$, we employ FS averages, keeping the angular dependence of $\mathcal{V}_{\boldsymbol{k}'\boldsymbol{k}}$, but ignoring any radial variation. The matrix elements are set to their value on the FS, for energies closer to the FS than the maximum magnon frequency $\omega_{\text{c}}$, i.e.~$|\epsilon_{\boldsymbol{k}}|,
|\epsilon_{\boldsymbol{k}'}| < \omega_{\text{c}}$. Otherwise, the coupling functions are set to zero. 
The resulting 
gap equation is \cite{Suppl}  
\begin{equation}
\label{eq:linFsavrgap}
   \lambda \boldsymbol{\Delta}(\phi) = -N_0 \langle \mathcal{V}(\phi', \phi) \boldsymbol{\Delta}(\phi') \rangle_{\text{FS},\phi'},
\end{equation}
where $N_0$ is the density of states per spin on the FS, and $\phi$ is the angle $\boldsymbol{k}$ makes with the $k_x$ axis. The dimensionless coupling constant $\lambda$ can be used to estimate the critical temperature. Given $\lambda \ll 1$, $k_{\text{B}}T_{\text{c}} = (2e^\gamma/\pi)\omega_{\text{c}}e^{-1/\lambda}$, where $\gamma$ is the Euler-Mascheroni constant \cite{SFsuperconductivity}. We solve Eq.~\eqref{eq:linFsavrgap} as a matrix eigenvalue problem. 
Then, $\lambda$ is the greatest eigenvalue and its corresponding eigenvectors give information about the structure of the gap \cite{Suppl}. 

Results for $\lambda$ are given in Fig.~\ref{fig:PDlam}(a). $\lambda$ increases with $\mu$, presumably because the FS becomes larger. $\lambda$ also increases toward the phase transition between SkX1 and SkX2 in the MML, since the magnon gap decreases, giving stronger electron-electron interactions. Figure \ref{fig:PDlam}(a) also acts as a phase diagram showing the symmetry of the gaps. 
The proceeding symmetry classifications are illustrated in Fig.~\ref{fig:PDlam}(b). 
In SkX1, for $K<K_t$, $\Delta_{\boldsymbol{k}\uparrow\downarrow}^{E(s)}$ decouples from the other gaps. In the green region, $\Delta_{\boldsymbol{k}\uparrow\downarrow}^{E(s)}$ has $p_x$-wave symmetry while the other gaps are zero. In the orange and blue regions, $\Delta_{\boldsymbol{k}\uparrow\downarrow}^{E(s)} = 0$. In the orange region, $\Delta_{\boldsymbol{k}\downarrow\downarrow}$ has $p_x+ip_y$-wave, i.e.~chiral $p$-wave symmetry, $\Delta_{\boldsymbol{k}\uparrow\uparrow} = -\Delta_{\boldsymbol{k}\downarrow\downarrow}^*$ and $\Delta_{\boldsymbol{k}\uparrow\downarrow}^{O(s)}$ is a small $d_{xy}$-wave gap for low $\mu$. For higher $\mu$, $\Delta_{\boldsymbol{k}\uparrow\downarrow}^{O(s)}$ becomes $g$-wave. In the blue region, $\Delta_{\boldsymbol{k}\downarrow\downarrow}$ has $f_y+if_x$-wave, i.e.~chiral $f$-wave symmetry, $\Delta_{\boldsymbol{k}\uparrow\uparrow} = -\Delta_{\boldsymbol{k}\downarrow\downarrow}^*$, while $\Delta_{\boldsymbol{k}\uparrow\downarrow}^{O(s)}$ shows $-d_{x^2-y^2}$-wave symmetry. The situation in SkX2 is similar, but all four gaps couple. While the other gap symmetries remain the same in the orange and blue regions, a comparatively small amplitude $p_x$-wave solution for $\Delta_{\boldsymbol{k}\uparrow\downarrow}^{E(s)}$ coexists. 


\begin{figure}
    \centering
    \includegraphics[width=0.85\linewidth]{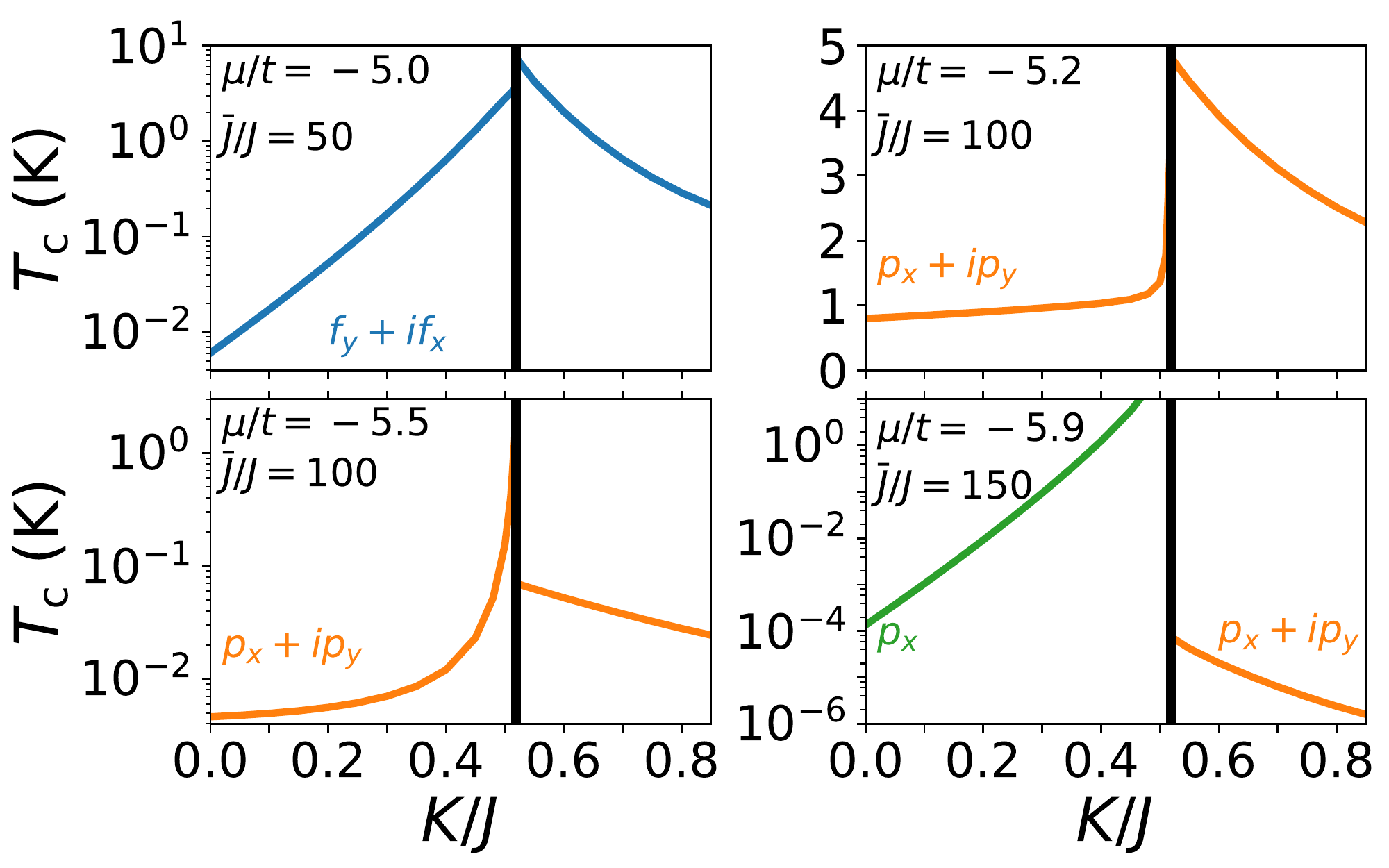}
    \caption{Estimates of $T_{\text{c}}$ as a function of the easy-axis anisotropy $K$ in the MML for different chemical potentials $\mu$. We have set an upper cutoff of 10~K since larger $T_{\text{c}}$ means that $\lambda \ll 1$ is no longer valid. The parameters are $t/J = 1000$, $D/J = 2.16$, $U/J = 0.35$, and $S=1$. In this figure only, $J = 1$~meV was set in order to find $T_{\text{c}}$ in Kelvin.}
    \label{fig:Tc}
\end{figure}

Figure \ref{fig:Tc} shows estimates of $T_{\text{c}}$ in Kelvin. Note from Fig.~\ref{fig:PDlam}(a) that $\lambda$ is too small in the orange and green regions to yield $T_{\text{c}} > 1$~mK when $\Bar{J}/J = 50$. Here, $\lambda \sim (\Bar{J}/t)^2$ and $T_{\text{c}}$ depends exponentially on $\lambda$. Hence, an increase of $\Bar{J}$ gives a major increase in $T_{\text{c}}$. We still keep $\Bar{J}\ll t$, so that the coupling to the magnet can be viewed as a perturbation. 

\paragraph*{Zero temperature.}
To study the low-temperature behavior, we derive a gap equation at zero temperature, where $\chi_{\boldsymbol{k}\eta} = 1/2E_{\boldsymbol{k}\eta}$. Using FS averages yields \cite{Suppl}
\begin{align}
\label{eq:GapT0}
    &\boldsymbol{\Delta}(\phi) = -N_0 \Bigg\langle \mathcal{V}(\phi',\phi)\sum_\eta \Big(\frac{\boldsymbol{\Delta}(\phi')}{2}+\eta \boldsymbol{B}(\phi')\Big) \nonumber \\
    &\times \text{arsinh} \Bigg[\frac{\sqrt{2}\omega_{\text{c}}}{\big(\Tr\hat{\Delta}(\phi') \hat{\Delta}^\dagger (\phi')+\eta \sqrt{A(\phi')}\big)^{1/2}}\Bigg] \Bigg\rangle_{\text{FS}, \phi'}.
\end{align}
In the original BCS treatment of phonon-mediated superconductivity, $2\Delta(0)/k_{\text{B}} T_{\text{c}} = 2\pi e^{-\gamma}$ where $\Delta(0)$ is the amplitude of the gap at zero temperature \cite{SFsuperconductivity}. We use this as the amplitude of an initial trial solution $\boldsymbol{\Delta}_0(\phi)$ with its structure around the FS guided by the solution close to $T_{\text{c}}$. Self-consistent iteration \cite{Suppl} is then used to obtain solutions of Eq.~\eqref{eq:GapT0} satisfying a convergence criterion, typically $10^{-4}$ of the amplitude of the gap vector. 
The amplitude is defined as the largest absolute value of the real or imaginary parts of the four gap functions.

\begin{figure}
    \centering
    \includegraphics[width=0.6\linewidth]{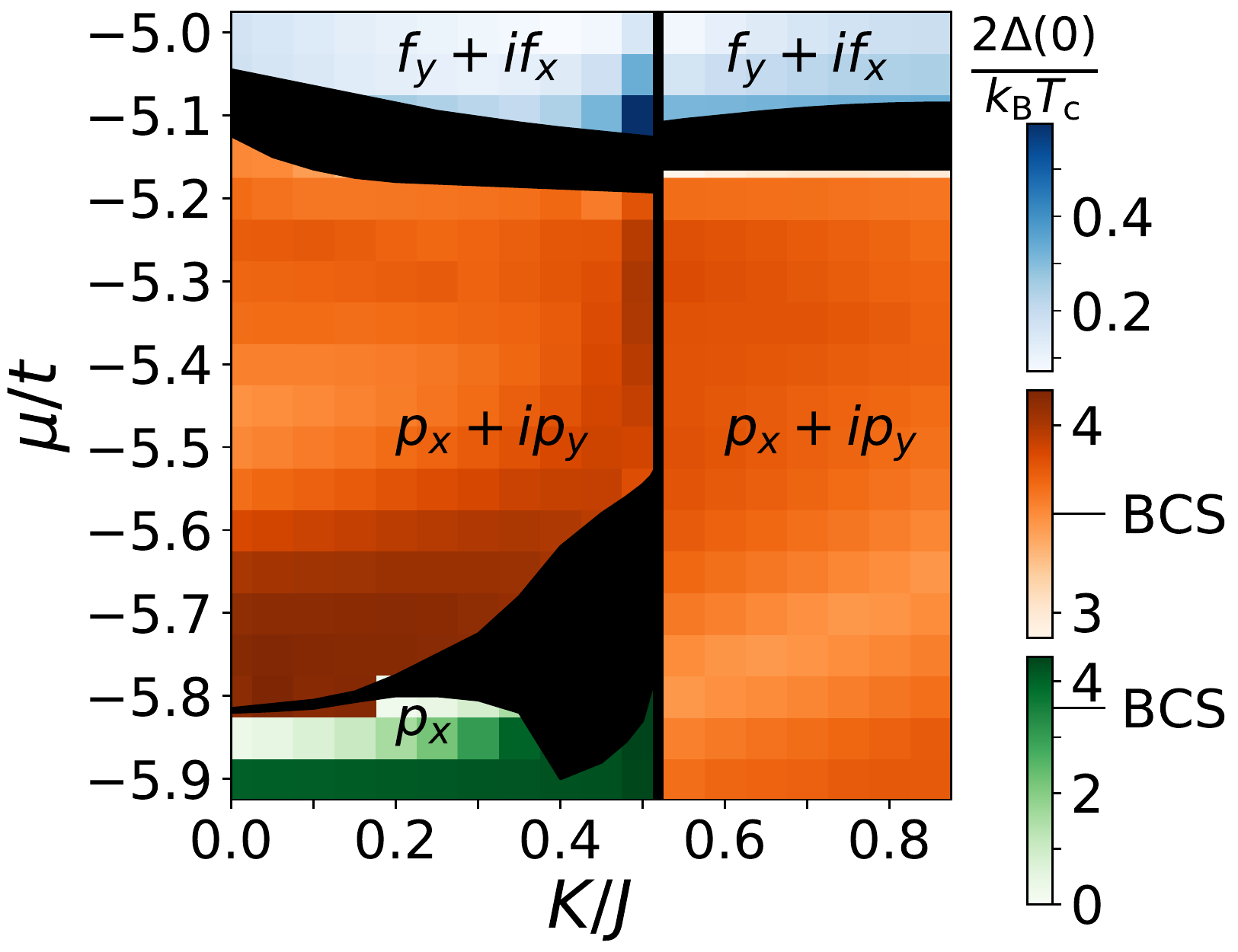}
    \caption{Phase diagram at zero temperature showing $2\Delta(0)/k_{\text{B}}T_{\text{c}} $ in color. Black regions show where two or more symmetries lead to convergence. Where applicable, the BCS result $2\Delta(0)/k_{\text{B}}T_{\text{c}} = 2\pi e^{-\gamma}$ is shown on the colorbars. The parameters are $t/J = 1000$, $D/J = 2.16$, $U/J = 0.35$, and $S=1$.}
    \label{fig:PDT0}
\end{figure}

Figure \ref{fig:PDT0} shows the resulting phase diagram at zero temperature. The black lines showing the phase transitions are wider, since there are regions where several symmetries can give solutions of the nonlinear Eq.~\eqref{eq:GapT0} \cite{Suppl}. Let $\Delta(0)$ be the maximum of the lowest energy band on the FS, $\Delta(0) \equiv \operatorname{max}_{\text{FS}} E_{\boldsymbol{k}-}$. This would be the largest gap found on the FS in an experiment. We see that $2\Delta(0)/k_{\text{B}} T_{\text{c}}$ is comparable to the BCS result in the $p_x$- and chiral $p$-wave phases, except that it drops to small values when approaching the phase transitions. In the chiral $f$-wave phases, $2\Delta(0)/k_{\text{B}} T_{\text{c}}$ is much smaller than the BCS result because $\Tr\hat{\Delta}_{\boldsymbol{k}} \hat{\Delta}_{\boldsymbol{k}}^\dagger \gtrsim \sqrt{A_{\boldsymbol{k}}}$ and so $E_{\boldsymbol{k}-}$ is small. This is due to the unconventional nature of the SC. 
The value of $2\Delta(0)/k_{\text{B}}T_{\text{c}}$ has a very weak dependence on $\Bar{J}$. 
We varied $\Bar{J}$ when obtaining Fig.~\ref{fig:PDT0} to ensure that the expected amplitude at zero temperature was larger than $10^{-5}J$.

\paragraph*{Topological superconductivity.}
The first requirement for (strong) TSC is a fully gapped bulk spectrum
\cite{TopoSCrevSato}. The chiral $f$-wave phases have a very small $E_{\boldsymbol{k}-}$, and 
we leave their topological classification as an open question. The $p_x$-wave phase is gapless and
topologically trivial. In the chiral $p$-wave phases the bulk spectrum $E_{\boldsymbol{k}\eta}$ is fully gapped.

To determine if the SC state is topologically nontrivial, we proceed by computing a bulk topological invariant. 
We define a Bogoliubov-de Gennes (BdG) Hamiltonian where the gaps $\Delta_{\boldsymbol{k}\uparrow\downarrow}$ and $\Delta_{\boldsymbol{k}\downarrow\uparrow}$ are multiplied by $1-x$. Using the gap functions in the chiral $p$-wave phases we find that tuning $x$ from 0 to 1 does not close the bulk gap \cite{Suppl}. Hence, the obtained SC is topologically equivalent to one where $\Delta_{\boldsymbol{k}\uparrow\downarrow} = \Delta_{\boldsymbol{k}\downarrow\uparrow} = 0$. For such a system, we can define two spin-decoupled BdG Hamiltonians $H_{\boldsymbol{k}\sigma} = \boldsymbol{d}_{\boldsymbol{k}\sigma} \cdot \boldsymbol{\sigma}$, where $\boldsymbol{d}_{\boldsymbol{k}\sigma} = (\Re\Delta_{\boldsymbol{k}\sigma\sigma}, -\Im\Delta_{\boldsymbol{k}\sigma\sigma}, \epsilon_{\boldsymbol{k}})$. The obtained SC states are time-reversal symmetric (TRS) \cite{Sigrist, Suppl}, and so the bulk topological invariant is defined as
\begin{equation}
    \nu_{\mathbb{Z}_2} = \frac{1}{2} (N_\uparrow - N_\downarrow)  \quad \text{mod } 2,
\end{equation}
with $N_\sigma = (1/8\pi) \int_{\text{EBZ}} d \boldsymbol{k} \epsilon_{ij} \hat{d}_{\boldsymbol{k}\sigma} \cdot (\partial_{k_i} \hat{d}_{\boldsymbol{k}\sigma} \cross \partial_{k_j} \hat{d}_{\boldsymbol{k}\sigma})$, $\epsilon_{ij}$ the Levi-Civita tensor, $i,j \in \{x,y\}$, and $\hat{d}_{\boldsymbol{k}\sigma}$ a unit vector along $\boldsymbol{d}_{\boldsymbol{k}\sigma}$ \cite{Bernevig2013}. Due to TRS, $N_\downarrow = -N_\uparrow$. Since the energy scale of $\epsilon_{\boldsymbol{k}}$ is much greater than the gaps $\Delta_{\boldsymbol{k}\sigma\sigma}$, the integrand is only nonzero close to the FS, where $\epsilon_{\boldsymbol{k}}$ is small. Hence, knowledge of the gaps close to the FS is sufficient to calculate the integral over the full EBZ. Using adaptive integration \cite{AdaptQuad, QSkQTPT}, $N_\downarrow$ approaches $1$ with increasing density of points. Therefore, $\nu_{\mathbb{Z}_2} = 1$, indicating a topologically nontrivial SC. 

The chiral $p$-wave phases of our SC are TRS 2D topological SCs.
In a finite geometry there will be two topologically protected, counterpropagating edge states which are Majorana fermions corresponding to each spin species \cite{Bernevig2013}. 
Additionally, there will be MBSs in the core of vortices \cite{Bernevig2013, TopoQuantumCompRevModPhys}. 

TSC requires spinless or spin polarized Cooper pairs \cite{Bernevig2013}. Previous studies of magnon-mediated superconductivity from colinear spin structures found only unpolarized Cooper pairs
\cite{ArneFMNM,ArneAFMNM_Umklapp,EirikNMAFM, EirikEliashberg}. 
The noncolinearity of the SkXs admits the creation of polarized Cooper pairs, making TSC possible. 
In heterostructures of chiral magnets and conventional SCs, it is found that the noncoplanar nature of skyrmions is essential to get a bulk gap and strong TSC \cite{SkTopoSCNagaosa, SkTopoSCMajoranaChen, TopoSCrevSato}. 
Unlike heterostructures involving conventional SCs, the pairing mechanism itself leads to TSC in the system we consider.
From Eq.~\eqref{eq:Hem_i} it is clear that noncolinearity is sufficient to facilitate polarized Cooper pairs \cite{Benestad}. 
Whether noncolinear, coplanar states would result in TSC requires detailed solutions of the gap equation and is left as an open question.

\paragraph*{Suggestions for materials.}
A SkX similar to SkX1 and SkX2 was observed in a MML of iron grown on top of the HM iridium \cite{HeinzeSkX}.  
To test our predictions, we suggest growing a NM which does not become a SC due to electron-phonon interactions, like copper, silver or gold \cite{Kittel2005}, on top of the MML. Producing an interfacial exchange interaction strong enough to bring $T_{\text{c}}$ to observable temperatures is a materials science challenge. 
The value of $\Bar{J}$ is associated with an overlap integral, depending on the chosen materials and how the interface is grown. 
See Ref.~\cite{Self-energy} and references therein for a discussion of the value of $\Bar{J}$.
Note that while we require $\Bar{J}/t \sim 0.1$ to get TSC with observable $T_{\text{c}}$, the studies of TSC in heterostructures of chiral magnets and conventional SCs often require $\Bar{J} \sim t$ for the system to enter a topologically nontrivial state \cite{SkTopoSCNagaosa, SkTopoSCMajoranaChen, SkTopoSCMajoranaLossFM, SkMajoranaRex, SkTopoSCMajoranaLossAFM, TopoSCandSkRev}.

\paragraph*{Conclusion.} 
We have studied an interface between a normal metal and an insulating magnetic monolayer hosting skyrmion crystals. 
The noncolinearity of the magnetic ground state allowed more exotic electron-electron interactions mediated by magnons than colinear magnetic ground states. 
In large parts of the phase diagram, we found topological superconductivity, with possible applications in topological quantum computing. 

\begin{acknowledgments}
\paragraph*{Acknowledgments.} 
We thank Jacob Benestad, Eirik Erlandsen, and Mathias Kl{\"a}ui for useful discussions.
We acknowledge funding from the Research Council of Norway (RCN) through its Centres of Excellence funding scheme, Project No.~262633, ``QuSpin," and RCN through Project No.~323766, ``Equilibrium and out-of-equilibrium quantum phenomena in superconducting hybrids with antiferromagnets and topological insulators."
\end{acknowledgments}

\bibliography{main.bbl}

\begin{thebibliography}{46}%
\makeatletter
\providecommand \@ifxundefined [1]{%
 \@ifx{#1\undefined}
}%
\providecommand \@ifnum [1]{%
 \ifnum #1\expandafter \@firstoftwo
 \else \expandafter \@secondoftwo
 \fi
}%
\providecommand \@ifx [1]{%
 \ifx #1\expandafter \@firstoftwo
 \else \expandafter \@secondoftwo
 \fi
}%
\providecommand \natexlab [1]{#1}%
\providecommand \enquote  [1]{``#1''}%
\providecommand \bibnamefont  [1]{#1}%
\providecommand \bibfnamefont [1]{#1}%
\providecommand \citenamefont [1]{#1}%
\providecommand \href@noop [0]{\@secondoftwo}%
\providecommand \href [0]{\begingroup \@sanitize@url \@href}%
\providecommand \@href[1]{\@@startlink{#1}\@@href}%
\providecommand \@@href[1]{\endgroup#1\@@endlink}%
\providecommand \@sanitize@url [0]{\catcode `\\12\catcode `\$12\catcode
  `\&12\catcode `\#12\catcode `\^12\catcode `\_12\catcode `\%12\relax}%
\providecommand \@@startlink[1]{}%
\providecommand \@@endlink[0]{}%
\providecommand \url  [0]{\begingroup\@sanitize@url \@url }%
\providecommand \@url [1]{\endgroup\@href {#1}{\urlprefix }}%
\providecommand \urlprefix  [0]{URL }%
\providecommand \Eprint [0]{\href }%
\providecommand \doibase [0]{https://doi.org/}%
\providecommand \selectlanguage [0]{\@gobble}%
\providecommand \bibinfo  [0]{\@secondoftwo}%
\providecommand \bibfield  [0]{\@secondoftwo}%
\providecommand \translation [1]{[#1]}%
\providecommand \BibitemOpen [0]{}%
\providecommand \bibitemStop [0]{}%
\providecommand \bibitemNoStop [0]{.\EOS\space}%
\providecommand \EOS [0]{\spacefactor3000\relax}%
\providecommand \BibitemShut  [1]{\csname bibitem#1\endcsname}%
\let\auto@bib@innerbib\@empty
\bibitem [{\citenamefont {Ladd}\ \emph {et~al.}(2010)\citenamefont {Ladd},
  \citenamefont {Jelezko}, \citenamefont {Laflamme}, \citenamefont {Nakamura},
  \citenamefont {Monroe},\ and\ \citenamefont
  {O’Brien}}]{ladd2010quantumcomp}%
  \BibitemOpen
  \bibfield  {author} {\bibinfo {author} {\bibfnamefont {T.~D.}\ \bibnamefont
  {Ladd}}, \bibinfo {author} {\bibfnamefont {F.}~\bibnamefont {Jelezko}},
  \bibinfo {author} {\bibfnamefont {R.}~\bibnamefont {Laflamme}}, \bibinfo
  {author} {\bibfnamefont {Y.}~\bibnamefont {Nakamura}}, \bibinfo {author}
  {\bibfnamefont {C.}~\bibnamefont {Monroe}},\ and\ \bibinfo {author}
  {\bibfnamefont {J.~L.}\ \bibnamefont {O’Brien}},\ }\bibfield  {title}
  {\bibinfo {title} {Quantum computers},\ }\href
  {https://doi.org/10.1038/nature08812} {\bibfield  {journal} {\bibinfo
  {journal} {Nature}\ }\textbf {\bibinfo {volume} {464}},\ \bibinfo {pages}
  {45} (\bibinfo {year} {2010})}\BibitemShut {NoStop}%
\bibitem [{\citenamefont {Nayak}\ \emph {et~al.}(2008)\citenamefont {Nayak},
  \citenamefont {Simon}, \citenamefont {Stern}, \citenamefont {Freedman},\ and\
  \citenamefont {Das~Sarma}}]{TopoQuantumCompRevModPhys}%
  \BibitemOpen
  \bibfield  {author} {\bibinfo {author} {\bibfnamefont {C.}~\bibnamefont
  {Nayak}}, \bibinfo {author} {\bibfnamefont {S.~H.}\ \bibnamefont {Simon}},
  \bibinfo {author} {\bibfnamefont {A.}~\bibnamefont {Stern}}, \bibinfo
  {author} {\bibfnamefont {M.}~\bibnamefont {Freedman}},\ and\ \bibinfo
  {author} {\bibfnamefont {S.}~\bibnamefont {Das~Sarma}},\ }\bibfield  {title}
  {\bibinfo {title} {{Non-Abelian} anyons and topological quantum
  computation},\ }\href {https://doi.org/10.1103/RevModPhys.80.1083} {\bibfield
   {journal} {\bibinfo  {journal} {Rev. Mod. Phys.}\ }\textbf {\bibinfo
  {volume} {80}},\ \bibinfo {pages} {1083} (\bibinfo {year}
  {2008})}\BibitemShut {NoStop}%
\bibitem [{\citenamefont {Bernevig}\ and\ \citenamefont
  {Hughes}(2013)}]{Bernevig2013}%
  \BibitemOpen
  \bibfield  {author} {\bibinfo {author} {\bibfnamefont {B.~A.}\ \bibnamefont
  {Bernevig}}\ and\ \bibinfo {author} {\bibfnamefont {T.~L.}\ \bibnamefont
  {Hughes}},\ }\href@noop {} {\emph {\bibinfo {title} {Topological Insulators
  and Topological Superconductors}}}\ (\bibinfo  {publisher} {Princeton
  University Press},\ \bibinfo {address} {Princeton, NJ},\ \bibinfo {year}
  {2013})\BibitemShut {NoStop}%
\bibitem [{\citenamefont {M{\'e}nard}\ \emph {et~al.}(2019)\citenamefont
  {M{\'e}nard}, \citenamefont {Mesaros}, \citenamefont {Brun}, \citenamefont
  {Debontridder}, \citenamefont {Roditchev}, \citenamefont {Simon},\ and\
  \citenamefont {Cren}}]{menard2019isolated}%
  \BibitemOpen
  \bibfield  {author} {\bibinfo {author} {\bibfnamefont {G.~C.}\ \bibnamefont
  {M{\'e}nard}}, \bibinfo {author} {\bibfnamefont {A.}~\bibnamefont {Mesaros}},
  \bibinfo {author} {\bibfnamefont {C.}~\bibnamefont {Brun}}, \bibinfo {author}
  {\bibfnamefont {F.}~\bibnamefont {Debontridder}}, \bibinfo {author}
  {\bibfnamefont {D.}~\bibnamefont {Roditchev}}, \bibinfo {author}
  {\bibfnamefont {P.}~\bibnamefont {Simon}},\ and\ \bibinfo {author}
  {\bibfnamefont {T.}~\bibnamefont {Cren}},\ }\bibfield  {title} {\bibinfo
  {title} {Isolated pairs of {Majorana} zero modes in a disordered
  superconducting lead monolayer},\ }\href
  {https://doi.org/10.1038/s41467-019-10397-5} {\bibfield  {journal} {\bibinfo
  {journal} {Nat. Commun.}\ }\textbf {\bibinfo {volume} {10}},\ \bibinfo
  {pages} {2587} (\bibinfo {year} {2019})}\BibitemShut {NoStop}%
\bibitem [{\citenamefont {Aghaee}\ \emph {et~al.}(2022)\citenamefont {Aghaee},
  \citenamefont {Akkala}, \citenamefont {Alam}, \citenamefont {Ali},
  \citenamefont {Alcaraz~Ramirez}, \citenamefont {Andrzejczuk}, \citenamefont
  {Antipov}, \citenamefont {Astafev}, \citenamefont {Bauer}, \citenamefont
  {Becker} \emph {et~al.}}]{MicrosoftTGP}%
  \BibitemOpen
  \bibfield  {author} {\bibinfo {author} {\bibfnamefont {M.}~\bibnamefont
  {Aghaee}}, \bibinfo {author} {\bibfnamefont {A.}~\bibnamefont {Akkala}},
  \bibinfo {author} {\bibfnamefont {Z.}~\bibnamefont {Alam}}, \bibinfo {author}
  {\bibfnamefont {R.}~\bibnamefont {Ali}}, \bibinfo {author} {\bibfnamefont
  {A.}~\bibnamefont {Alcaraz~Ramirez}}, \bibinfo {author} {\bibfnamefont
  {M.}~\bibnamefont {Andrzejczuk}}, \bibinfo {author} {\bibfnamefont {A.~E.}\
  \bibnamefont {Antipov}}, \bibinfo {author} {\bibfnamefont {M.}~\bibnamefont
  {Astafev}}, \bibinfo {author} {\bibfnamefont {B.}~\bibnamefont {Bauer}},
  \bibinfo {author} {\bibfnamefont {J.}~\bibnamefont {Becker}}, \emph
  {et~al.},\ }\bibfield  {title} {\bibinfo {title} {{InAs-Al Hybrid Devices
  Passing the Topological Gap Protocol}},\ }\href
  {https://arxiv.org/abs/2207.02472} {\bibfield  {journal} {\bibinfo  {journal}
  {arXiv:2207.02472}\ } (\bibinfo {year} {2022})}\BibitemShut {NoStop}%
\bibitem [{\citenamefont {Sato}\ and\ \citenamefont
  {Ando}(2017)}]{TopoSCrevSato}%
  \BibitemOpen
  \bibfield  {author} {\bibinfo {author} {\bibfnamefont {M.}~\bibnamefont
  {Sato}}\ and\ \bibinfo {author} {\bibfnamefont {Y.}~\bibnamefont {Ando}},\
  }\bibfield  {title} {\bibinfo {title} {Topological superconductors: a
  review},\ }\href {https://doi.org/10.1088/1361-6633/aa6ac7} {\bibfield
  {journal} {\bibinfo  {journal} {Rep. Prog. Phys.}\ }\textbf {\bibinfo
  {volume} {80}},\ \bibinfo {pages} {076501} (\bibinfo {year}
  {2017})}\BibitemShut {NoStop}%
\bibitem [{\citenamefont {Zlotnikov}\ \emph {et~al.}(2021)\citenamefont
  {Zlotnikov}, \citenamefont {Shustin},\ and\ \citenamefont
  {Fedoseev}}]{TopoSCandSkRev}%
  \BibitemOpen
  \bibfield  {author} {\bibinfo {author} {\bibfnamefont {A.~O.}\ \bibnamefont
  {Zlotnikov}}, \bibinfo {author} {\bibfnamefont {M.~S.}\ \bibnamefont
  {Shustin}},\ and\ \bibinfo {author} {\bibfnamefont {A.~D.}\ \bibnamefont
  {Fedoseev}},\ }\bibfield  {title} {\bibinfo {title} {{Aspects of Topological
  Superconductivity in 2D Systems: Noncollinear Magnetism, Skyrmions, and
  Higher-order Topology}},\ }\href {https://doi.org/10.1007/s10948-021-06029-z}
  {\bibfield  {journal} {\bibinfo  {journal} {J. Supercond. Nov. Magn.}\
  }\textbf {\bibinfo {volume} {34}},\ \bibinfo {pages} {3053} (\bibinfo {year}
  {2021})}\BibitemShut {NoStop}%
\bibitem [{\citenamefont {Nakosai}\ \emph {et~al.}(2013)\citenamefont
  {Nakosai}, \citenamefont {Tanaka},\ and\ \citenamefont
  {Nagaosa}}]{SkTopoSCNagaosa}%
  \BibitemOpen
  \bibfield  {author} {\bibinfo {author} {\bibfnamefont {S.}~\bibnamefont
  {Nakosai}}, \bibinfo {author} {\bibfnamefont {Y.}~\bibnamefont {Tanaka}},\
  and\ \bibinfo {author} {\bibfnamefont {N.}~\bibnamefont {Nagaosa}},\
  }\bibfield  {title} {\bibinfo {title} {Two-dimensional $p$-wave
  superconducting states with magnetic moments on a conventional $s$-wave
  superconductor},\ }\href {https://doi.org/10.1103/PhysRevB.88.180503}
  {\bibfield  {journal} {\bibinfo  {journal} {Phys. Rev. B}\ }\textbf {\bibinfo
  {volume} {88}},\ \bibinfo {pages} {180503(R)} (\bibinfo {year}
  {2013})}\BibitemShut {NoStop}%
\bibitem [{\citenamefont {Chen}\ and\ \citenamefont
  {Schnyder}(2015)}]{SkTopoSCMajoranaChen}%
  \BibitemOpen
  \bibfield  {author} {\bibinfo {author} {\bibfnamefont {W.}~\bibnamefont
  {Chen}}\ and\ \bibinfo {author} {\bibfnamefont {A.~P.}\ \bibnamefont
  {Schnyder}},\ }\bibfield  {title} {\bibinfo {title} {Majorana edge states in
  superconductor-noncollinear magnet interfaces},\ }\href
  {https://doi.org/10.1103/PhysRevB.92.214502} {\bibfield  {journal} {\bibinfo
  {journal} {Phys. Rev. B}\ }\textbf {\bibinfo {volume} {92}},\ \bibinfo
  {pages} {214502} (\bibinfo {year} {2015})}\BibitemShut {NoStop}%
\bibitem [{\citenamefont {Yang}\ \emph {et~al.}(2016)\citenamefont {Yang},
  \citenamefont {Stano}, \citenamefont {Klinovaja},\ and\ \citenamefont
  {Loss}}]{SkTopoSCMajoranaLossFM}%
  \BibitemOpen
  \bibfield  {author} {\bibinfo {author} {\bibfnamefont {G.}~\bibnamefont
  {Yang}}, \bibinfo {author} {\bibfnamefont {P.}~\bibnamefont {Stano}},
  \bibinfo {author} {\bibfnamefont {J.}~\bibnamefont {Klinovaja}},\ and\
  \bibinfo {author} {\bibfnamefont {D.}~\bibnamefont {Loss}},\ }\bibfield
  {title} {\bibinfo {title} {Majorana bound states in magnetic skyrmions},\
  }\href {https://doi.org/10.1103/PhysRevB.93.224505} {\bibfield  {journal}
  {\bibinfo  {journal} {Phys. Rev. B}\ }\textbf {\bibinfo {volume} {93}},\
  \bibinfo {pages} {224505} (\bibinfo {year} {2016})}\BibitemShut {NoStop}%
\bibitem [{\citenamefont {Rex}\ \emph {et~al.}(2019)\citenamefont {Rex},
  \citenamefont {Gornyi},\ and\ \citenamefont {Mirlin}}]{SkMajoranaRex}%
  \BibitemOpen
  \bibfield  {author} {\bibinfo {author} {\bibfnamefont {S.}~\bibnamefont
  {Rex}}, \bibinfo {author} {\bibfnamefont {I.~V.}\ \bibnamefont {Gornyi}},\
  and\ \bibinfo {author} {\bibfnamefont {A.~D.}\ \bibnamefont {Mirlin}},\
  }\bibfield  {title} {\bibinfo {title} {Majorana bound states in magnetic
  skyrmions imposed onto a superconductor},\ }\href
  {https://doi.org/10.1103/PhysRevB.100.064504} {\bibfield  {journal} {\bibinfo
   {journal} {Phys. Rev. B}\ }\textbf {\bibinfo {volume} {100}},\ \bibinfo
  {pages} {064504} (\bibinfo {year} {2019})}\BibitemShut {NoStop}%
\bibitem [{\citenamefont {D\'{\i}az}\ \emph {et~al.}(2021)\citenamefont
  {D\'{\i}az}, \citenamefont {Klinovaja}, \citenamefont {Loss},\ and\
  \citenamefont {Hoffman}}]{SkTopoSCMajoranaLossAFM}%
  \BibitemOpen
  \bibfield  {author} {\bibinfo {author} {\bibfnamefont {S.~A.}\ \bibnamefont
  {D\'{\i}az}}, \bibinfo {author} {\bibfnamefont {J.}~\bibnamefont
  {Klinovaja}}, \bibinfo {author} {\bibfnamefont {D.}~\bibnamefont {Loss}},\
  and\ \bibinfo {author} {\bibfnamefont {S.}~\bibnamefont {Hoffman}},\
  }\bibfield  {title} {\bibinfo {title} {Majorana bound states induced by
  antiferromagnetic skyrmion textures},\ }\href
  {https://doi.org/10.1103/PhysRevB.104.214501} {\bibfield  {journal} {\bibinfo
   {journal} {Phys. Rev. B}\ }\textbf {\bibinfo {volume} {104}},\ \bibinfo
  {pages} {214501} (\bibinfo {year} {2021})}\BibitemShut {NoStop}%
\bibitem [{\citenamefont {Mohanta}\ \emph {et~al.}(2021)\citenamefont
  {Mohanta}, \citenamefont {Okamoto},\ and\ \citenamefont
  {Dagotto}}]{SkTopoSCMajoranaDagotto}%
  \BibitemOpen
  \bibfield  {author} {\bibinfo {author} {\bibfnamefont {N.}~\bibnamefont
  {Mohanta}}, \bibinfo {author} {\bibfnamefont {S.}~\bibnamefont {Okamoto}},\
  and\ \bibinfo {author} {\bibfnamefont {E.}~\bibnamefont {Dagotto}},\
  }\bibfield  {title} {\bibinfo {title} {{Skyrmion control of Majorana states
  in planar Josephson junctions}},\ }\href
  {https://doi.org/10.1038/s42005-021-00666-5} {\bibfield  {journal} {\bibinfo
  {journal} {Commun. Phys.}\ }\textbf {\bibinfo {volume} {4}},\ \bibinfo
  {pages} {163} (\bibinfo {year} {2021})}\BibitemShut {NoStop}%
\bibitem [{\citenamefont {Garnier}\ \emph {et~al.}(2019)\citenamefont
  {Garnier}, \citenamefont {Mesaros},\ and\ \citenamefont
  {Simon}}]{SkTopoSCgarnier}%
  \BibitemOpen
  \bibfield  {author} {\bibinfo {author} {\bibfnamefont {M.}~\bibnamefont
  {Garnier}}, \bibinfo {author} {\bibfnamefont {A.}~\bibnamefont {Mesaros}},\
  and\ \bibinfo {author} {\bibfnamefont {P.}~\bibnamefont {Simon}},\ }\bibfield
   {title} {\bibinfo {title} {Topological superconductivity with deformable
  magnetic skyrmions},\ }\href {https://doi.org/10.1038/s42005-019-0226-5}
  {\bibfield  {journal} {\bibinfo  {journal} {Commun. Phys.}\ }\textbf
  {\bibinfo {volume} {2}},\ \bibinfo {pages} {126} (\bibinfo {year}
  {2019})}\BibitemShut {NoStop}%
\bibitem [{\citenamefont {Mascot}\ \emph {et~al.}(2021)\citenamefont {Mascot},
  \citenamefont {Bedow}, \citenamefont {Graham}, \citenamefont {Rachel},\ and\
  \citenamefont {Morr}}]{SkTopoSCMajoranaMascot}%
  \BibitemOpen
  \bibfield  {author} {\bibinfo {author} {\bibfnamefont {E.}~\bibnamefont
  {Mascot}}, \bibinfo {author} {\bibfnamefont {J.}~\bibnamefont {Bedow}},
  \bibinfo {author} {\bibfnamefont {M.}~\bibnamefont {Graham}}, \bibinfo
  {author} {\bibfnamefont {S.}~\bibnamefont {Rachel}},\ and\ \bibinfo {author}
  {\bibfnamefont {D.~K.}\ \bibnamefont {Morr}},\ }\bibfield  {title} {\bibinfo
  {title} {Topological superconductivity in skyrmion lattices},\ }\href
  {https://doi.org/10.1038/s41535-020-00299-x} {\bibfield  {journal} {\bibinfo
  {journal} {npj Quantum Mater.}\ }\textbf {\bibinfo {volume} {6}},\ \bibinfo
  {pages} {6} (\bibinfo {year} {2021})}\BibitemShut {NoStop}%
\bibitem [{\citenamefont {Petrovi\ifmmode~\acute{c}\else \'{c}\fi{}}\ \emph
  {et~al.}(2021)\citenamefont {Petrovi\ifmmode~\acute{c}\else \'{c}\fi{}},
  \citenamefont {Raju}, \citenamefont {Tee}, \citenamefont {Louat},
  \citenamefont {Maggio-Aprile}, \citenamefont {Menezes}, \citenamefont
  {Wyszy\ifmmode~\acute{n}\else \'{n}\fi{}ski}, \citenamefont {Duong},
  \citenamefont {Reznikov}, \citenamefont {Renner}, \citenamefont {Milo\ifmmode
  \check{s}\else \v{s}\fi{}evi\ifmmode~\acute{c}\else \'{c}\fi{}},\ and\
  \citenamefont {Panagopoulos}}]{ExpSkHeterostructure}%
  \BibitemOpen
  \bibfield  {author} {\bibinfo {author} {\bibfnamefont {A.~P.}\ \bibnamefont
  {Petrovi\ifmmode~\acute{c}\else \'{c}\fi{}}}, \bibinfo {author}
  {\bibfnamefont {M.}~\bibnamefont {Raju}}, \bibinfo {author} {\bibfnamefont
  {X.~Y.}\ \bibnamefont {Tee}}, \bibinfo {author} {\bibfnamefont
  {A.}~\bibnamefont {Louat}}, \bibinfo {author} {\bibfnamefont
  {I.}~\bibnamefont {Maggio-Aprile}}, \bibinfo {author} {\bibfnamefont {R.~M.}\
  \bibnamefont {Menezes}}, \bibinfo {author} {\bibfnamefont {M.~J.}\
  \bibnamefont {Wyszy\ifmmode~\acute{n}\else \'{n}\fi{}ski}}, \bibinfo {author}
  {\bibfnamefont {N.~K.}\ \bibnamefont {Duong}}, \bibinfo {author}
  {\bibfnamefont {M.}~\bibnamefont {Reznikov}}, \bibinfo {author}
  {\bibfnamefont {C.}~\bibnamefont {Renner}}, \bibinfo {author} {\bibfnamefont
  {M.~V.}\ \bibnamefont {Milo\ifmmode \check{s}\else
  \v{s}\fi{}evi\ifmmode~\acute{c}\else \'{c}\fi{}}},\ and\ \bibinfo {author}
  {\bibfnamefont {C.}~\bibnamefont {Panagopoulos}},\ }\bibfield  {title}
  {\bibinfo {title} {{Skyrmion-(Anti)Vortex Coupling in a Chiral
  Magnet-Superconductor Heterostructure}},\ }\href
  {https://doi.org/10.1103/PhysRevLett.126.117205} {\bibfield  {journal}
  {\bibinfo  {journal} {Phys. Rev. Lett.}\ }\textbf {\bibinfo {volume} {126}},\
  \bibinfo {pages} {117205} (\bibinfo {year} {2021})}\BibitemShut {NoStop}%
\bibitem [{\citenamefont {Nagaosa}\ and\ \citenamefont
  {Tokura}(2013)}]{nagaosaRev}%
  \BibitemOpen
  \bibfield  {author} {\bibinfo {author} {\bibfnamefont {N.}~\bibnamefont
  {Nagaosa}}\ and\ \bibinfo {author} {\bibfnamefont {Y.}~\bibnamefont
  {Tokura}},\ }\bibfield  {title} {\bibinfo {title} {Topological properties and
  dynamics of magnetic skyrmions},\ }\href
  {https://doi.org/10.1038/nnano.2013.243} {\bibfield  {journal} {\bibinfo
  {journal} {Nat. Nanotechnol.}\ }\textbf {\bibinfo {volume} {8}},\ \bibinfo
  {pages} {899} (\bibinfo {year} {2013})}\BibitemShut {NoStop}%
\bibitem [{\citenamefont {Finocchio}\ \emph {et~al.}(2016)\citenamefont
  {Finocchio}, \citenamefont {Büttner}, \citenamefont {Tomasello},
  \citenamefont {Carpentieri},\ and\ \citenamefont {Kläui}}]{KlauiRev2016}%
  \BibitemOpen
  \bibfield  {author} {\bibinfo {author} {\bibfnamefont {G.}~\bibnamefont
  {Finocchio}}, \bibinfo {author} {\bibfnamefont {F.}~\bibnamefont {Büttner}},
  \bibinfo {author} {\bibfnamefont {R.}~\bibnamefont {Tomasello}}, \bibinfo
  {author} {\bibfnamefont {M.}~\bibnamefont {Carpentieri}},\ and\ \bibinfo
  {author} {\bibfnamefont {M.}~\bibnamefont {Kläui}},\ }\bibfield  {title}
  {\bibinfo {title} {Magnetic skyrmions: from fundamental to applications},\
  }\href {https://doi.org/10.1088/0022-3727/49/42/423001} {\bibfield  {journal}
  {\bibinfo  {journal} {J. Phys. D: Appl. Phys.}\ }\textbf {\bibinfo {volume}
  {49}},\ \bibinfo {pages} {423001} (\bibinfo {year} {2016})}\BibitemShut
  {NoStop}%
\bibitem [{\citenamefont {Kargarian}\ \emph {et~al.}(2016)\citenamefont
  {Kargarian}, \citenamefont {Efimkin},\ and\ \citenamefont
  {Galitski}}]{KargarianFMTI}%
  \BibitemOpen
  \bibfield  {author} {\bibinfo {author} {\bibfnamefont {M.}~\bibnamefont
  {Kargarian}}, \bibinfo {author} {\bibfnamefont {D.~K.}\ \bibnamefont
  {Efimkin}},\ and\ \bibinfo {author} {\bibfnamefont {V.}~\bibnamefont
  {Galitski}},\ }\bibfield  {title} {\bibinfo {title} {{Amperean Pairing at the
  Surface of Topological Insulators}},\ }\href
  {https://doi.org/10.1103/PhysRevLett.117.076806} {\bibfield  {journal}
  {\bibinfo  {journal} {Phys. Rev. Lett.}\ }\textbf {\bibinfo {volume} {117}},\
  \bibinfo {pages} {076806} (\bibinfo {year} {2016})}\BibitemShut {NoStop}%
\bibitem [{\citenamefont {Hugdal}\ \emph {et~al.}(2018)\citenamefont {Hugdal},
  \citenamefont {Rex}, \citenamefont {Nogueira},\ and\ \citenamefont
  {Sudb{\o}}}]{HugdalTIFMAFM}%
  \BibitemOpen
  \bibfield  {author} {\bibinfo {author} {\bibfnamefont {H.~G.}\ \bibnamefont
  {Hugdal}}, \bibinfo {author} {\bibfnamefont {S.}~\bibnamefont {Rex}},
  \bibinfo {author} {\bibfnamefont {F.~S.}\ \bibnamefont {Nogueira}},\ and\
  \bibinfo {author} {\bibfnamefont {A.}~\bibnamefont {Sudb{\o}}},\ }\bibfield
  {title} {\bibinfo {title} {{Magnon-induced superconductivity in a topological
  insulator coupled to ferromagnetic and antiferromagnetic insulators}},\
  }\href {https://doi.org/10.1103/PhysRevB.97.195438} {\bibfield  {journal}
  {\bibinfo  {journal} {Phys. Rev. B}\ }\textbf {\bibinfo {volume} {97}},\
  \bibinfo {pages} {195438} (\bibinfo {year} {2018})}\BibitemShut {NoStop}%
\bibitem [{\citenamefont {Erlandsen}\ \emph {et~al.}(2020)\citenamefont
  {Erlandsen}, \citenamefont {Brataas},\ and\ \citenamefont
  {Sudb\o{}}}]{EirikTIAFM}%
  \BibitemOpen
  \bibfield  {author} {\bibinfo {author} {\bibfnamefont {E.}~\bibnamefont
  {Erlandsen}}, \bibinfo {author} {\bibfnamefont {A.}~\bibnamefont {Brataas}},\
  and\ \bibinfo {author} {\bibfnamefont {A.}~\bibnamefont {Sudb\o{}}},\
  }\bibfield  {title} {\bibinfo {title} {Magnon-mediated superconductivity on
  the surface of a topological insulator},\ }\href
  {https://doi.org/10.1103/PhysRevB.101.094503} {\bibfield  {journal} {\bibinfo
   {journal} {Phys. Rev. B}\ }\textbf {\bibinfo {volume} {101}},\ \bibinfo
  {pages} {094503} (\bibinfo {year} {2020})}\BibitemShut {NoStop}%
\bibitem [{\citenamefont {Rohling}\ \emph {et~al.}(2018)\citenamefont
  {Rohling}, \citenamefont {Fj\ae{}rbu},\ and\ \citenamefont
  {Brataas}}]{ArneFMNM}%
  \BibitemOpen
  \bibfield  {author} {\bibinfo {author} {\bibfnamefont {N.}~\bibnamefont
  {Rohling}}, \bibinfo {author} {\bibfnamefont {E.~L.}\ \bibnamefont
  {Fj\ae{}rbu}},\ and\ \bibinfo {author} {\bibfnamefont {A.}~\bibnamefont
  {Brataas}},\ }\bibfield  {title} {\bibinfo {title} {Superconductivity induced
  by interfacial coupling to magnons},\ }\href
  {https://doi.org/10.1103/PhysRevB.97.115401} {\bibfield  {journal} {\bibinfo
  {journal} {Phys. Rev. B}\ }\textbf {\bibinfo {volume} {97}},\ \bibinfo
  {pages} {115401} (\bibinfo {year} {2018})}\BibitemShut {NoStop}%
\bibitem [{\citenamefont {Fj\ae{}rbu}\ \emph {et~al.}(2019)\citenamefont
  {Fj\ae{}rbu}, \citenamefont {Rohling},\ and\ \citenamefont
  {Brataas}}]{ArneAFMNM_Umklapp}%
  \BibitemOpen
  \bibfield  {author} {\bibinfo {author} {\bibfnamefont {E.~L.}\ \bibnamefont
  {Fj\ae{}rbu}}, \bibinfo {author} {\bibfnamefont {N.}~\bibnamefont
  {Rohling}},\ and\ \bibinfo {author} {\bibfnamefont {A.}~\bibnamefont
  {Brataas}},\ }\bibfield  {title} {\bibinfo {title} {Superconductivity at
  metal-antiferromagnetic insulator interfaces},\ }\href
  {https://doi.org/10.1103/PhysRevB.100.125432} {\bibfield  {journal} {\bibinfo
   {journal} {Phys. Rev. B}\ }\textbf {\bibinfo {volume} {100}},\ \bibinfo
  {pages} {125432} (\bibinfo {year} {2019})}\BibitemShut {NoStop}%
\bibitem [{\citenamefont {Erlandsen}\ \emph {et~al.}(2019)\citenamefont
  {Erlandsen}, \citenamefont {Kamra}, \citenamefont {Brataas},\ and\
  \citenamefont {Sudb\o{}}}]{EirikNMAFM}%
  \BibitemOpen
  \bibfield  {author} {\bibinfo {author} {\bibfnamefont {E.}~\bibnamefont
  {Erlandsen}}, \bibinfo {author} {\bibfnamefont {A.}~\bibnamefont {Kamra}},
  \bibinfo {author} {\bibfnamefont {A.}~\bibnamefont {Brataas}},\ and\ \bibinfo
  {author} {\bibfnamefont {A.}~\bibnamefont {Sudb\o{}}},\ }\bibfield  {title}
  {\bibinfo {title} {Enhancement of superconductivity mediated by
  antiferromagnetic squeezed magnons},\ }\href
  {https://doi.org/10.1103/PhysRevB.100.100503} {\bibfield  {journal} {\bibinfo
   {journal} {Phys. Rev. B}\ }\textbf {\bibinfo {volume} {100}},\ \bibinfo
  {pages} {100503(R)} (\bibinfo {year} {2019})}\BibitemShut {NoStop}%
\bibitem [{\citenamefont {Thingstad}\ \emph {et~al.}(2021)\citenamefont
  {Thingstad}, \citenamefont {Erlandsen},\ and\ \citenamefont
  {Sudb\o{}}}]{EirikEliashberg}%
  \BibitemOpen
  \bibfield  {author} {\bibinfo {author} {\bibfnamefont {E.}~\bibnamefont
  {Thingstad}}, \bibinfo {author} {\bibfnamefont {E.}~\bibnamefont
  {Erlandsen}},\ and\ \bibinfo {author} {\bibfnamefont {A.}~\bibnamefont
  {Sudb\o{}}},\ }\bibfield  {title} {\bibinfo {title} {Eliashberg study of
  superconductivity induced by interfacial coupling to antiferromagnets},\
  }\href {https://doi.org/10.1103/PhysRevB.104.014508} {\bibfield  {journal}
  {\bibinfo  {journal} {Phys. Rev. B}\ }\textbf {\bibinfo {volume} {104}},\
  \bibinfo {pages} {014508} (\bibinfo {year} {2021})}\BibitemShut {NoStop}%
\bibitem [{\citenamefont {Gong}\ \emph {et~al.}(2017)\citenamefont {Gong},
  \citenamefont {Kargarian}, \citenamefont {Stern}, \citenamefont {Yue},
  \citenamefont {Zhou}, \citenamefont {Jin}, \citenamefont {Galitski},
  \citenamefont {Yakovenko},\ and\ \citenamefont
  {Xia}}]{ExpMagnonInducedHeterostruct}%
  \BibitemOpen
  \bibfield  {author} {\bibinfo {author} {\bibfnamefont {X.}~\bibnamefont
  {Gong}}, \bibinfo {author} {\bibfnamefont {M.}~\bibnamefont {Kargarian}},
  \bibinfo {author} {\bibfnamefont {A.}~\bibnamefont {Stern}}, \bibinfo
  {author} {\bibfnamefont {D.}~\bibnamefont {Yue}}, \bibinfo {author}
  {\bibfnamefont {H.}~\bibnamefont {Zhou}}, \bibinfo {author} {\bibfnamefont
  {X.}~\bibnamefont {Jin}}, \bibinfo {author} {\bibfnamefont {V.~M.}\
  \bibnamefont {Galitski}}, \bibinfo {author} {\bibfnamefont {V.~M.}\
  \bibnamefont {Yakovenko}},\ and\ \bibinfo {author} {\bibfnamefont
  {J.}~\bibnamefont {Xia}},\ }\bibfield  {title} {\bibinfo {title}
  {Time-reversal symmetry-breaking superconductivity in epitaxial
  bismuth/nickel bilayers},\ }\href {https://doi.org/10.1126/sciadv.1602579}
  {\bibfield  {journal} {\bibinfo  {journal} {Sci. Adv.}\ }\textbf {\bibinfo
  {volume} {3}},\ \bibinfo {pages} {e1602579} (\bibinfo {year}
  {2017})}\BibitemShut {NoStop}%
\bibitem [{\citenamefont {M\ae{}land}\ and\ \citenamefont
  {Sudb\o{}}(2022{\natexlab{a}})}]{QSkOP}%
  \BibitemOpen
  \bibfield  {author} {\bibinfo {author} {\bibfnamefont {K.}~\bibnamefont
  {M\ae{}land}}\ and\ \bibinfo {author} {\bibfnamefont {A.}~\bibnamefont
  {Sudb\o{}}},\ }\bibfield  {title} {\bibinfo {title} {Quantum fluctuations in
  the order parameter of quantum skyrmion crystals},\ }\href
  {https://doi.org/10.1103/PhysRevB.105.224416} {\bibfield  {journal} {\bibinfo
   {journal} {Phys. Rev. B}\ }\textbf {\bibinfo {volume} {105}},\ \bibinfo
  {pages} {224416} (\bibinfo {year} {2022}{\natexlab{a}})}\BibitemShut
  {NoStop}%
\bibitem [{\citenamefont {M\ae{}land}\ and\ \citenamefont
  {Sudb\o{}}(2022{\natexlab{b}})}]{QSkQTPT}%
  \BibitemOpen
  \bibfield  {author} {\bibinfo {author} {\bibfnamefont {K.}~\bibnamefont
  {M\ae{}land}}\ and\ \bibinfo {author} {\bibfnamefont {A.}~\bibnamefont
  {Sudb\o{}}},\ }\bibfield  {title} {\bibinfo {title} {Quantum topological
  phase transitions in skyrmion crystals},\ }\href
  {https://doi.org/10.1103/PhysRevResearch.4.L032025} {\bibfield  {journal}
  {\bibinfo  {journal} {Phys. Rev. Res.}\ }\textbf {\bibinfo {volume} {4}},\
  \bibinfo {pages} {L032025} (\bibinfo {year}
  {2022}{\natexlab{b}})}\BibitemShut {NoStop}%
\bibitem [{Sup()}]{Suppl}%
  \BibitemOpen
  \href@noop {} {}\bibinfo {note} {See Supplemental Material on page
  \pageref{sec:Suppl} for (i) the model for the magnetic monolayer, (ii) a
  derivation of the effective electron-electron interaction, (iii) derivations
  of the gap equations, (iv) plots of the gap functions, and (v) details of the
  calculation of the bulk topological invariant, which includes
  Refs.~\cite{H4PRB, SkHallBeyond, HProtation_2009, dosSantosPRB,
  TopoMagnonSC}.}\BibitemShut {Stop}%
\bibitem [{\citenamefont {Albaridy}\ \emph {et~al.}(2020)\citenamefont
  {Albaridy}, \citenamefont {Manchon},\ and\ \citenamefont
  {Schwingenschlögl}}]{TuneK}%
  \BibitemOpen
  \bibfield  {author} {\bibinfo {author} {\bibfnamefont {R.}~\bibnamefont
  {Albaridy}}, \bibinfo {author} {\bibfnamefont {A.}~\bibnamefont {Manchon}},\
  and\ \bibinfo {author} {\bibfnamefont {U.}~\bibnamefont
  {Schwingenschlögl}},\ }\bibfield  {title} {\bibinfo {title} {{Tunable
  magnetic anisotropy in Cr{\textendash}trihalide Janus monolayers}},\ }\href
  {https://doi.org/10.1088/1361-648x/ab8986} {\bibfield  {journal} {\bibinfo
  {journal} {J. Phys.: Condens. Matter}\ }\textbf {\bibinfo {volume} {32}},\
  \bibinfo {pages} {355702} (\bibinfo {year} {2020})}\BibitemShut {NoStop}%
\bibitem [{\citenamefont {Webster}\ and\ \citenamefont
  {Yan}(2018)}]{TuneK_PRB}%
  \BibitemOpen
  \bibfield  {author} {\bibinfo {author} {\bibfnamefont {L.}~\bibnamefont
  {Webster}}\ and\ \bibinfo {author} {\bibfnamefont {J.-A.}\ \bibnamefont
  {Yan}},\ }\bibfield  {title} {\bibinfo {title} {{Strain-tunable magnetic
  anisotropy in monolayer ${\text{CrCl}}_{3}$, ${\text{CrBr}}_{3}$, and
  ${\text{CrI}}_{3}$}},\ }\href {https://doi.org/10.1103/PhysRevB.98.144411}
  {\bibfield  {journal} {\bibinfo  {journal} {Phys. Rev. B}\ }\textbf {\bibinfo
  {volume} {98}},\ \bibinfo {pages} {144411} (\bibinfo {year}
  {2018})}\BibitemShut {NoStop}%
\bibitem [{\citenamefont {Heinze}\ \emph {et~al.}(2011)\citenamefont {Heinze},
  \citenamefont {von Bergmann}, \citenamefont {Menzel}, \citenamefont {Brede},
  \citenamefont {Kubetzka}, \citenamefont {Wiesendanger}, \citenamefont
  {Bihlmayer},\ and\ \citenamefont {Bl{\"u}gel}}]{HeinzeSkX}%
  \BibitemOpen
  \bibfield  {author} {\bibinfo {author} {\bibfnamefont {S.}~\bibnamefont
  {Heinze}}, \bibinfo {author} {\bibfnamefont {K.}~\bibnamefont {von
  Bergmann}}, \bibinfo {author} {\bibfnamefont {M.}~\bibnamefont {Menzel}},
  \bibinfo {author} {\bibfnamefont {J.}~\bibnamefont {Brede}}, \bibinfo
  {author} {\bibfnamefont {A.}~\bibnamefont {Kubetzka}}, \bibinfo {author}
  {\bibfnamefont {R.}~\bibnamefont {Wiesendanger}}, \bibinfo {author}
  {\bibfnamefont {G.}~\bibnamefont {Bihlmayer}},\ and\ \bibinfo {author}
  {\bibfnamefont {S.}~\bibnamefont {Bl{\"u}gel}},\ }\bibfield  {title}
  {\bibinfo {title} {Spontaneous atomic-scale magnetic skyrmion lattice in two
  dimensions},\ }\href {https://doi.org/10.1038/nphys2045} {\bibfield
  {journal} {\bibinfo  {journal} {Nat. Phys.}\ }\textbf {\bibinfo {volume}
  {7}},\ \bibinfo {pages} {713} (\bibinfo {year} {2011})}\BibitemShut {NoStop}%
\bibitem [{\citenamefont {Kajiwara}\ \emph {et~al.}(2010)\citenamefont
  {Kajiwara}, \citenamefont {Harii}, \citenamefont {Takahashi}, \citenamefont
  {Ohe}, \citenamefont {Uchida}, \citenamefont {Mizuguchi}, \citenamefont
  {Umezawa}, \citenamefont {Kawai}, \citenamefont {Ando}, \citenamefont
  {Takanashi}, \citenamefont {Maekawa},\ and\ \citenamefont
  {Saitoh}}]{ExpInterfaceExchange}%
  \BibitemOpen
  \bibfield  {author} {\bibinfo {author} {\bibfnamefont {Y.}~\bibnamefont
  {Kajiwara}}, \bibinfo {author} {\bibfnamefont {K.}~\bibnamefont {Harii}},
  \bibinfo {author} {\bibfnamefont {S.}~\bibnamefont {Takahashi}}, \bibinfo
  {author} {\bibfnamefont {J.}~\bibnamefont {Ohe}}, \bibinfo {author}
  {\bibfnamefont {K.}~\bibnamefont {Uchida}}, \bibinfo {author} {\bibfnamefont
  {M.}~\bibnamefont {Mizuguchi}}, \bibinfo {author} {\bibfnamefont
  {H.}~\bibnamefont {Umezawa}}, \bibinfo {author} {\bibfnamefont
  {H.}~\bibnamefont {Kawai}}, \bibinfo {author} {\bibfnamefont
  {K.}~\bibnamefont {Ando}}, \bibinfo {author} {\bibfnamefont {K.}~\bibnamefont
  {Takanashi}}, \bibinfo {author} {\bibfnamefont {S.}~\bibnamefont {Maekawa}},\
  and\ \bibinfo {author} {\bibfnamefont {E.}~\bibnamefont {Saitoh}},\
  }\bibfield  {title} {\bibinfo {title} {Transmission of electrical signals by
  spin-wave interconversion in a magnetic insulator},\ }\href
  {https://doi.org/10.1038/nature08876} {\bibfield  {journal} {\bibinfo
  {journal} {Nature}\ }\textbf {\bibinfo {volume} {464}},\ \bibinfo {pages}
  {262} (\bibinfo {year} {2010})}\BibitemShut {NoStop}%
\bibitem [{\citenamefont {M\ae{}land}\ \emph {et~al.}(2021)\citenamefont
  {M\ae{}land}, \citenamefont {R\o{}st}, \citenamefont {Wells},\ and\
  \citenamefont {Sudb\o{}}}]{Self-energy}%
  \BibitemOpen
  \bibfield  {author} {\bibinfo {author} {\bibfnamefont {K.}~\bibnamefont
  {M\ae{}land}}, \bibinfo {author} {\bibfnamefont {H.~I.}\ \bibnamefont
  {R\o{}st}}, \bibinfo {author} {\bibfnamefont {J.~W.}\ \bibnamefont {Wells}},\
  and\ \bibinfo {author} {\bibfnamefont {A.}~\bibnamefont {Sudb\o{}}},\
  }\bibfield  {title} {\bibinfo {title} {Electron-magnon coupling and
  quasiparticle lifetimes on the surface of a topological insulator},\ }\href
  {https://doi.org/10.1103/PhysRevB.104.125125} {\bibfield  {journal} {\bibinfo
   {journal} {Phys. Rev. B}\ }\textbf {\bibinfo {volume} {104}},\ \bibinfo
  {pages} {125125} (\bibinfo {year} {2021})}\BibitemShut {NoStop}%
\bibitem [{\citenamefont {Colpa}(1978)}]{COLPA}%
  \BibitemOpen
  \bibfield  {author} {\bibinfo {author} {\bibfnamefont {J.~H.~P.}\
  \bibnamefont {Colpa}},\ }\bibfield  {title} {\bibinfo {title}
  {{Diagonalization of the quadratic boson Hamiltonian}},\ }\href
  {https://doi.org/10.1016/0378-4371(78)90160-7} {\bibfield  {journal}
  {\bibinfo  {journal} {Physica}\ }\textbf {\bibinfo {volume} {93A}},\ \bibinfo
  {pages} {327} (\bibinfo {year} {1978})}\BibitemShut {NoStop}%
\bibitem [{\citenamefont {Schrieffer}\ and\ \citenamefont
  {Wolff}(1966)}]{SchriefferWolff}%
  \BibitemOpen
  \bibfield  {author} {\bibinfo {author} {\bibfnamefont {J.~R.}\ \bibnamefont
  {Schrieffer}}\ and\ \bibinfo {author} {\bibfnamefont {P.~A.}\ \bibnamefont
  {Wolff}},\ }\bibfield  {title} {\bibinfo {title} {{Relation between the
  Anderson and Kondo Hamiltonians}},\ }\href
  {https://doi.org/10.1103/PhysRev.149.491} {\bibfield  {journal} {\bibinfo
  {journal} {Phys. Rev.}\ }\textbf {\bibinfo {volume} {149}},\ \bibinfo {pages}
  {491} (\bibinfo {year} {1966})}\BibitemShut {NoStop}%
\bibitem [{\citenamefont {Sigrist}\ and\ \citenamefont {Ueda}(1991)}]{Sigrist}%
  \BibitemOpen
  \bibfield  {author} {\bibinfo {author} {\bibfnamefont {M.}~\bibnamefont
  {Sigrist}}\ and\ \bibinfo {author} {\bibfnamefont {K.}~\bibnamefont {Ueda}},\
  }\bibfield  {title} {\bibinfo {title} {Phenomenological theory of
  unconventional superconductivity},\ }\href
  {https://doi.org/10.1103/RevModPhys.63.239} {\bibfield  {journal} {\bibinfo
  {journal} {Rev. Mod. Phys.}\ }\textbf {\bibinfo {volume} {63}},\ \bibinfo
  {pages} {239} (\bibinfo {year} {1991})}\BibitemShut {NoStop}%
\bibitem [{\citenamefont {Fossheim}\ and\ \citenamefont
  {Sudb{\o}}(2004)}]{SFsuperconductivity}%
  \BibitemOpen
  \bibfield  {author} {\bibinfo {author} {\bibfnamefont {K.}~\bibnamefont
  {Fossheim}}\ and\ \bibinfo {author} {\bibfnamefont {A.}~\bibnamefont
  {Sudb{\o}}},\ }\href@noop {} {\emph {\bibinfo {title} {Superconductivity:
  Physics and Applications}}}\ (\bibinfo  {publisher} {Wiley, Chichester,
  England},\ \bibinfo {year} {2004})\BibitemShut {NoStop}%
\bibitem [{\citenamefont {Mousavi}\ \emph {et~al.}(2012)\citenamefont
  {Mousavi}, \citenamefont {Pask},\ and\ \citenamefont {Sukumar}}]{AdaptQuad}%
  \BibitemOpen
  \bibfield  {author} {\bibinfo {author} {\bibfnamefont {S.~E.}\ \bibnamefont
  {Mousavi}}, \bibinfo {author} {\bibfnamefont {J.~E.}\ \bibnamefont {Pask}},\
  and\ \bibinfo {author} {\bibfnamefont {N.}~\bibnamefont {Sukumar}},\
  }\bibfield  {title} {\bibinfo {title} {Efficient adaptive integration of
  functions with sharp gradients and cusps in $n$-dimensional
  parallelepipeds},\ }\href {https://doi.org/10.1002/nme.4267} {\bibfield
  {journal} {\bibinfo  {journal} {Int. J. Numer. Methods Eng.}\ }\textbf
  {\bibinfo {volume} {91}},\ \bibinfo {pages} {343} (\bibinfo {year}
  {2012})}\BibitemShut {NoStop}%
\bibitem [{\citenamefont {Benestad}(2022)}]{Benestad}%
  \BibitemOpen
  \bibfield  {author} {\bibinfo {author} {\bibfnamefont {J.}~\bibnamefont
  {Benestad}},\ }\bibfield  {title} {\bibinfo {title} {Electron-magnon coupling
  and magnon-induced superconductivity in hybrid structures of metals and
  magnets with non-collinear ground states},\ }\href
  {https://ntnuopen.ntnu.no/ntnu-xmlui/handle/11250/3015244} {\bibfield
  {journal} {\bibinfo  {journal} {Master's thesis, Norwegian University of
  Science and Technology}\ } (\bibinfo {year} {2022})}\BibitemShut {NoStop}%
\bibitem [{\citenamefont {Kittel}(2005)}]{Kittel2005}%
  \BibitemOpen
  \bibfield  {author} {\bibinfo {author} {\bibfnamefont {C.}~\bibnamefont
  {Kittel}},\ }\href@noop {} {\emph {\bibinfo {title} {{Introduction to Solid
  State Physics}}}},\ \bibinfo {edition} {8th}\ ed.\ (\bibinfo  {publisher}
  {Wiley},\ \bibinfo {address} {New York},\ \bibinfo {year} {2005})\BibitemShut
  {NoStop}%
\bibitem [{\citenamefont {MacDonald}\ \emph {et~al.}(1988)\citenamefont
  {MacDonald}, \citenamefont {Girvin},\ and\ \citenamefont {Yoshioka}}]{H4PRB}%
  \BibitemOpen
  \bibfield  {author} {\bibinfo {author} {\bibfnamefont {A.~H.}\ \bibnamefont
  {MacDonald}}, \bibinfo {author} {\bibfnamefont {S.~M.}\ \bibnamefont
  {Girvin}},\ and\ \bibinfo {author} {\bibfnamefont {D.}~\bibnamefont
  {Yoshioka}},\ }\bibfield  {title} {\bibinfo {title} {{$t/U$ expansion for the
  Hubbard model}},\ }\href {https://doi.org/10.1103/PhysRevB.37.9753}
  {\bibfield  {journal} {\bibinfo  {journal} {Phys. Rev. B}\ }\textbf {\bibinfo
  {volume} {37}},\ \bibinfo {pages} {9753} (\bibinfo {year}
  {1988})}\BibitemShut {NoStop}%
\bibitem [{\citenamefont {G{\"o}bel}\ \emph {et~al.}(2021)\citenamefont
  {G{\"o}bel}, \citenamefont {Mertig},\ and\ \citenamefont
  {Tretiakov}}]{SkHallBeyond}%
  \BibitemOpen
  \bibfield  {author} {\bibinfo {author} {\bibfnamefont {B.}~\bibnamefont
  {G{\"o}bel}}, \bibinfo {author} {\bibfnamefont {I.}~\bibnamefont {Mertig}},\
  and\ \bibinfo {author} {\bibfnamefont {O.~A.}\ \bibnamefont {Tretiakov}},\
  }\bibfield  {title} {\bibinfo {title} {Beyond skyrmions: Review and
  perspectives of alternative magnetic quasiparticles},\ }\href
  {https://doi.org/10.1016/j.physrep.2020.10.001} {\bibfield  {journal}
  {\bibinfo  {journal} {Phys. Rep.}\ }\textbf {\bibinfo {volume} {895}},\
  \bibinfo {pages} {1} (\bibinfo {year} {2021})}\BibitemShut {NoStop}%
\bibitem [{\citenamefont {Haraldsen}\ and\ \citenamefont
  {Fishman}(2009)}]{HProtation_2009}%
  \BibitemOpen
  \bibfield  {author} {\bibinfo {author} {\bibfnamefont {J.~T.}\ \bibnamefont
  {Haraldsen}}\ and\ \bibinfo {author} {\bibfnamefont {R.~S.}\ \bibnamefont
  {Fishman}},\ }\bibfield  {title} {\bibinfo {title} {{Spin rotation technique
  for non-collinear magnetic systems: application to the generalized Villain
  model}},\ }\href {https://doi.org/10.1088/0953-8984/21/21/216001} {\bibfield
  {journal} {\bibinfo  {journal} {J. Phys.: Condens. Matter}\ }\textbf
  {\bibinfo {volume} {21}},\ \bibinfo {pages} {216001} (\bibinfo {year}
  {2009})}\BibitemShut {NoStop}%
\bibitem [{\citenamefont {dos Santos}\ \emph {et~al.}(2018)\citenamefont {dos
  Santos}, \citenamefont {dos Santos~Dias}, \citenamefont {Guimar\~aes},
  \citenamefont {Bouaziz},\ and\ \citenamefont {Lounis}}]{dosSantosPRB}%
  \BibitemOpen
  \bibfield  {author} {\bibinfo {author} {\bibfnamefont {F.~J.}\ \bibnamefont
  {dos Santos}}, \bibinfo {author} {\bibfnamefont {M.}~\bibnamefont {dos
  Santos~Dias}}, \bibinfo {author} {\bibfnamefont {F.~S.~M.}\ \bibnamefont
  {Guimar\~aes}}, \bibinfo {author} {\bibfnamefont {J.}~\bibnamefont
  {Bouaziz}},\ and\ \bibinfo {author} {\bibfnamefont {S.}~\bibnamefont
  {Lounis}},\ }\bibfield  {title} {\bibinfo {title} {Spin-resolved inelastic
  electron scattering by spin waves in noncollinear magnets},\ }\href
  {https://doi.org/10.1103/PhysRevB.97.024431} {\bibfield  {journal} {\bibinfo
  {journal} {Phys. Rev. B}\ }\textbf {\bibinfo {volume} {97}},\ \bibinfo
  {pages} {024431} (\bibinfo {year} {2018})}\BibitemShut {NoStop}%
\bibitem [{\citenamefont {Laurell}\ and\ \citenamefont
  {Fiete}(2017)}]{TopoMagnonSC}%
  \BibitemOpen
  \bibfield  {author} {\bibinfo {author} {\bibfnamefont {P.}~\bibnamefont
  {Laurell}}\ and\ \bibinfo {author} {\bibfnamefont {G.~A.}\ \bibnamefont
  {Fiete}},\ }\bibfield  {title} {\bibinfo {title} {{Topological Magnon Bands
  and Unconventional Superconductivity in Pyrochlore Iridate Thin Films}},\
  }\href {https://doi.org/10.1103/PhysRevLett.118.177201} {\bibfield  {journal}
  {\bibinfo  {journal} {Phys. Rev. Lett.}\ }\textbf {\bibinfo {volume} {118}},\
  \bibinfo {pages} {177201} (\bibinfo {year} {2017})}\BibitemShut {NoStop}%
\end{thebibliography}%

\clearpage
\onecolumngrid
\allowdisplaybreaks

\renewcommand{\thefigure}{S\arabic{figure}}
\renewcommand{\theHfigure}{S\arabic{figure}}
\setcounter{figure}{0}  

\renewcommand{\thetable}{S\Roman{table}}
\setcounter{table}{0}  

\renewcommand{\theequation}{S\arabic{equation}}
\setcounter{equation}{0}  

\setcounter{secnumdepth}{2}

\renewcommand{\thesection}{S\arabic{section}}
\renewcommand{\thesubsection}{\thesection.\arabic{subsection}}
\renewcommand{\thesubsubsection}{\thesubsection.\arabic{subsubsection}}
\makeatletter
\renewcommand{\p@subsection}{}
\renewcommand{\p@subsubsection}{}
\makeatother


\phantomsection
\begin{large}
\begin{center}
    \textbf{Supplemental Material for ``Topological Superconductivity Mediated by Skyrmionic Magnons''} \label{sec:Suppl}
\end{center}
\end{large}

\title{Supplemental Material for ``Topological Superconductivity Mediated by Skyrmionic Magnons''}


\maketitle


\section{Introduction}
This Supplemental Material contains details of calculations that are left out of the main text. Section \ref{sec:MML} covers the magnetic monolayer (MML) while Sec.~\ref{sec:NM} covers the normal metal (NM). Section \ref{sec:EMC} concerns the electron-magnon interaction across the interface and gives a derivation of the effective electron-electron interaction mediated by the skyrmionic magnons. Section \ref{sec:SC} presents the generalized BCS theory, derivations of the gap equations, and shows some plots of the resultant gap functions. Finally, Sec.~\ref{sec:TSC} gives details about the calculation of the topological invariant that are left out of the main text.

\section{Magnetic monolayer} \label{sec:MML}
\subsection{Model and ground states}
As in Refs.~\cite{QSkOP, QSkQTPT}, we use the time-reversal-symmetric (TRS) Hamiltonian
\begin{equation}
\label{eq:H}
    H_{\text{MML}} = H_{\text{ex}} + H_{\text{DM}} + H_{\text{A}} + H_{4},
\end{equation}
where
\begin{align}
    H_{\text{ex}}  =& -J\sum_{\langle ij \rangle} \boldsymbol{S}_{i}\cdot \boldsymbol{S}_{j},  \qquad H_{\text{DM}} = \sum_{\langle ij \rangle} \boldsymbol{D}_{ij} \cdot (\boldsymbol{S}_i \cross \boldsymbol{S}_j), \qquad H_{\text{A}} = - K\sum_i S_{iz}^2,    \\
    H_4 =& U\sum_{ijkl}^\diamond \big[(\boldsymbol{S}_i \cdot \boldsymbol{S}_j)(\boldsymbol{S}_k \cdot \boldsymbol{S}_l) + (\boldsymbol{S}_i \cdot \boldsymbol{S}_l)(\boldsymbol{S}_j \cdot \boldsymbol{S}_k)-(\boldsymbol{S}_i \cdot \boldsymbol{S}_k)(\boldsymbol{S}_j \cdot \boldsymbol{S}_l)\big].
\end{align}
Here, $\boldsymbol{S}_i$, with magnitude $S$, is the spin operator at lattice site $i$ on the triangular lattice. The DMI vector is set to $\boldsymbol{D}_{ij} = D \hat{r}_{ij} \cross \hat{z}$ for nearest neighbors, where $\hat{r}_{ij}$ is a unit vector from site $i$ to site $j$.  The four-spin interaction $H_4$ results from higher order expansions of the Hubbard model \cite{H4PRB}, which are nonnegligible if the nearest-neighbor exchange interaction, $J$, is weak \cite{HeinzeSkX}. It involves four spins that are located on counterclockwise diamonds of minimal area \cite{HeinzeSkX, H4PRB}.

The periodicity of the ground state (GS) will depend on the parameters in the model. Let $\lambda_x$ be the periodicity in the $x$ direction in terms of lattice sites, while $\lambda_y$ is the periodicity in the $y$ direction in terms of lattice chains. We chose to work with a commensurate SkX with $\lambda_x = 5, \lambda_y = 6$ as the preferred periodicity, which requires $D/J = 2.16$ and $U/J = 0.35$ with $S=1$. We employed a self-consistent iteration approach to find the classical ground states \cite{dosSantosPRB, QSkOP, QSkQTPT}.

\subsection{Net magnetization and emergent magnetic field}
The classical spin is denoted $\boldsymbol{m}_i$ and is a unit vector.
The GS net magnetizations are
$
    \overline{m}_\alpha = \frac{1}{N}\sum_i m_{i\alpha}.
$
In SkX1 $\overline{m}_x = \overline{m}_y = 0$ and $\overline{m}_z \approx -0.002$. In SkX2 $\overline{m}_y = \overline{m}_z = 0$ and $\overline{m}_x \approx -0.014$.
With such small net magnetizations, we neglect orbital effects on the NM when studying superconductivity. 
Two-dimensional (2D) skyrmions also set up an emergent magnetic field, $\boldsymbol{b}$, defined in continuum models as \cite{nagaosaRev, SkHallBeyond}
\begin{equation}
    \boldsymbol{b} = b_z \hat{z} = \boldsymbol{m}(\boldsymbol{r}) \cdot [\partial_x \boldsymbol{m}(\boldsymbol{r}) \times \partial_y \boldsymbol{m}(\boldsymbol{r})] \hat{z}.
\end{equation}
By inspecting the SkX1 and SkX2 states it is clear that this field will be staggered, unlike larger skyrmions stabilized in magnetic fields \cite{nagaosaRev, SkHallBeyond}.
In the weak-coupling case, the Cooper pairs are much larger than the skyrmions, so the staggered emergent magnetic field is also assumed to have negligible effects on the superconducting state.

\subsection{Holstein-Primakoff approach} \label{sec:HProtation}
Based on the classical GS, a local orthonormal frame $\{\hat{e}_1^i, \hat{e}_2^i, \hat{e}_3^i\}$ with $\hat{e}_3^i = \boldsymbol{m}_i$ is introduced \cite{HProtation_2009}. Here, $\boldsymbol{m}_i = (\sin\theta_i\cos\phi_i, \sin\theta_i\sin\phi_i, \cos\theta_i)$, $\hat{e}_1^i = (\cos\theta_i\cos\phi_i, \cos\theta_i\sin\phi_i, -\sin\theta_i)$, and $\hat{e}_2^i = (-\sin\phi_i, \cos\phi_i, 0)$. The polar and azimuthal angles $\{\theta_i, \phi_i \}$ are specified by the classical GS. 
With $\hat{r}_\alpha$ the Cartesian coordinates, we define a rotation matrix through, $\hat{r}_\alpha = R_{\alpha\beta}^i \hat{e}_\beta^i$.
The local rotations can then be inserted in the Hamiltonian. The Holstein-Primakoff (HP) transformation is performed by $S_{i3}=\boldsymbol{S}_i \cdot \hat{e}_3^i = S - a_i^\dagger a_i$, $S_{i\pm} = \boldsymbol{S}_i \cdot \hat{e}_\pm^i $, $\hat{e}_\pm^i = \hat{e}_1^i \pm i\hat{e}_2^i$, $S_{i+} = \sqrt{2S}a_i,$ and $S_{i-} = \sqrt{2S}a_i^\dagger$. We truncate at second order in magnon operators assuming small spin fluctuations, which should be valid at low temperatures \cite{HProtation_2009, QSkOP, QSkQTPT}.

If site $i$ is located on sublattice $r$ we have the Fourier transform (FT) $a_{i} = \frac{1}{\sqrt{N'}} \sum_{\boldsymbol{q}\in \text{mBZ}} a_{\boldsymbol{q}}^{(r)} e^{i\boldsymbol{q}\cdot\boldsymbol{r}_i}$, where $N'$ is the number of magnetic unit cells, i.e., $N' = N/15$ for SkX1 and SkX2.
The 15 sublattices in SkX1 and SkX2 are all centered rectangular lattices with primitive vectors $\boldsymbol{a}_1 = (5/2, -3\sqrt{3}/2)$ and $\boldsymbol{a}_2 = (5/2, 3\sqrt{3}/2)$. The magnetic first Brillouin zone (mBZ) is a nonregular hexagon with vertices at $(\pm 52\pi/135, 0)$ and $(\pm 2\pi/135,\pm 2\pi/3\sqrt{3})$. 

The part of the Hamiltonian that is quadratic in magnon operators can be written \cite{QSkOP, QSkQTPT}
$
    H_2 = (1/2)\sum_{\boldsymbol{q}}  \boldsymbol{a}_{\boldsymbol{q}}^\dagger M_{\boldsymbol{q}}  \boldsymbol{a}_{\boldsymbol{q}},
$
where $\boldsymbol{a}_{\boldsymbol{q}}^\dagger = (a_{\boldsymbol{q}}^{(1)\dagger}, a_{\boldsymbol{q}}^{(2)\dagger}, \dots, a_{\boldsymbol{q}}^{(15)\dagger}, a_{-\boldsymbol{q}}^{(1)}, \dots, a_{-\boldsymbol{q}}^{(15)})$, and a detailed derivation of the matrix elements of $M_{\boldsymbol{q}}$ can be found in the Supplemental Material of Ref.~\cite{QSkQTPT}.
The magnon operators are transformed to their diagonal basis through a paraunitary matrix $T_{\boldsymbol{q}}$ with left block matrices $U_{\boldsymbol{q}}, V_{\boldsymbol{q}}$ \cite{COLPA, QSkOP, QSkQTPT}, yielding
\begin{equation}
H_{\text{MML}} = \sum_{\boldsymbol{q}\in \text{mBZ},n} \omega_{\boldsymbol{q}n} b_{\boldsymbol{q}n}^\dagger b_{\boldsymbol{q}n}.
\end{equation}
Here, $b_{\boldsymbol{q}n}$ annihilates a magnon in mode $n$ with momentum $\boldsymbol{q}$.
Detailed plots and analyses of the 15 magnon bands $\omega_{\boldsymbol{q}n}$ can be found in Refs.~\cite{QSkOP, QSkQTPT}. The magnon gap, $\omega_0 = \operatorname{min}_{\boldsymbol{q}} \omega_{\boldsymbol{q},15}$ varies between $\omega_0 \approx 0.67J$ and $\omega_0 \approx 0.07J$, see Fig.~2(c) of Ref.~\cite{QSkOP}. It takes on its lowest values closest to the phase transition between SkX1 and SkX2. 

\section{Normal metal} \label{sec:NM}
\subsection{Tight binding Hamiltonian}
The normal metal is described by a hopping term, $t$, and a controllable chemical potential, $\mu$,
\begin{equation}
    H_{\text{NM}} = -t \sum_{\langle i,j\rangle \sigma} c_{i\sigma}^\dagger c_{j\sigma} - \mu \sum_{i\sigma} c_{i\sigma}^\dagger c_{i\sigma},
\end{equation}
where $c_{i\sigma}$ annihilates an electron with spin $\sigma$ at site $i$.
This is diagonalized by an FT to get the electron energy $\epsilon_{\boldsymbol{k}}$, as described in the main text. The Fermi surface (FS) is defined by $\epsilon_{\boldsymbol{k}_{\text{F}}} = 0$, with two examples shown in Fig.~1(a) of the main text along with the electron first Brillouin zone (eBZ).

\subsection{Density of states}
The density of states is
\begin{equation}
    D(\epsilon) = \sum_{\boldsymbol{k}\sigma} \delta(\epsilon-\epsilon_{\boldsymbol{k}}) = 2\sum_{\boldsymbol{k}} \delta(\epsilon-\epsilon_{\boldsymbol{k}}).
\end{equation}
We follow the same route as Ref.~\cite{Self-energy} to calculate this.
The sum is rewritten into an integral over polar coordinates as follows,
\begin{equation}
    \sum_{\boldsymbol{k}} \to \frac{N}{A_{\text{eBZ}}}\int_{\text{eBZ}} dk_x \int dk_y \to \frac{N}{A_{\text{eBZ}}}\int_{-\pi}^\pi d\theta \int_0^{c(\theta)} dk k,
\end{equation}
where $k_x = k\cos\theta$, $k_y = k\sin\theta$, $\theta = \operatorname{atan2}(k_y, k_x)$, $A_{\text{eBZ}}$ is the area of the eBZ, and the upper cutoff $c(\theta)$ ensures that the integral is limited to the eBZ. The cutoff function is
\begin{equation}
    c(\theta) = \frac{4\pi/3}{1+\frac{2/\sqrt{3}-1}{\sqrt{2}-1}\pqty{\abs{\sin\frac{3\theta}{2}}+\abs{\cos\frac{3\theta}{2}}-1}}.
\end{equation}
Then,
\begin{equation}
    D(\epsilon) = \frac{2N}{A_{\text{eBZ}}}\int_{-\pi}^\pi d\theta \int_0^{c(\theta)} dk k \delta(\epsilon-\epsilon_{k,\theta}).
\end{equation}
For a function $f(k)$ with roots $k_i$ and $f'(k_i) \neq 0$,
$
    \delta[f(k)] = \sum_i \delta(k-k_i)/|f'(k_i)|.
$
The role of the chemical potential in this context will just be to set the range of $\epsilon_{\boldsymbol{k}}$, and hence the range of $\epsilon$ values where the DOS is nonzero. Therefore, $\mu = 0$ is chosen. Here,
$
    f(k) = \epsilon +2t[\cos(k\cos\theta) +2\cos(k\cos\theta/2)\cos (\sqrt{3}k\sin\theta/2)].
$
For each $\theta$ we include all roots $k_i < c(\theta)$ to get
\begin{equation}
\label{eq:numDOS}
    D(\epsilon) = \frac{2N}{A_{\text{eBZ}}}\int_{-\pi}^\pi d\theta \sum_i \frac{k_i(\theta)}{|f'[k_i(\theta)]|}.
\end{equation}

\begin{figure}
    \centering
    \includegraphics[width = 0.5\textwidth]{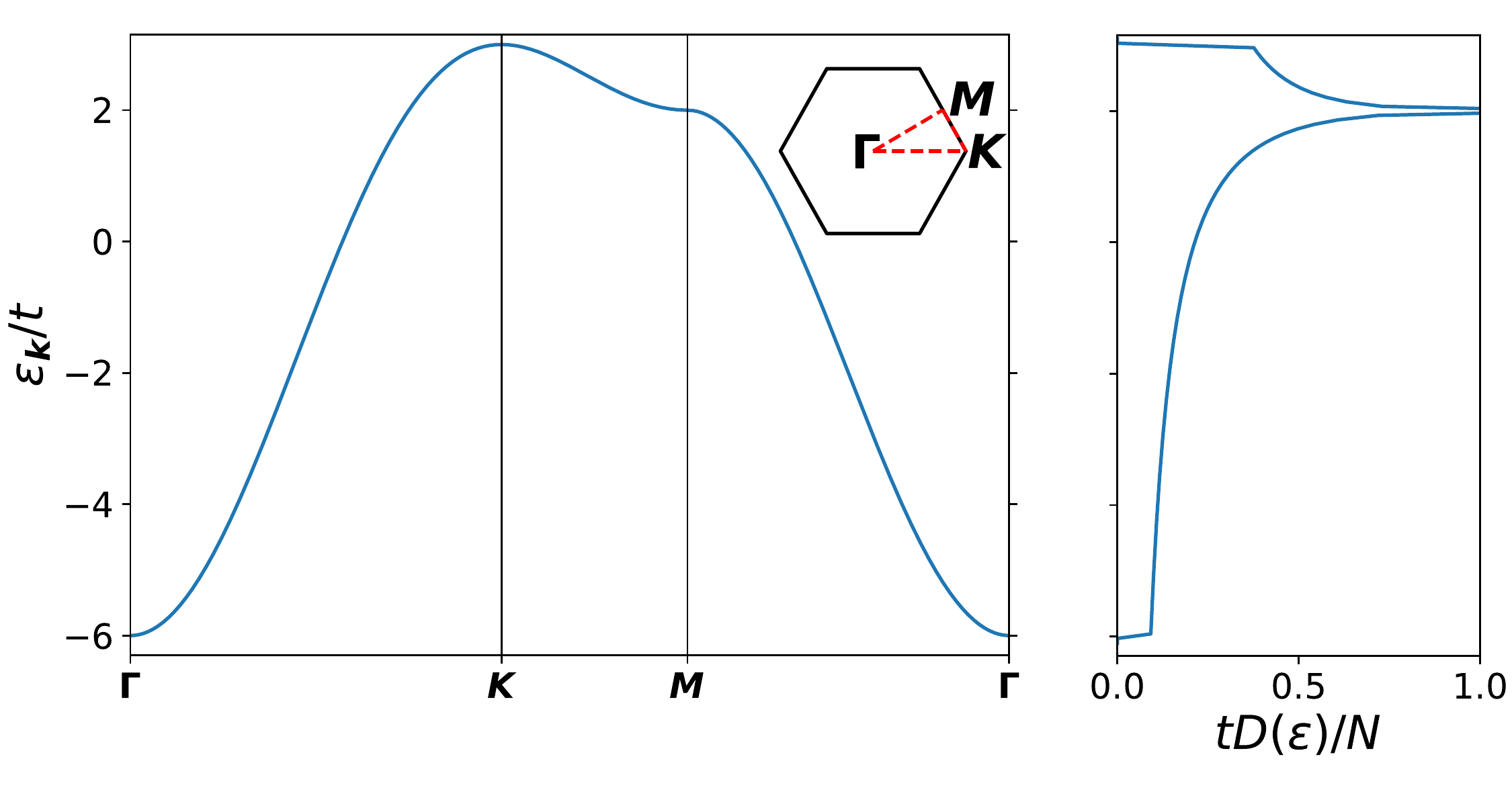}
    \caption{The electron energy and the density of states with $\mu = 0$. $\boldsymbol{k}$ follows the path in the first Brillouin zone sketched in the inset.}
    \label{fig:DOS}
\end{figure}

The density of states, computed numerically from Eq.~\eqref{eq:numDOS}, is shown in Fig.~\ref{fig:DOS}. We find that for $\epsilon \gtrsim -6t$ the DOS is approximately constant, $tD(\epsilon)/N = \sqrt{3}/6\pi \approx 0.092$, which can be proved analytically. Also, the integral over the DOS, $\int d\epsilon D(\epsilon)/N = 2,$ as expected for a spin-degenerate system. The van Hove singularity is located at $\epsilon = 2t$. If we are interested in the DOS at the FS, what we need is $D(\epsilon = 0)$ from a calculation starting from the relevant value of $\mu$. Due to the similar way $\epsilon$ and $\mu$ enter the preceding equations, we can instead use the result from setting $\mu = 0$, and find that the DOS on the FS for some nonzero $\mu$ is given by $D(\epsilon = \mu) \equiv D_0$.

\subsection{Symmetries of gap solutions}
\begin{table}[ht]
    \centering
    \caption{Some named basis functions for the triangular lattice. The value of $l$ controls the parity in $\boldsymbol{k}$ and the number of zeros along a circle around the origin. The name is given from $l$ and the position of the zeros. \label{tab:landnames}}
    \begin{tabular}{ccccc}
    \hline
    \hline
        $l$ & name & \#zeros & parity & zeros along $k_x = 0$, $k_y = 0$, or $k_x = \pm k_y$ \\
    \hline
        0 & $s$ & 0 & even &  \\
        1 & $p_x$ & 2 & odd & $k_x = 0$ \\
        1 & $p_y$ & 2 & odd & $k_y = 0$ \\
        2 & $d_{xy}$ & 4 & even & $k_x = 0, k_y = 0$ \\
        2 & $d_{x^2-y^2}$ & 4 & even & $k_x = \pm k_y$ \\
        3 & $f_x$ & 6 & odd & $k_x = 0$ \\
        3 & $f_y$ &  6 & odd & $k_y = 0$ \\
        4 & $g$ & 8 & even & $k_x = 0, k_y = 0, k_x = \pm k_y$ \\
        \hline
        \hline
    \end{tabular}
\end{table}
A selection of basis function for the triangular lattice can be found in Ref.~\cite{Benestad}. 
The superconductor (SC) gap can be expressed as a linear combination of all basis functions. Hence, checking which of a chosen set of basis functions the gap has its largest overlap with may not be the best way to determine its symmetry. Instead, we use the number and position of zeros on the FS to define the gaps in terms of the symmetries $s, p_x, p_y, d_{xy}, d_{x^2-y^2}, f_x, f_y, g$, see Table \ref{tab:landnames}. The names originate with atomic orbitals, where $l$ is the angular momentum quantum number. What we name $f_x$ is in fact a cubic function, $f_x = f_{x(x^2-3y^2)} \sim k_x(k_x^2-3k_y^2)$ and $f_y = f_{y(3x^2-y^2)} \sim k_y(3k_x^2-k_y^2)$. Simple representations of the basis functions can be defined in terms of the angle $\phi = \operatorname{atan2}(k_y,k_x)$ for $\boldsymbol{k}$ on the FS; $s:$ constant, $p_x: \cos\phi$, $p_y: \sin\phi$, $d_{xy}: -\sin2\phi$, $d_{x^2-y^2}: -\cos2\phi$, $f_x: -\cos3\phi$, $f_y: \sin3\phi$, $g: \sin4\phi$. I.e.~$\pm\cos l\phi, \pm\sin l\phi$ are simplified lattice harmonics, where signs are chosen in order to comply to the triangular lattice basis functions given in Ref.~\cite{Benestad}.

\section{Electron-magnon interaction} \label{sec:EMC}
\subsection{Rotation to local spin axis and Fourier transform}
In the interfacial exchange coupling,
\begin{equation}
    \boldsymbol{c}_i^\dagger \boldsymbol{\sigma} \boldsymbol{c}_i = (c_{i\uparrow}^\dagger c_{i\downarrow} + c_{i\downarrow}^\dagger c_{i\uparrow}, -ic_{i\uparrow}^\dagger c_{i\downarrow} + ic_{i\downarrow}^\dagger c_{i\uparrow}, c_{i\uparrow}^\dagger c_{i\uparrow} - c_{i\downarrow}^\dagger c_{i\downarrow}),
\end{equation}
\begin{equation}
    S_{ix} = \boldsymbol{S}_i \cdot \hat{x} = \sum_\beta \boldsymbol{S}_i \cdot \hat{e}_\beta^i R_{1\beta}^i = \sum_{\beta} S_{i\beta} R_{1\beta}^i, \qquad S_{iy} = \sum_{\beta} S_{i\beta} R_{2\beta}^i, \qquad S_{iz} = \sum_{\beta} S_{i\beta} R_{3\beta}^i.
\end{equation}
The HP transformation,
\begin{equation}
    S_{i1} = \frac{1}{2}(S_{i+}+S_{i-}) = \sqrt{\frac{S}{2}}(a_i + a_i^\dagger), \qquad S_{i2} = \frac{1}{2i}(S_{i+}-S_{i-}) = i\sqrt{\frac{S}{2}}(a_i^\dagger - a_i), \qquad S_{i3} = S - a_i^\dagger a_i,
\end{equation}
can then be inserted,
\begin{align}
    \boldsymbol{c}_i^\dagger \boldsymbol{\sigma} \boldsymbol{c}_i \cdot \boldsymbol{S}_i =& a_i c_{i\uparrow}^\dagger c_{i\downarrow} \sqrt{\frac{S}{2}} (\underbrace{R_{11}^i - iR_{21}^i}_{\cos\theta_i e^{-i\phi_i}} \underbrace{-R_{22}^i-iR_{12}^i}_{-e^{-i\phi_i}}) + \text{H.c.} +a_i c_{i\downarrow}^\dagger c_{i\uparrow} \sqrt{\frac{S}{2}} (\underbrace{R_{11}^i + iR_{21}^i}_{\cos\theta_i e^{i\phi_i}} \underbrace{+R_{22}^i-iR_{12}^i}_{+e^{i\phi_i}}) + \text{H.c.}  \nonumber \\
    &+a_i c_{i\uparrow}^\dagger c_{i\uparrow} \sqrt{\frac{S}{2}} \underbrace{R_{31}^i}_{-\sin\theta_i} + \text{H.c.} + a_i c_{i\downarrow}^\dagger c_{i\downarrow} \sqrt{\frac{S}{2}} \underbrace{(-R_{31}^i)}_{+\sin\theta_i} + \text{H.c.} \nonumber \\
    &+c_{i\uparrow}^\dagger c_{i\downarrow} (S-a_i^\dagger a_i) \underbrace{(R_{13}^i-iR_{23}^i)}_{\sin\theta_i e^{-i\phi_i}} + \text{H.c.} + c_{i\uparrow}^\dagger c_{i\uparrow}(S-a_i^\dagger a_i) \underbrace{R_{33}^i}_{\cos\theta_i} + c_{i\downarrow}^\dagger c_{i\downarrow} (S-a_i^\dagger a_i) \underbrace{(-R_{33}^i)}_{-\cos\theta_i}.
\end{align}
We ignore the terms containing two magnon operators in these electron-magnon interactions. At low temperature with few magnons present, the electron-magnon coupling (EMC) terms with only one magnon will dominate. 
The above can be rewritten more succinctly as
\begin{align}
    H_{\text{em}} =& -2\Bar{J}\sqrt{\frac{S}{2}} \sum_{i\sigma} [e^{-i\sigma\phi_i}(\cos\theta_i-\sigma)a_i c_{i\sigma}^\dagger c_{i,-\sigma} + \text{H.c.}] +2\Bar{J}\sqrt{\frac{S}{2}}\sum_{i\sigma}(\sigma \sin\theta_i a_i c_{i\sigma}^\dagger c_{i\sigma} + \text{H.c.}) \nonumber \\
    &-2\Bar{J}S\sum_{i\sigma} \sin\theta_i e^{-i\sigma\phi_i} c_{i\sigma}^\dagger c_{i,-\sigma} -2\Bar{J}S\sum_{i\sigma} \sigma \cos\theta_i c_{i\sigma}^\dagger c_{i,\sigma} .
\end{align}
The terms in the last line do not contain magnons, and are neglected, to be justified in Sec.~\ref{sec:effectiveint}.

\begin{figure}
    \begin{minipage}{.5\textwidth}
      \includegraphics[width=0.7\linewidth]{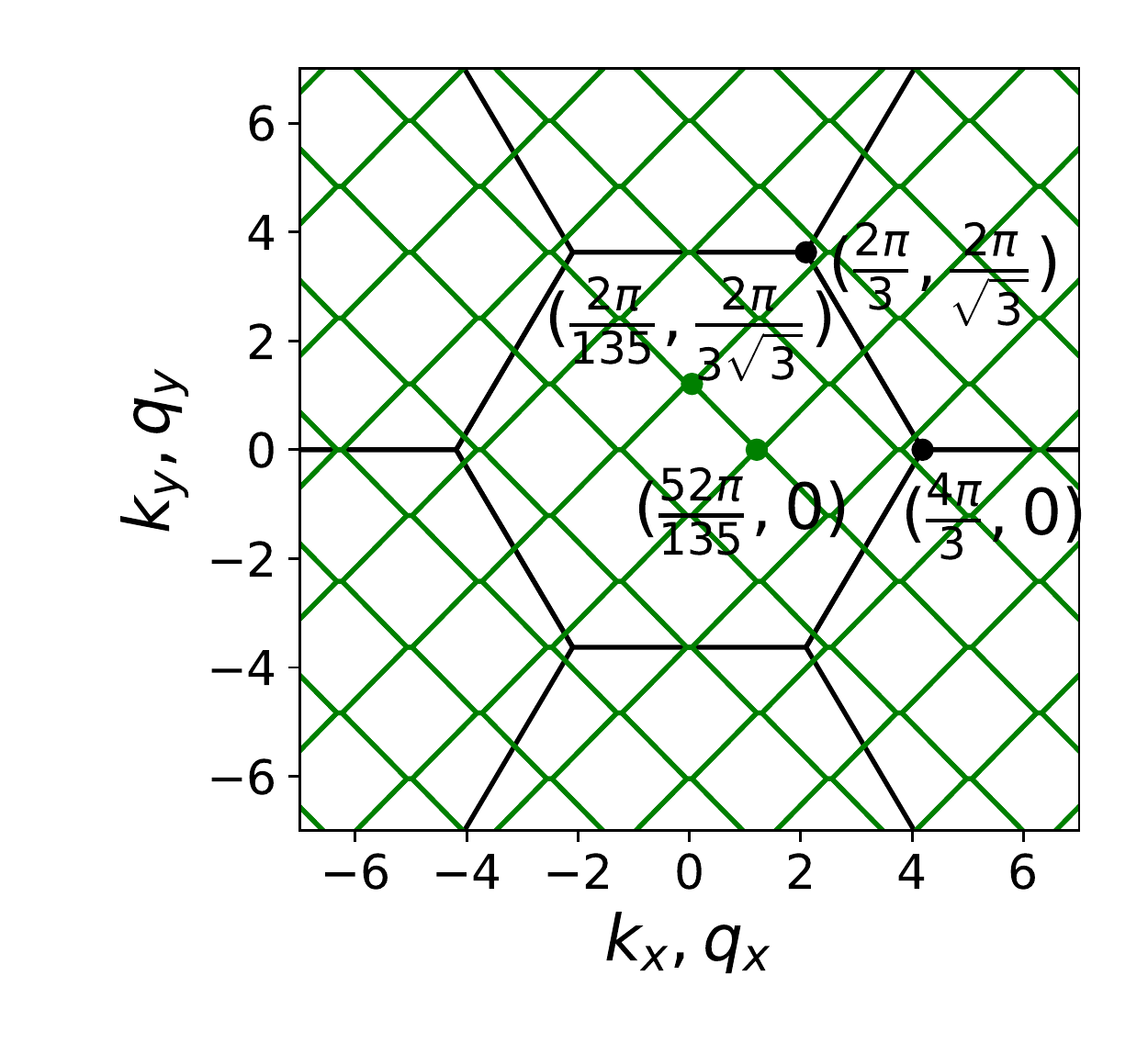}
    \end{minipage}%
    \begin{minipage}{.5\textwidth}
      \includegraphics[width=0.7\linewidth]{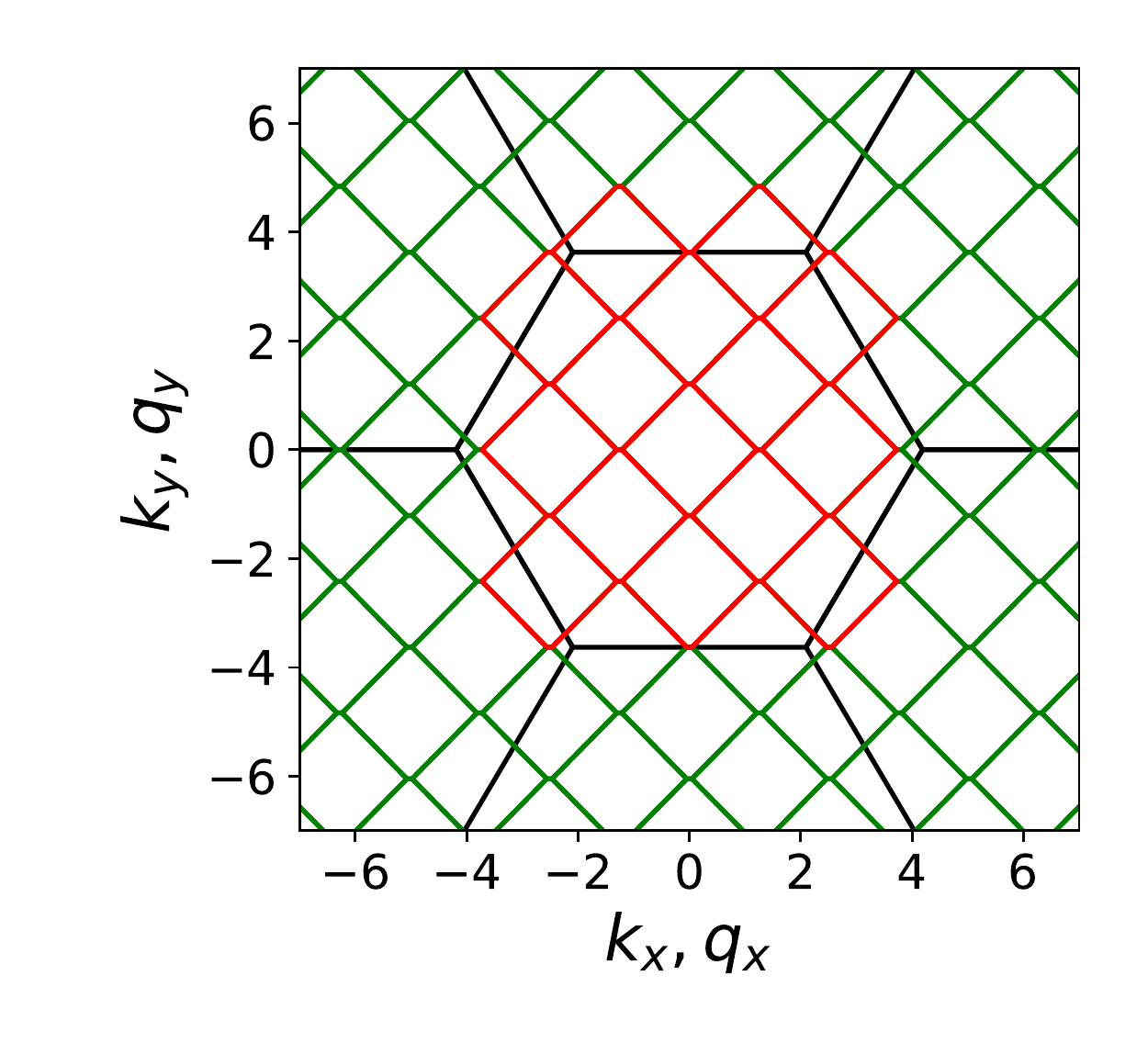}
    \end{minipage}%
    \caption{Left: Repeated eBZ in black and mBZ in green with some vertices annotated. Right: The 15 red mBZs cover, due to periodicity, the eBZ. \label{fig:BZ}}
\end{figure}

\begin{figure}
    \begin{minipage}{.5\textwidth}
      \includegraphics[width=0.7\linewidth]{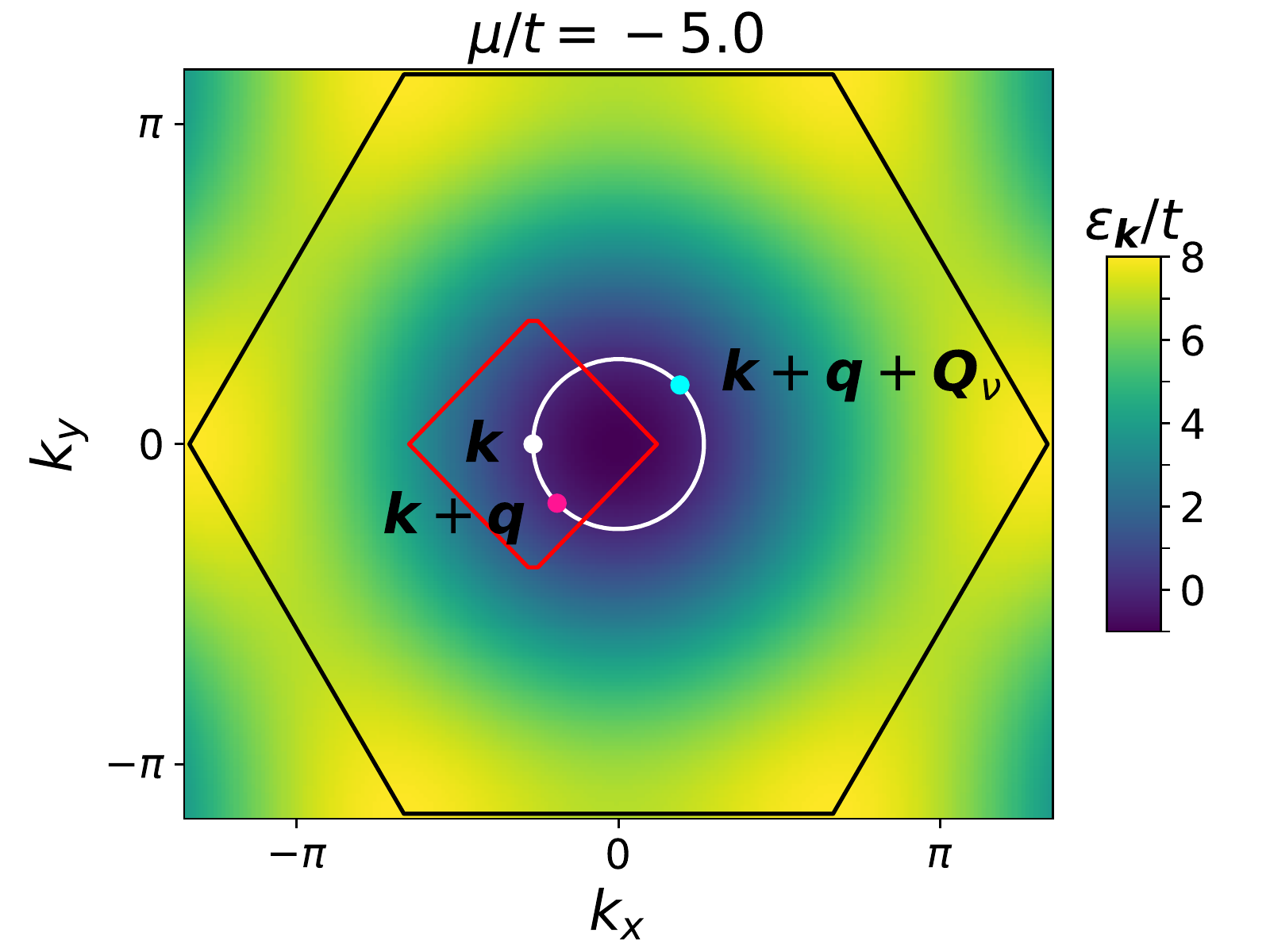}
    \end{minipage}%
    \begin{minipage}{.5\textwidth}
      \includegraphics[width=0.7\linewidth]{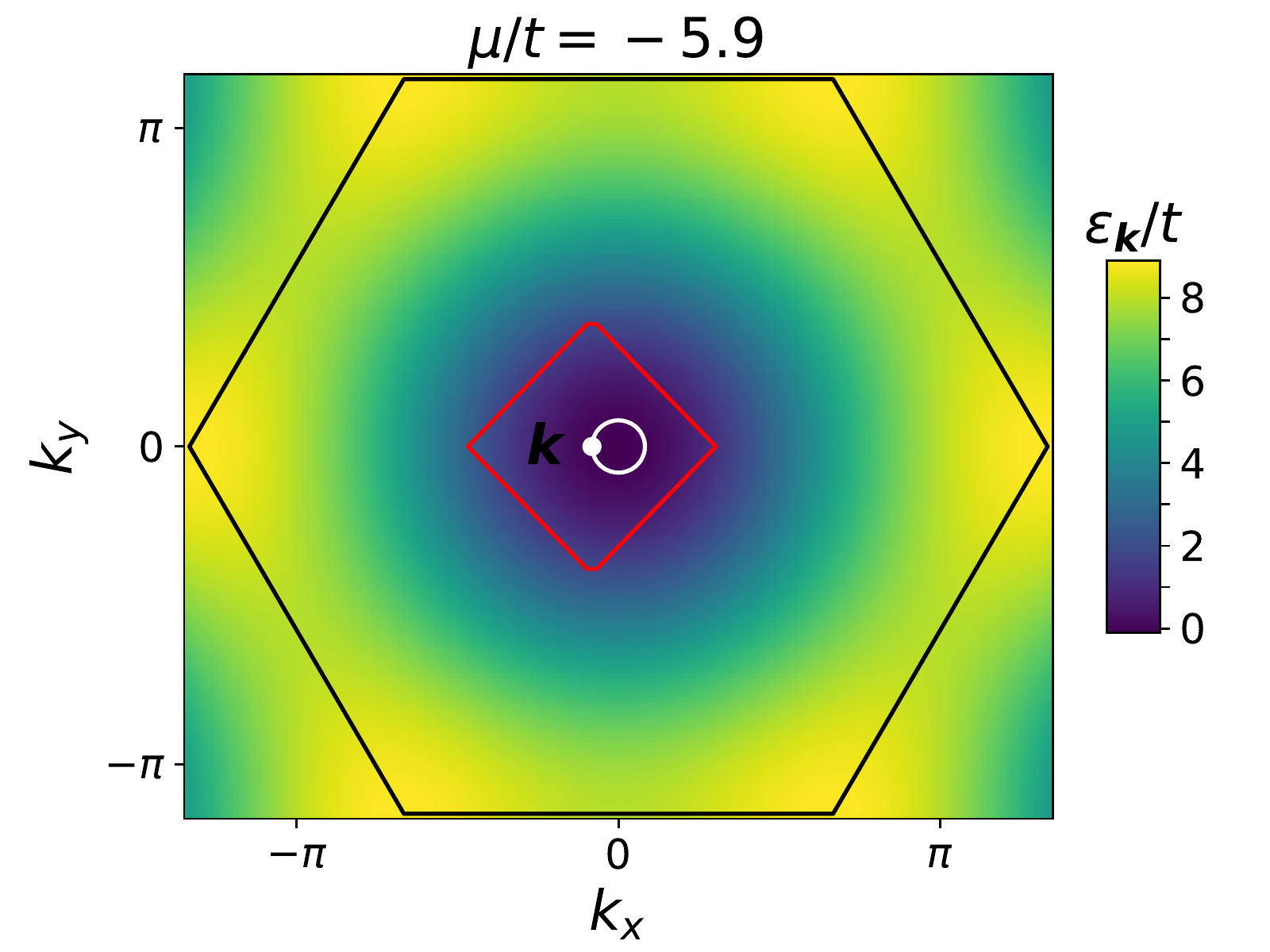}
    \end{minipage}%
    \caption{Illustration of the Fermi surface (FS) in white at different $\mu$ and to what extent regular processes (pink) cover the FS. The mBZ is drawn in red centered at a chosen vector $\boldsymbol{k}$ on the FS. In the left figure $\mu/t = -5$ and we see that Umklapp processes (light blue) are needed to reach all points on the FS. Note that $\boldsymbol{k}+\boldsymbol{q}$ need not be on the FS for Umklapp processes, as long as $\boldsymbol{k}+\boldsymbol{q}+\boldsymbol{Q}_\nu$ is. In the right figure $\mu/t = -5.9$ and it is clear that regular processes will be sufficient to reach any point on the FS. \label{fig:RUmpklapp}}
\end{figure}

The eBZ and mBZ are shown together in Fig.~\ref{fig:BZ}. We show how 15 mBZs cover the eBZ, similar to how 2 reduced Brillouin zones cover the first Brillouin zone (1BZ) of the square lattice in antiferromagnet normal metal structures \cite{EirikNMAFM, ArneAFMNM_Umklapp, EirikEliashberg}. The areas of the mBZ and eBZ are related by $A_{\text{mBZ}}/A_{\text{eBZ}} = 15$, so the preceding statement is exact. This guides the implementation of Umklapp processes.

The importance of Umklapp processes in the context of superconductivity is decided by the relative size of the FS to the mBZ. If the FS is much smaller than the mBZ, Umklapp processes are likely suppressed. We plot $\epsilon_{\boldsymbol{k}}$, the eBZ, the mBZ and the Fermi surface for different choices of $\mu$ in Fig.~1(a) in the main text. To complement this, Fig. \ref{fig:RUmpklapp} shows that for $\mu/t = -5.9$ the FS is sufficiently small that regular processes $\boldsymbol{k} \to \boldsymbol{k}+\boldsymbol{q}$ with $\boldsymbol{q} \in $~mBZ cover the FS. For $\mu/t = -5$, the FS has a size such that regular processes do not cover the entire FS. Umklapp processes $\boldsymbol{k} \to \boldsymbol{k} + \boldsymbol{q} + \boldsymbol{Q}_\nu$ are required to reach all points on the FS. Here $\boldsymbol{Q}_\nu$ is a set of 15 reciprocal lattice vectors,
\begin{align}
    \boldsymbol{Q}_\nu =& \bigg\{ (0,0), \pqty{\pm \frac{2\pi}{5}, \frac{2\pi}{3\sqrt{3}}}, \pqty{\pm \frac{2\pi}{5}, -\frac{2\pi}{3\sqrt{3}}},   \pqty{\pm \frac{4\pi}{5}, 0}, \pqty{0, \pm\frac{4\pi}{3\sqrt{3}}}, \nonumber \\
    & \pqty{\pm \frac{2\pi}{5}, \frac{2\pi}{\sqrt{3}}}, \pqty{\pm \frac{4\pi}{5}, \frac{4\pi}{3\sqrt{3}}}, \pqty{\pm \frac{4\pi}{5}, -\frac{4\pi}{3\sqrt{3}}} \bigg\}.
\end{align}
Umklapp processes become relevant for $\mu/t \gtrsim -5.7$, where the diameter of the approximately circular FS becomes larger than the distance between the $\boldsymbol{\Gamma}$ and $\boldsymbol{M}$ points in the mBZ \cite{QSkOP, QSkQTPT}.

We introduce Umklapp processes explicitly by rewriting the FT of the electron operators
\begin{equation}
    c_{i\sigma} = \frac{1}{\sqrt{N}} \sum_{\boldsymbol{k}\in \text{eBZ}} c_{\boldsymbol{k}\sigma} e^{i\boldsymbol{k}\cdot\boldsymbol{r}_i} = \frac{1}{\sqrt{N}} \sum_{\boldsymbol{k}\in \text{mBZ}} \sum_\nu c_{\boldsymbol{k}+\boldsymbol{Q}_\nu, \sigma} e^{i(\boldsymbol{k}+\boldsymbol{Q}_\nu) \cdot\boldsymbol{r}_i} .
\end{equation}

That way, $\sum_{\boldsymbol{k}\in \text{eBZ}}$ covers the area enclosed by the black eBZ in the right part of Fig. \ref{fig:BZ}, while $\sum_{\boldsymbol{k}\in \text{mBZ}} \sum_\nu$ covers the area enclosed by the 15 red mBZ in the right part of Fig. \ref{fig:BZ}. Due to periodicity, this is equivalent. 


The FT is performed as
\begin{align}
    \sum_i a_i c_{i\sigma}^\dagger c_{i\sigma'} =& \sum_r \sum_{i\in r} \frac{1}{\sqrt{N'}}  \sum_{\boldsymbol{q}\in \text{mBZ}} a_{\boldsymbol{q}}^{(r)} e^{i\boldsymbol{q}\cdot\boldsymbol{r}_i} \frac{1}{\sqrt{N}}\sum_{\boldsymbol{k}'\in \text{mBZ}} \sum_{\nu'} c_{\boldsymbol{k}'+\boldsymbol{Q}_{\nu'}, \sigma}^\dagger e^{-i(\boldsymbol{k}'+\boldsymbol{Q}_{\nu'}) \cdot\boldsymbol{r}_i} \nonumber \\
    &\cross \frac{1}{\sqrt{N}}\sum_{\boldsymbol{k}\in \text{mBZ}} \sum_\nu c_{\boldsymbol{k}+\boldsymbol{Q}_\nu, \sigma'} e^{i(\boldsymbol{k}+\boldsymbol{Q}_\nu) \cdot\boldsymbol{r}_i} e^{i\boldsymbol{k}\cdot\boldsymbol{r}_i}.
\end{align}
We separate a factor $e^{-i(\boldsymbol{Q}_{\nu'}-\boldsymbol{Q}_{\nu})\cdot \boldsymbol{r}_i} = e^{-i(\boldsymbol{Q}_{\nu'}-\boldsymbol{Q}_{\nu})\cdot \boldsymbol{r}_r}$ since it is the same for any $i\in r$. Then it can be taken outside the sum over $i$, and $\sum_{i\in r} e^{i(\boldsymbol{k}+\boldsymbol{q}-\boldsymbol{k}')\cdot \boldsymbol{r}_i} = N' \delta_{\boldsymbol{k}+\boldsymbol{q},\boldsymbol{k}'}$. This gives
\begin{equation}
    \sum_i a_i c_{i\sigma}^\dagger c_{i\sigma'} =\sum_r \frac{\sqrt{N'}}{N} \sum_{\boldsymbol{k},\boldsymbol{q} \in \text{mBZ}} \sum_{\nu\nu'} e^{-i(\boldsymbol{Q}_{\nu'}-\boldsymbol{Q}_{\nu})\cdot \boldsymbol{r}_r} a_{\boldsymbol{q}}^{(r)}c_{\boldsymbol{k}+\boldsymbol{q}+\boldsymbol{Q}_{\nu'}, \sigma}^\dagger c_{\boldsymbol{k}+\boldsymbol{Q}_\nu, \sigma'}.
\end{equation}
Consider the case that $\boldsymbol{Q}_{\nu'}-\boldsymbol{Q}_{\nu} = \boldsymbol{Q}_{\nu''}+\boldsymbol{G}$, where $\boldsymbol{G}$ is a shorthand for the set of reciprocal lattice vectors for the electrons. Some examples are $\boldsymbol{G} = (\pm 2\pi, \pm 2\pi/\sqrt{3}), (0, \pm 4\pi/\sqrt{3})$. Adding $\boldsymbol{G}$ has no effect in the subscript of electron operators or when multiplying with site positions $\boldsymbol{r}_r$ which all lie on the triangular lattice (by definition, $e^{i\boldsymbol{G}\cdot \boldsymbol{r}_i} = 1$). Hence, we can ignore it. In this way, all possible $\nu, \nu'$ are considered by summing over $\nu$ and $\nu''$,
\begin{equation}
    \sum_i a_i c_{i\sigma}^\dagger c_{i\sigma'} =\sum_r \frac{\sqrt{N'}}{N} \sum_{\boldsymbol{k},\boldsymbol{q} \in \text{mBZ}} \sum_{\nu\nu''} e^{-i\boldsymbol{Q}_{\nu''} \cdot \boldsymbol{r}_r} a_{\boldsymbol{q}}^{(r)}c_{\boldsymbol{k}+\boldsymbol{q}+\boldsymbol{Q}_\nu + \boldsymbol{Q}_{\nu''}, \sigma}^\dagger c_{\boldsymbol{k}+\boldsymbol{Q}_\nu, \sigma'}.
\end{equation}
Both operators have a momentum $\boldsymbol{Q}_\nu$ added, so the sum over $\nu$ entails moving around in momentum space, covering the eBZ. The sum over $\nu$ can be absorbed by extending the range of $\boldsymbol{k}$ to the eBZ,
\begin{equation}
    \sum_i a_i c_{i\sigma}^\dagger c_{i\sigma'} =\sum_r \frac{\sqrt{N'}}{N} \sum_{\boldsymbol{k}\in \text{eBZ}, \boldsymbol{q} \in \text{mBZ}} \sum_{\nu} e^{-i\boldsymbol{Q}_{\nu} \cdot \boldsymbol{r}_r} a_{\boldsymbol{q}}^{(r)}c_{\boldsymbol{k}+\boldsymbol{q}+\boldsymbol{Q}_\nu, \sigma}^\dagger c_{\boldsymbol{k}, \sigma'},
\end{equation}
where we renamed $\nu'' \to \nu$.

It total, this yields
\begin{equation}
    H_{\text{em}} = \sum_{\boldsymbol{k}\in \text{eBZ},\boldsymbol{q}\in \text{mBZ}}\sum_{\nu r\sigma\sigma'} (g_{\nu r}^{\sigma\sigma'}a_{\boldsymbol{q}}^{(r)} c_{\boldsymbol{k}+\boldsymbol{q}+\boldsymbol{Q}_\nu,\sigma}^\dagger c_{\boldsymbol{k},\sigma'} + \text{H.c.}),
\end{equation}
with
\begin{equation}
    g_{\nu r}^{\sigma,-\sigma} = -2\Bar{J}\sqrt{\frac{S}{2}} \frac{\sqrt{N'}}{N}e^{-i\sigma\phi_r}(\cos\theta_r-\sigma)  e^{-i\boldsymbol{Q}_{\nu} \cdot \boldsymbol{r}_r},
\qquad
    g_{\nu r}^{\sigma,\sigma} = 2\Bar{J}\sqrt{\frac{S}{2}} \frac{\sqrt{N'}}{N}\sigma \sin\theta_r  e^{-i\boldsymbol{Q}_{\nu} \cdot \boldsymbol{r}_r}.
\end{equation}
Writing out the H.c.~gives,
\begin{equation}
    H_{\text{em}} = \sum_{\boldsymbol{k}\in \text{eBZ},\boldsymbol{q}\in \text{mBZ}}\sum_{\nu r\sigma\sigma'} (g_{\nu r}^{\sigma\sigma'}a_{\boldsymbol{q}}^{(r)} c_{\boldsymbol{k}+\boldsymbol{q}+\boldsymbol{Q}_\nu,\sigma}^\dagger c_{\boldsymbol{k},\sigma'} + g_{\nu r}^{\sigma\sigma' *}a_{\boldsymbol{q}}^{(r)\dagger} c_{\boldsymbol{k},\sigma'}^\dagger c_{\boldsymbol{k}+\boldsymbol{q}+\boldsymbol{Q}_\nu,\sigma}).
\end{equation}
In the second term, the following rewrites are performed, $\boldsymbol{q} \to -\boldsymbol{q}$, $\nu \to \overline{\nu}$ (where $\boldsymbol{Q}_{\overline{\nu}} = -\boldsymbol{Q}_\nu$), $\sigma \leftrightarrow \sigma'$, and $\boldsymbol{k} \to \boldsymbol{k} + \boldsymbol{q} + \boldsymbol{Q}_\nu$. This gives
\begin{equation}
    H_{\text{em}} = \sum_{\boldsymbol{k}\in \text{eBZ},\boldsymbol{q}\in \text{mBZ}}\sum_{\nu r\sigma\sigma'} (g_{\nu r}^{\sigma\sigma'}a_{\boldsymbol{q}}^{(r)} c_{\boldsymbol{k}+\boldsymbol{q}+\boldsymbol{Q}_\nu,\sigma}^\dagger c_{\boldsymbol{k},\sigma'} + g_{\overline{\nu} r}^{\sigma'\sigma*}a_{-\boldsymbol{q}}^{(r)\dagger} c_{\boldsymbol{k}+\boldsymbol{q}+\boldsymbol{Q}_\nu,\sigma}^\dagger c_{\boldsymbol{k},\sigma'}).
\end{equation}
Inserting diagonalized bosonic operators yields \cite{QSkOP, QSkQTPT},
\begin{align}
    H_{\text{em}} = \sum_{\boldsymbol{k}\in \text{eBZ},\boldsymbol{q}\in \text{mBZ}} \sum_{\nu, r,n,\sigma,\sigma'}  &\Big[ g_{\nu r}^{\sigma\sigma'}\pqty{U_{\boldsymbol{q},r,n}^\dagger b_{\boldsymbol{q},n} -V_{\boldsymbol{q},r,n}^\dagger b_{-\boldsymbol{q},n}^\dagger } c_{\boldsymbol{k}+\boldsymbol{q}+\boldsymbol{Q}_\nu,\sigma}^\dagger c_{\boldsymbol{k},\sigma'} \nonumber \\
    &+  g_{\overline{\nu}  r}^{\sigma'\sigma *}\pqty{-V_{-\boldsymbol{q},r,n}^T b_{\boldsymbol{q},n} +U_{-\boldsymbol{q},r,n}^T b_{-\boldsymbol{q},n}^\dagger } c_{\boldsymbol{k}+\boldsymbol{q}+\boldsymbol{Q}_\nu,\sigma}^\dagger c_{\boldsymbol{k},\sigma'} \Big].
\end{align}

\subsection{Effective interaction} \label{sec:effectiveint}

Let $H = H_0 +\eta H_1$, with
\begin{equation}
    H_0 = \sum_{\boldsymbol{k}\in \text{eBZ},\sigma} \epsilon_{\boldsymbol{k}} c_{\boldsymbol{k}\sigma}^\dagger c_{\boldsymbol{k}\sigma} + \sum_{\boldsymbol{q} \in \text{mBZ}, n} \omega_{\boldsymbol{q},n} b_{\boldsymbol{q},n}^\dagger b_{\boldsymbol{q},n},
\end{equation}
$\eta H_1 = H_{\text{em}}$. 
By considering $\eta H_1$ as a perturbation, an effective electron-electron interaction is derived using the Schrieffer-Wolff transformation \cite{SchriefferWolff},
$
    H_{\text{eff}} = e^{-\eta S} H e^{-\eta S}.
$
The Baker–Campbell–Hausdorff expansion can be used to rewrite this as a perturbation series
\begin{equation}
    H_{\text{eff}} = H_0 + \eta H_{1} + \eta [H_0, S] + \eta^2 [H_{1}, S] + \frac{1}{2} \eta^2  [[H_0, S],S] + \order{\eta^3}.
\end{equation}
Treating $\eta$ as a smallness parameter, terms of order $\eta^3$ and higher are ignored.
To eliminate terms containing two fermion operators in the effective Hamiltonian, terms of order $\eta$ should be zero. Hence, the generator $\eta S$ should be chosen such that
$
    \eta H_{1} + \eta [H_0, S] = 0.
$
Then,
$
    H_{\text{eff}} \approx H_0 + \frac{1}{2} [\eta H_{1}, \eta S].
$
The resulting terms will consist of four electron operators, giving an effective electron-electron interaction.

We use the ansatz
\begin{align}
    \eta S = \sum_{\boldsymbol{k}\in \text{eBZ},\boldsymbol{q}\in \text{mBZ}} \sum_{\nu, r,n,\sigma,\sigma'}  &\Big[ g_{\nu r}^{\sigma\sigma'}\pqty{x_{\boldsymbol{k}\boldsymbol{q},n\nu}^{\sigma\sigma'}U_{\boldsymbol{q},r,n}^\dagger b_{\boldsymbol{q},n} -y_{\boldsymbol{k}\boldsymbol{q},n\nu}^{\sigma\sigma'}V_{\boldsymbol{q},r,n}^\dagger b_{-\boldsymbol{q},n}^\dagger } c_{\boldsymbol{k}+\boldsymbol{q}+\boldsymbol{Q}_\nu,\sigma}^\dagger c_{\boldsymbol{k},\sigma'} \nonumber \\
    &+  g_{\overline{\nu}  r}^{\sigma'\sigma *}\pqty{-z_{\boldsymbol{k}\boldsymbol{q},n\nu}^{\sigma\sigma'}V_{-\boldsymbol{q},r,n}^T b_{\boldsymbol{q},n} +w_{\boldsymbol{k}\boldsymbol{q},n\nu}^{\sigma\sigma'}U_{-\boldsymbol{q},r,n}^T b_{-\boldsymbol{q},n}^\dagger } c_{\boldsymbol{k}+\boldsymbol{q}+\boldsymbol{Q}_\nu,\sigma}^\dagger c_{\boldsymbol{k},\sigma'} \Big],
\end{align}
and write this as
$
    \eta S = \sum_r (\eta S_x^{(r)} + \eta S_y^{(r)} + \eta S_z^{(r)} + \eta S_w^{(r)}),
$
where we have split it into the parts at a specific $r$ that include the $x_{\boldsymbol{k}\boldsymbol{q},n\nu}^{\sigma\sigma'}$ factor  and so on. The corresponding terms in $\eta H_1$ are named $\eta H_{1x}^{(r)}$. 

We want $\eta H_1 + [H_0, \eta S] = 0$, i.e., $H_0 \eta S -\eta S H_0 = -\eta H_1$.  Our approach is to calculate $H_0 \eta S -\eta S H_0$ by putting it inside left and right eigenstates containing magnons and electrons of specified types, $\bra{n}H_0 \eta S -\eta S H_0 \ket{m}$. To determine $x_{\boldsymbol{k}\boldsymbol{q},n\nu}^{\sigma\sigma'}$, the states $\bra{n}$ and $\ket{m}$ need to be chosen such that $\bra{n} \eta S_x^{(r)} \ket{m} \neq 0$. Note that $\eta S_x^{(r)} \ket{m} \sim \ket{n}$ and $\bra{n} \eta S_x^{(r)} \sim \bra{m}$ are required for $\bra{n} \eta S_x^{(r)} \ket{m} \neq 0$. 
Due to the Pauli principle the choice of electrons is unique. For magnons, we could in principle add an arbitrary number of magnons to the states $\bra{n}$ and $\ket{m}$, with the constraint that we add the exact same states in both $\bra{n}$ and $\ket{m}$, to keep $\bra{n} \eta S_x^{(r)} \ket{m} \neq 0$. The energies of these additional magnons would simply drop out in the following equations, and have no effect on $x_{\boldsymbol{k}\boldsymbol{q},n\nu}^{\sigma\sigma'}$. Choosing
$
    \bra{n} = \bra{\boldsymbol{k}+\boldsymbol{q}+\boldsymbol{Q}_\nu,\sigma ; 0}, \ket{m} = \ket{\boldsymbol{k}, \sigma' ; \boldsymbol{q}, n},
$
gives
\begin{equation}
    \bra{n} H_0 \eta S_x^{(r)} -\eta S_x^{(r)} H_0 \ket{m} = (\epsilon_{\boldsymbol{k}+\boldsymbol{q}+\boldsymbol{Q}_\nu,\sigma} - \epsilon_{\boldsymbol{k},\sigma'} - \omega_{\boldsymbol{q},n})\bra{n} \eta S_x^{(r)} \ket{m}.
\end{equation}
Now, we have two expressions for $\bra{n} \eta S_x^{(r)} \ket{m}$, 
\begin{equation}
    \bra{n} \eta S_x^{(r)} \ket{m} = x_{\boldsymbol{k}\boldsymbol{q},n\nu}^{\sigma\sigma'}\bra{n} \eta H_{1x}^{(r)} \ket{m} = -\frac{\bra{n} \eta H_{1x}^{(r)} \ket{m}}{\epsilon_{\boldsymbol{k}+\boldsymbol{q}+\boldsymbol{Q}_\nu,\sigma} - \epsilon_{\boldsymbol{k},\sigma'} - \omega_{\boldsymbol{q},n}}.
\end{equation}
Both are equal if
$
    x_{\boldsymbol{k}\boldsymbol{q},n\nu}^{\sigma\sigma'} = 1/(\epsilon_{\boldsymbol{k},\sigma'} -\epsilon_{\boldsymbol{k}+\boldsymbol{q}+\boldsymbol{Q}_\nu,\sigma} + \omega_{\boldsymbol{q},n}).
$
Similarly, $z_{\boldsymbol{k}\boldsymbol{q},n\nu}^{\sigma\sigma'} = x_{\boldsymbol{k}\boldsymbol{q},n\nu}^{\sigma\sigma'}$.

For $\bra{n} \eta S_y^{(r)} \ket{m} \neq 0$, we choose
$
    \bra{n} = \bra{\boldsymbol{k}+\boldsymbol{q}+\boldsymbol{Q}_\nu,\sigma ; -\boldsymbol{q}, n}, \ket{m} = \ket{\boldsymbol{k}, \sigma' ; 0}.
$
Then,
\begin{equation}
    \bra{n} H_0 \eta S_y^{(r)} -\eta S_y^{(r)} H_0 \ket{m} = (\epsilon_{\boldsymbol{k}+\boldsymbol{q}+\boldsymbol{Q}_\nu,\sigma} - \epsilon_{\boldsymbol{k},\sigma'} + \omega_{-\boldsymbol{q},n})\bra{n} \eta S_y^{(r)} \ket{m}, 
\end{equation}
\begin{equation}
    \bra{n} \eta S_y^{(r)} \ket{m} = y_{\boldsymbol{k}\boldsymbol{q},n\nu}^{\sigma\sigma'}\bra{n} \eta H_{1y}^{(r)} \ket{m} = -\frac{\bra{n} \eta H_{1y}^{(r)} \ket{m}}{\epsilon_{\boldsymbol{k}+\boldsymbol{q}+\boldsymbol{Q}_\nu,\sigma} - \epsilon_{\boldsymbol{k},\sigma'} + \omega_{-\boldsymbol{q},n}}.
\end{equation}
This is achieved for
$
    y_{\boldsymbol{k}\boldsymbol{q},n\nu}^{\sigma\sigma'} = 1/(\epsilon_{\boldsymbol{k},\sigma'} -\epsilon_{\boldsymbol{k}+\boldsymbol{q}+\boldsymbol{Q}_\nu,\sigma} - \omega_{-\boldsymbol{q},n}).
$
Similarly, $w_{\boldsymbol{k}\boldsymbol{q},n\nu}^{\sigma\sigma'} = y_{\boldsymbol{k}\boldsymbol{q},n\nu}^{\sigma\sigma'}$. 
Inserting $\epsilon_{\boldsymbol{k},\sigma} = \epsilon_{\boldsymbol{k}}$ gives
\begin{equation}
    x_{\boldsymbol{k}\boldsymbol{q}n\nu} = \frac{1}{\epsilon_{\boldsymbol{k}} -\epsilon_{\boldsymbol{k}+\boldsymbol{q}+\boldsymbol{Q}_\nu} + \omega_{\boldsymbol{q},n}},
\qquad
    y_{\boldsymbol{k}\boldsymbol{q}n\nu} = \frac{1}{\epsilon_{\boldsymbol{k}} -\epsilon_{\boldsymbol{k}+\boldsymbol{q}+\boldsymbol{Q}_\nu} - \omega_{-\boldsymbol{q},n}}.
\end{equation}

With $\eta H_1 + [H_0, \eta S] = 0$ now ensured, we have
$
    H_{\text{eff}} =  H_0 -  [\eta S, \eta H_1]/2.
$
Let $a$ and $c$ stand for magnon operators and $B$ and $D$ stand for fermion operator pairs. Hence, lowercase letters always commute with capital letters. We are considering commutators on the form
$
    [aB, cD] = [a,c]BD + ca[B,D].
$
If the number $[a,c] \neq 0$, then $BD$ are the pairing terms we are interested in, involving 4 fermion operators. 
Other possible terms involve e.g.~two fermion operators or two magnon operators in $H_{\text{eff}}$. These represent corrections that are higher order in perturbation theory than the pairing terms we will focus on.

To get pairing terms with 4 fermion operators, requires commutators of the form
\begin{equation}
    [b_{\boldsymbol{q},n} c_{\boldsymbol{k}+\boldsymbol{q}+\boldsymbol{Q}_\nu,\alpha}^\dagger c_{\boldsymbol{k},\alpha'}, b_{\boldsymbol{q},n}^\dagger c_{\boldsymbol{k}'-\boldsymbol{q}+\boldsymbol{Q}_{\nu'},\beta}^\dagger c_{\boldsymbol{k}',\beta'}] = \underbrace{[b_{\boldsymbol{q},n}, b_{\boldsymbol{q},n}^\dagger]}_{=1} c_{\boldsymbol{k}+\boldsymbol{q}+\boldsymbol{Q}_\nu,\alpha}^\dagger c_{\boldsymbol{k},\alpha'}c_{\boldsymbol{k}'-\boldsymbol{q}+\boldsymbol{Q}_{\nu'},\beta}^\dagger c_{\boldsymbol{k}',\beta'} + \underbrace{\dots}_{\text{neglected}},
\end{equation}
and
\begin{equation}
    [b_{-\boldsymbol{q},n}^\dagger c_{\boldsymbol{k}+\boldsymbol{q}+\boldsymbol{Q}_\nu,\alpha}^\dagger c_{\boldsymbol{k},\alpha'}, b_{-\boldsymbol{q},n} c_{\boldsymbol{k}'-\boldsymbol{q}+\boldsymbol{Q}_{\nu'},\beta}^\dagger c_{\boldsymbol{k}',\beta'}] = \underbrace{[b_{-\boldsymbol{q},n}^\dagger, b_{-\boldsymbol{q},n}]}_{=-1} c_{\boldsymbol{k}+\boldsymbol{q}+\boldsymbol{Q}_\nu,\alpha}^\dagger c_{\boldsymbol{k},\alpha'}c_{\boldsymbol{k}'-\boldsymbol{q}+\boldsymbol{Q}_{\nu'},\beta}^\dagger c_{\boldsymbol{k}',\beta'} +\underbrace{\dots}_{\text{neglected}}.
\end{equation}
Keeping only pairing terms we write
\begin{equation}
\label{eq:pairing}
    H_{\text{Pair}} = - \frac{1}{2} [\eta S, \eta H_1] = \sum_{\boldsymbol{k}\boldsymbol{k}' \in \text{eBZ}, \boldsymbol{q} \in \text{mBZ}} \sum_{\nu\nu' }\sum_{\alpha\alpha'\beta\beta'} V_{\boldsymbol{k}\boldsymbol{q}\nu\nu'}^{\alpha\alpha'\beta\beta'} c_{\boldsymbol{k}+\boldsymbol{q}+\boldsymbol{Q}_\nu,\alpha}^\dagger c_{\boldsymbol{k},\alpha'}c_{\boldsymbol{k}'-\boldsymbol{q}+\boldsymbol{Q}_{\nu'},\beta}^\dagger c_{\boldsymbol{k}',\beta'},
\end{equation}
with
\begin{align}
\label{eq:couplingfunction}
    V_{\boldsymbol{k}\boldsymbol{q}\nu\nu'}^{\alpha\alpha'\beta\beta'} &= -\frac{1}{2}\sum_{rr' n} \Big[ g_{\nu r}^{\alpha \alpha'} g_{\nu' r'}^{\beta \beta'} U_{\boldsymbol{q},r,n}^\dagger (-V_{-\boldsymbol{q},r',n}^\dagger)x_{\boldsymbol{k}\boldsymbol{q}n\nu}+g_{\nu r}^{\alpha \alpha'} g_{\overline{\nu}' r'}^{\beta' \beta*} U_{\boldsymbol{q},r,n}^\dagger U_{\boldsymbol{q},r',n}^T x_{\boldsymbol{k}\boldsymbol{q}n\nu} \nonumber \\
    &+g_{\overline{\nu} r}^{\alpha' \alpha*} g_{\nu' r'}^{\beta \beta'} (-V_{-\boldsymbol{q},r,n}^T) (-V_{-\boldsymbol{q},r',n}^\dagger)x_{\boldsymbol{k}\boldsymbol{q}n\nu} +g_{\overline{\nu} r}^{\alpha' \alpha*} g_{\overline{\nu}' r'}^{\beta' \beta*} (-V_{-\boldsymbol{q},r,n}^T) U_{\boldsymbol{q},r',n}^T x_{\boldsymbol{k}\boldsymbol{q}n\nu} \nonumber \\
    &-g_{\nu r}^{\alpha \alpha'} g_{\nu' r'}^{\beta \beta'} (-V_{\boldsymbol{q},r,n}^\dagger) U_{-\boldsymbol{q},r',n}^\dagger y_{\boldsymbol{k}\boldsymbol{q}n\nu} -g_{\nu r}^{\alpha \alpha'} g_{\overline{\nu}' r'}^{\beta' \beta*} (-V_{\boldsymbol{q},r,n}^\dagger) (-V_{\boldsymbol{q},r',n}^T) y_{\boldsymbol{k}\boldsymbol{q}n\nu} \nonumber \\
    &-g_{\overline{\nu} r}^{\alpha' \alpha*} g_{\nu' r'}^{\beta \beta'} U_{-\boldsymbol{q},r,n}^T U_{-\boldsymbol{q},r',n}^\dagger y_{\boldsymbol{k}\boldsymbol{q}n\nu} -g_{\overline{\nu} r}^{\alpha' \alpha*} g_{\overline{\nu}' r'}^{\beta' \beta*} U_{-\boldsymbol{q},r,n}^T (-V_{\boldsymbol{q},r',n}^T) y_{\boldsymbol{k}\boldsymbol{q}n\nu} \Big].
\end{align}
Note that $V_{\boldsymbol{k}\boldsymbol{q}\nu\nu'}^{\alpha\alpha'\beta\beta'}/t$ is of order $(\Bar{J}/t)^2$. Meanwhile, the electron and magnon dispersions are in the denominator. Inserting any renormalizations of these, of order $\Bar{J}/t$ from the HP transformation of $H_{\text{em}}$, or of order $(\Bar{J}/t)^2$ from neglected terms in the Schrieffer-Wolff transformation, would lead to higher order terms than $(\Bar{J}/t)^2$ in $V_{\boldsymbol{k}\boldsymbol{q}\nu\nu'}^{\alpha\alpha'\beta\beta'}/t$. We consider weak-coupling, $\Bar{J}\ll t$, and limit $V_{\boldsymbol{k}\boldsymbol{q}\nu\nu'}^{\alpha\alpha'\beta\beta'}/t$ to order $(\Bar{J}/t)^2$. That is why all renormalizations of the electron and magnon bands due to electron-magnon coupling have been ignored.
From 
\begin{equation}
    T_{\boldsymbol{q}}^{-1} = \begin{pmatrix} U_{\boldsymbol{q}}^\dagger & -V_{\boldsymbol{q}}^\dagger \\ -V_{-\boldsymbol{q}}^T & U_{-\boldsymbol{q}}^T    \end{pmatrix},
\end{equation}
we see that all combinations $UU, UV, VU, VV$ in $V_{\boldsymbol{k}\boldsymbol{q}\nu\nu'}^{\alpha\alpha'\beta\beta'}$ come from the same column of $T_{\boldsymbol{q}}^{-1}$, only that one of the terms are complex conjugated. Hence, any arbitrary phase factor that has been multiplied to a column of $T_{\boldsymbol{q}}^{-1}$ is canceled in $V_{\boldsymbol{k}\boldsymbol{q}\nu\nu'}^{\alpha\alpha'\beta\beta'}$. It can be shown that the pairing Hamiltonian $H_{\text{Pair}}$ is Hermitian, as expected.

\section{Superconductivity} \label{sec:SC}
\subsection{BCS-type pairing}
We now insert BCS-type momentum pairing in Eq.~\eqref{eq:pairing}. We set $\boldsymbol{k}' = -\boldsymbol{k}$, and at each $\nu$ we choose $\nu' = \Bar{\nu}$ such that $\boldsymbol{Q}_{\Bar{\nu}} = -\boldsymbol{Q}_{\nu}$ modulo some reciprocal lattice vector $\boldsymbol{G}$ from the triangular lattice. Defining a new $\boldsymbol{k}' = \boldsymbol{k}+\boldsymbol{q}+\boldsymbol{Q}_\nu$ gives
\begin{equation}
    H = \sum_{\boldsymbol{k}\boldsymbol{k}'\in \text{eBZ}} \sum_{\sigma_1 \sigma_2 \sigma_3 \sigma_4} V_{\boldsymbol{k}\boldsymbol{k}'}^{\sigma_1 \sigma_2 \sigma_3 \sigma_4} c_{\boldsymbol{k}',\sigma_1}^\dagger c_{-\boldsymbol{k}',\sigma_2}^\dagger c_{-\boldsymbol{k},\sigma_3}c_{\boldsymbol{k},\sigma_4}.
\end{equation}
Here, $V_{\boldsymbol{k}\boldsymbol{k}'}^{\sigma_1 \sigma_2 \sigma_3 \sigma_4} = V_{\boldsymbol{k}\boldsymbol{q}\nu\Bar{\nu}}^{\sigma_1 \sigma_4 \sigma_2 \sigma_3}$. The momentum $\boldsymbol{q} \in \text{mBZ}$ and Umklapp process $\nu$ are found from $\boldsymbol{k}'-\boldsymbol{k}$. $\nu$ is determined from which $\boldsymbol{Q}_\nu$ is closest to $\boldsymbol{k}'-\boldsymbol{k}$. Then, $\boldsymbol{q} = \boldsymbol{k}'-\boldsymbol{k}-\boldsymbol{Q}_\nu$.

We instead want
\begin{equation}
\label{eq:Vbargen}
    H = \frac{1}{2} \sum_{\boldsymbol{k}\boldsymbol{k}'\in \text{eBZ}} \sum_{\sigma_1 \sigma_2 \sigma_3 \sigma_4} \bar{V}_{\boldsymbol{k}\boldsymbol{k}'}^{\sigma_1 \sigma_2 \sigma_3 \sigma_4} c_{\boldsymbol{k}',\sigma_1}^\dagger c_{-\boldsymbol{k}',\sigma_2}^\dagger c_{-\boldsymbol{k},\sigma_3}c_{\boldsymbol{k},\sigma_4},
\end{equation}
where $\bar{V}_{\boldsymbol{k}\boldsymbol{k}'}^{\sigma_1 \sigma_2 \sigma_3 \sigma_4} $, unlike our original $V_{\boldsymbol{k}\boldsymbol{k}'}^{\sigma_1 \sigma_2 \sigma_3 \sigma_4}$, obeys the symmetries \cite{Sigrist}
\begin{equation}
\label{eq:Sigristsymm}
    \bar{V}_{\boldsymbol{k}\boldsymbol{k}'}^{\sigma_1 \sigma_2 \sigma_3 \sigma_4} = \bar{V}_{-\boldsymbol{k},-\boldsymbol{k}'}^{\sigma_2 \sigma_1 \sigma_4 \sigma_3} = -\bar{V}_{-\boldsymbol{k},\boldsymbol{k}'}^{\sigma_1 \sigma_2 \sigma_4 \sigma_3} = -\bar{V}_{\boldsymbol{k},-\boldsymbol{k}'}^{\sigma_2 \sigma_1 \sigma_3 \sigma_4}.
\end{equation}
Additionally, $\bar{V}_{\boldsymbol{k}\boldsymbol{k}'}^{\sigma_1 \sigma_2 \sigma_3 \sigma_4} = (\bar{V}_{\boldsymbol{k}'\boldsymbol{k}}^{\sigma_4 \sigma_3 \sigma_2 \sigma_1})^*$ automatically from Hermiticity.

Imagine we sum over $\boldsymbol{k}'', \boldsymbol{k}''', \sigma'_1 \sigma'_2 \sigma'_3 \sigma'_4$ in the original expression and write down the 4 terms that can give the operator structure $c_{\boldsymbol{k}',\sigma_1}^\dagger c_{-\boldsymbol{k}',\sigma_2}^\dagger c_{-\boldsymbol{k},\sigma_3}c_{\boldsymbol{k},\sigma_4}$ after applying anticommutators,
\begin{align}
    &V_{\boldsymbol{k}\boldsymbol{k}'}^{\sigma_1 \sigma_2 \sigma_3 \sigma_4} c_{\boldsymbol{k}',\sigma_1}^\dagger c_{-\boldsymbol{k}',\sigma_2}^\dagger c_{-\boldsymbol{k},\sigma_3}c_{\boldsymbol{k},\sigma_4} +V_{-\boldsymbol{k},-\boldsymbol{k}'}^{\sigma_2 \sigma_1 \sigma_4 \sigma_3}  \underbrace{c_{-\boldsymbol{k}',\sigma_2}^\dagger c_{\boldsymbol{k}',\sigma_1}^\dagger}_{\text{anticommute}, -1} \underbrace{c_{\boldsymbol{k},\sigma_4}c_{-\boldsymbol{k},\sigma_3}}_{\text{anticommute}, -1} \nonumber \\
    &+V_{-\boldsymbol{k},\boldsymbol{k}'}^{\sigma_1 \sigma_2 \sigma_4 \sigma_3} c_{\boldsymbol{k}',\sigma_1}^\dagger c_{-\boldsymbol{k}',\sigma_2}^\dagger \underbrace{c_{\boldsymbol{k},\sigma_4}c_{-\boldsymbol{k},\sigma_3}}_{\text{anticommute}, -1} +V_{\boldsymbol{k},-\boldsymbol{k}'}^{\sigma_2 \sigma_1 \sigma_3 \sigma_4}\underbrace{c_{-\boldsymbol{k}',\sigma_2}^\dagger c_{\boldsymbol{k}',\sigma_1}^\dagger}_{\text{anticommute}, -1} c_{-\boldsymbol{k},\sigma_3}c_{\boldsymbol{k},\sigma_4} \nonumber \\
    =&(V_{\boldsymbol{k}\boldsymbol{k}'}^{\sigma_1 \sigma_2 \sigma_3 \sigma_4}+V_{-\boldsymbol{k},-\boldsymbol{k}'}^{\sigma_2 \sigma_1 \sigma_4 \sigma_3}-V_{-\boldsymbol{k},\boldsymbol{k}'}^{\sigma_1 \sigma_2 \sigma_4 \sigma_3}-V_{\boldsymbol{k},-\boldsymbol{k}'}^{\sigma_2 \sigma_1 \sigma_3 \sigma_4})c_{\boldsymbol{k}',\sigma_1}^\dagger c_{-\boldsymbol{k}',\sigma_2}^\dagger c_{-\boldsymbol{k},\sigma_3}c_{\boldsymbol{k},\sigma_4} \nonumber \\
    \equiv&\tilde{V}_{\boldsymbol{k}\boldsymbol{k}'}^{\sigma_1 \sigma_2 \sigma_3 \sigma_4}c_{\boldsymbol{k}',\sigma_1}^\dagger c_{-\boldsymbol{k}',\sigma_2}^\dagger c_{-\boldsymbol{k},\sigma_3}c_{\boldsymbol{k},\sigma_4}.
\end{align}
Now, define $\bar{V}$ through
\begin{equation}
    \bar{V}_{\boldsymbol{k}\boldsymbol{k}'}^{\sigma_1 \sigma_2 \sigma_3 \sigma_4} + \bar{V}_{-\boldsymbol{k},-\boldsymbol{k}'}^{\sigma_2 \sigma_1 \sigma_4 \sigma_3}  -\bar{V}_{-\boldsymbol{k},\boldsymbol{k}'}^{\sigma_1 \sigma_2 \sigma_4 \sigma_3}  -\bar{V}_{\boldsymbol{k},-\boldsymbol{k}'}^{\sigma_2 \sigma_1 \sigma_3 \sigma_4} = 2\tilde{V}_{\boldsymbol{k}\boldsymbol{k}'}^{\sigma_1 \sigma_2 \sigma_3 \sigma_4}.
\end{equation}
The factor 2 is used to comply to the factor $1/2$ in front of the sum in Eq.~\eqref{eq:Vbargen}.
Using the symmetries in Eq.~\eqref{eq:Sigristsymm} we define
\begin{equation}
\label{eq:Vbardef}
    \bar{V}_{\boldsymbol{k}\boldsymbol{k}'}^{\sigma_1 \sigma_2 \sigma_3 \sigma_4} = \frac{\tilde{V}_{\boldsymbol{k}\boldsymbol{k}'}^{\sigma_1 \sigma_2 \sigma_3 \sigma_4}}{2} = \frac{V_{\boldsymbol{k}\boldsymbol{k}'}^{\sigma_1 \sigma_2 \sigma_3 \sigma_4}+V_{-\boldsymbol{k},-\boldsymbol{k}'}^{\sigma_2 \sigma_1 \sigma_4 \sigma_3}-V_{-\boldsymbol{k},\boldsymbol{k}'}^{\sigma_1 \sigma_2 \sigma_4 \sigma_3}-V_{\boldsymbol{k},-\boldsymbol{k}'}^{\sigma_2 \sigma_1 \sigma_3 \sigma_4}}{2}.
\end{equation}

\subsection{Generalized BCS theory}
We follow the generalized BCS theory presented in Ref.~\cite{Sigrist}. Having performed the Schrieffer-Wolff transformation, we can now describe the system using only electron operators,
\begin{equation}
    H = \sum_{\boldsymbol{k}\sigma}\epsilon_{\boldsymbol{k}}c_{\boldsymbol{k},\sigma}^\dagger c_{\boldsymbol{k},\sigma} + \frac{1}{2} \sum_{\boldsymbol{k}\boldsymbol{k}'\in \text{eBZ}} \sum_{\sigma_1 \sigma_2 \sigma_3 \sigma_4} \bar{V}_{\boldsymbol{k}\boldsymbol{k}'}^{\sigma_1 \sigma_2 \sigma_3 \sigma_4} c_{\boldsymbol{k}',\sigma_1}^\dagger c_{-\boldsymbol{k}',\sigma_2}^\dagger c_{-\boldsymbol{k},\sigma_3}c_{\boldsymbol{k},\sigma_4}.
\end{equation}
The Cooper pair mean field is defined as an ensemble average,
\begin{equation}
    b_{\boldsymbol{k}\sigma\sigma'} = \langle c_{-\boldsymbol{k},\sigma}c_{\boldsymbol{k},\sigma'} \rangle, \qquad  b_{\boldsymbol{k}\sigma\sigma'}^\dagger = \langle c_{\boldsymbol{k},\sigma'}^\dagger c_{-\boldsymbol{k},\sigma}^\dagger \rangle.
\end{equation}
Let
$
    c_{-\boldsymbol{k},\sigma}c_{\boldsymbol{k},\sigma'} = b_{\boldsymbol{k}\sigma\sigma'} + \delta b_{\boldsymbol{k}\sigma\sigma'},
$
with $\delta b_{\boldsymbol{k}\sigma\sigma'} = c_{-\boldsymbol{k},\sigma}c_{\boldsymbol{k},\sigma'} - \langle c_{-\boldsymbol{k},\sigma}c_{\boldsymbol{k},\sigma'} \rangle$. Assuming the fluctuations around mean field are small, $(\delta b)^2$ is ignored. Inserting gives
\begin{equation}
    c_{\boldsymbol{k}',\sigma_1}^\dagger c_{-\boldsymbol{k}',\sigma_2}^\dagger c_{-\boldsymbol{k},\sigma_3}c_{\boldsymbol{k},\sigma_4} = b_{\boldsymbol{k}\sigma_3\sigma_4}c_{\boldsymbol{k}',\sigma_1}^\dagger c_{-\boldsymbol{k}',\sigma_2}^\dagger + b_{\boldsymbol{k}'\sigma_2\sigma_1}^\dagger c_{-\boldsymbol{k},\sigma_3}c_{\boldsymbol{k},\sigma_4} - b_{\boldsymbol{k}\sigma_3\sigma_4}b_{\boldsymbol{k}'\sigma_2\sigma_1}^\dagger.
\end{equation}
Then,
\begin{align}
    H =& \sum_{\boldsymbol{k}\sigma}\epsilon_{\boldsymbol{k}}c_{\boldsymbol{k},\sigma}^\dagger c_{\boldsymbol{k},\sigma} + \frac{1}{2} \sum_{\boldsymbol{k}\sigma_1 \sigma_2 } \Delta_{\boldsymbol{k}\sigma_1\sigma_2}  b_{\boldsymbol{k}\sigma_2\sigma_1}^\dagger -\frac{1}{2}\sum_{\boldsymbol{k}\sigma_1 \sigma_2 } (\Delta_{\boldsymbol{k}\sigma_1\sigma_2}
    c_{\boldsymbol{k},\sigma_1}^\dagger c_{-\boldsymbol{k},\sigma_2}^\dagger + \Delta_{\boldsymbol{k}\sigma_2\sigma_1}^\dagger c_{-\boldsymbol{k},\sigma_1}c_{\boldsymbol{k},\sigma_2}).
\end{align}
The gap functions are defined as
\begin{align}
    \Delta_{\boldsymbol{k}\sigma_1\sigma_2} &= -\sum_{\boldsymbol{k}'\sigma_3\sigma_4} \bar{V}_{\boldsymbol{k}'\boldsymbol{k}}^{\sigma_1 \sigma_2 \sigma_3 \sigma_4} b_{\boldsymbol{k}' \sigma_3 \sigma_4}, \qquad \Delta_{\boldsymbol{k}\sigma_2\sigma_1}^\dagger = -\sum_{\boldsymbol{k}'\sigma_3\sigma_4} \bar{V}_{\boldsymbol{k}\boldsymbol{k}'}^{\sigma_3 \sigma_4 \sigma_1 \sigma_2} b_{\boldsymbol{k}' \sigma_4 \sigma_3}^\dagger. 
\end{align}

We write this as a matrix diagonalization problem through
\begin{equation}
    H = H_0 +\frac{1}{2}\sum_{\boldsymbol{k}} \boldsymbol{c}_{\boldsymbol{k}}^\dagger H_{\boldsymbol{k}} \boldsymbol{c}_{\boldsymbol{k}}, 
\end{equation}
with $\boldsymbol{c}_{\boldsymbol{k}}^\dagger = (c_{\boldsymbol{k}\uparrow}^\dagger, c_{\boldsymbol{k}\downarrow}^\dagger, c_{-\boldsymbol{k}\uparrow}, c_{-\boldsymbol{k}\downarrow})$ and
\begin{equation}
    H_{\boldsymbol{k}} = 
    \begin{pmatrix}
    \epsilon_{\boldsymbol{k}}  & 0 & \Delta_{\boldsymbol{k}\uparrow\uparrow} & \Delta_{\boldsymbol{k}\uparrow\downarrow}  \\
    0 & \epsilon_{\boldsymbol{k}} & \Delta_{\boldsymbol{k}\downarrow\uparrow} & \Delta_{\boldsymbol{k}\downarrow\downarrow}\\
    \Delta_{\boldsymbol{k}\uparrow\uparrow}^\dagger & \Delta_{\boldsymbol{k}\downarrow\uparrow}^\dagger & -\epsilon_{\boldsymbol{k}} & 0 \\
    \Delta_{\boldsymbol{k}\uparrow\downarrow}^\dagger & \Delta_{\boldsymbol{k}\downarrow\downarrow}^\dagger & 0 & -\epsilon_{\boldsymbol{k}}  
    \end{pmatrix}.
\end{equation}
Rewriting the Hamiltonian in this way leads to a shift of $\sum_{\boldsymbol{k}} \epsilon_{\boldsymbol{k}}$.
$H_{\boldsymbol{k}}$ is diagonalized by a unitary matrix 
through $U_{\boldsymbol{k}}^\dagger H_{\boldsymbol{k}} U_{\boldsymbol{k}} = \hat{E}_{\boldsymbol{k}}$, $\hat{E}_{\boldsymbol{k}} = \operatorname{diag}(E_{\boldsymbol{k}+}, E_{\boldsymbol{k}-}, -E_{-\boldsymbol{k}+}, -E_{-\boldsymbol{k}-})$. The diagonalized operators are $\boldsymbol{\gamma}_{\boldsymbol{k}}^\dagger = (\gamma_{\boldsymbol{k}+}^\dagger, \gamma_{\boldsymbol{k}-}^\dagger, \gamma_{-\boldsymbol{k}+}, \gamma_{-\boldsymbol{k}-})$. 
The energies can be written
\begin{equation}
    E_{\boldsymbol{k}\pm} = \sqrt{\epsilon_{\boldsymbol{k}}^2 + \frac{1}{2}\Tr\hat{\Delta}_{\boldsymbol{k}} \hat{\Delta}_{\boldsymbol{k}}^\dagger \pm \frac{1}{2}\sqrt{A_{\boldsymbol{k}}}},
\end{equation}
\begin{equation}
    \frac{1}{2}\Tr\hat{\Delta}_{\boldsymbol{k}} \hat{\Delta}_{\boldsymbol{k}}^\dagger =\frac{1}{2} (|\Delta_{\boldsymbol{k}\uparrow\uparrow}|^2 + |\Delta_{\boldsymbol{k}\uparrow\downarrow}|^2 + |\Delta_{\boldsymbol{k}\downarrow\uparrow}|^2 + |\Delta_{\boldsymbol{k}\downarrow\downarrow}|^2   ),
\end{equation}
\begin{align}
    A_{\boldsymbol{k}} =& (|\Delta_{\boldsymbol{k}\uparrow\uparrow}|^2-|\Delta_{\boldsymbol{k}\downarrow\downarrow}|^2)^2 + (|\Delta_{\boldsymbol{k}\uparrow\downarrow}|^2 - |\Delta_{\boldsymbol{k}\downarrow\uparrow}|^2)^2 +2(|\Delta_{\boldsymbol{k}\uparrow\uparrow}|^2 +|\Delta_{\boldsymbol{k}\downarrow\downarrow}|^2)(|\Delta_{\boldsymbol{k}\uparrow\downarrow}|^2 + |\Delta_{\boldsymbol{k}\downarrow\uparrow}|^2)\nonumber \\ &+4\Delta_{\boldsymbol{k}\uparrow\uparrow}\Delta_{\boldsymbol{k}\downarrow\downarrow}\Delta_{\boldsymbol{k}\uparrow\downarrow}^\dagger \Delta_{\boldsymbol{k}\downarrow\uparrow}^\dagger + 4\Delta_{\boldsymbol{k}\uparrow\uparrow}^\dagger\Delta_{\boldsymbol{k}\downarrow\downarrow}^\dagger\Delta_{\boldsymbol{k}\uparrow\downarrow} \Delta_{\boldsymbol{k}\downarrow\uparrow}.
\end{align}
The diagonalized Hamiltonian is
\begin{align}
    H =& H_0 + \sum_{\boldsymbol{k}\eta} E_{\boldsymbol{k}\eta}\gamma_{\boldsymbol{k}\eta}^\dagger\gamma_{\boldsymbol{k}\eta}.
\end{align}
with $H_0 = \sum_{\boldsymbol{k}} \epsilon_{\boldsymbol{k}}-\frac{1}{2}\sum_{\boldsymbol{k}\eta} E_{\boldsymbol{k}\eta} + \frac{1}{2} \sum_{\boldsymbol{k}}\sum_{\sigma_1\sigma_2} \Delta_{\boldsymbol{k}\sigma_1 \sigma_2}b_{\boldsymbol{k}\sigma_2 \sigma_1}^\dagger$. 

\subsubsection{Gap equation}
For a Hamiltonian on the above form, the grand canonical partition function is \cite{SFsuperconductivity}
\begin{equation}
    Z = e^{-\beta F} = e^{-\beta H_0} \prod_{\boldsymbol{k}}(1+e^{-\beta E_{\boldsymbol{k}+}})(1+e^{-\beta E_{\boldsymbol{k}-}}).
\end{equation}
The free energy is
\begin{equation}
    F = \sum_{\boldsymbol{k}} \epsilon_{\boldsymbol{k}}-\frac{1}{2}\sum_{\boldsymbol{k}\eta} E_{\boldsymbol{k}\eta} + \frac{1}{2} \sum_{\boldsymbol{k}}\sum_{\sigma_1\sigma_2} \Delta_{\boldsymbol{k}\sigma_1 \sigma_2} b_{\boldsymbol{k}\sigma_2 \sigma_1}^\dagger -\frac{1}{\beta} \sum_{\boldsymbol{k}\eta} \ln (1+e^{-\beta E_{\boldsymbol{k}\eta}}).
\end{equation}
To obtain the gap equation we minimize $F$ to get an equation for $b_{\boldsymbol{k}\sigma_2 \sigma_1}^\dagger$,
\begin{align}
    \pdv{F}{\Delta_{\boldsymbol{k}\sigma_1\sigma_2}} =& -\frac{1}{2}\sum_{\eta}\pdv{E_{\boldsymbol{k}\eta}}{\Delta_{\boldsymbol{k}\sigma_1\sigma_2}}  +\frac{1}{2}b_{\boldsymbol{k}\sigma_2 \sigma_1}^\dagger-\frac{1}{\beta} \sum_{\eta}\frac{1}{1+e^{-\beta E_{\boldsymbol{k}\eta}}}(-\beta)\pdv{E_{\boldsymbol{k}\eta}}{\Delta_{\boldsymbol{k}\sigma_1\sigma_2}}e^{-\beta E_{\boldsymbol{k}\eta}}  = 0 ,
\end{align}
\begin{equation}
    b_{\boldsymbol{k}\sigma_2 \sigma_1}^\dagger = \sum_{\eta} \pdv{E_{\boldsymbol{k}\eta}}{\Delta_{\boldsymbol{k}\sigma_1\sigma_2}} \underbrace{\pqty{1-2\frac{e^{-x_\eta}}{1+e^{-x_\eta}}}}_{=\frac{1-e^{-x_\eta}}{1+e^{-x_\eta}}=\frac{e^{x_\eta}-1}{e^{x_\eta}+1}=\tanh\frac{x_\eta}{2}},
\end{equation}
with $x_\eta = \beta E_{\boldsymbol{k}\eta}$. The derivative is
\begin{equation}
    \pdv{E_{\boldsymbol{k}\pm}}{\Delta_{\boldsymbol{k}\sigma_1\sigma_2}} = \frac{1}{2E_{\boldsymbol{k}\pm}}\pqty{\frac{1}{2}\Delta_{\boldsymbol{k}\sigma_1\sigma_2}^\dagger \pm \frac{1}{4\sqrt{A_{\boldsymbol{k}}}}\pdv{A_{\boldsymbol{k}}}{\Delta_{\boldsymbol{k}\sigma_1\sigma_2}}  }.
\end{equation}
Define $B_{\boldsymbol{k}\sigma_1\sigma_2}^\dagger \equiv \frac{1}{4\sqrt{A_{\boldsymbol{k}}}}\pdv{A_{\boldsymbol{k}}}{\Delta_{\boldsymbol{k}\sigma_1\sigma_2}}$.
Then,
\begin{equation}
    b_{\boldsymbol{k}\sigma_2 \sigma_1}^\dagger = \sum_{\eta} \pqty{\frac{1}{2}\Delta_{\boldsymbol{k}\sigma_1\sigma_2}^\dagger + \eta B_{\boldsymbol{k}\sigma_1\sigma_2}^\dagger } \underbrace{\frac{\tanh\frac{\beta E_{\boldsymbol{k}\eta}}{2}}{2E_{\boldsymbol{k}\eta}}}_{\equiv\chi_{\boldsymbol{k}\eta}},
\end{equation}
which is inserted in the gap function to get the gap equation,
\begin{equation}
    \label{eq:gapeq}
    \Delta_{\boldsymbol{k}\sigma_1\sigma_2} = -\sum_{\boldsymbol{k}'\sigma_3\sigma_4} \bar{V}_{\boldsymbol{k}'\boldsymbol{k}}^{\sigma_1\sigma_2\sigma_3\sigma_4} \sum_{\eta} \pqty{\frac{1}{2}\Delta_{\boldsymbol{k}'\sigma_4\sigma_3} + \eta B_{\boldsymbol{k}'\sigma_4\sigma_3} } \chi_{\boldsymbol{k}'\eta}.
\end{equation}

It is convenient to introduce new gap functions for singlet [odd in spin, $O(s)$] and unpolarized triplet [even in spin, $E(s)$],
\begin{equation}
    \Delta_{\boldsymbol{k}\uparrow\downarrow}^{O(s)} = \frac{\Delta_{\boldsymbol{k}\uparrow\downarrow}-\Delta_{\boldsymbol{k}\downarrow\uparrow}}{2}, \qquad \Delta_{\boldsymbol{k}\uparrow\downarrow}^{E(s)} = \frac{\Delta_{\boldsymbol{k}\uparrow\downarrow}+\Delta_{\boldsymbol{k}\downarrow\uparrow}}{2}.
\end{equation}
Similarly, we define
\begin{equation}
    B_{\boldsymbol{k}\uparrow\downarrow}^{O(s)} = \frac{B_{\boldsymbol{k}\uparrow\downarrow}-B_{\boldsymbol{k}\downarrow\uparrow}}{2}, \qquad B_{\boldsymbol{k}\uparrow\downarrow}^{E(s)} = \frac{B_{\boldsymbol{k}\uparrow\downarrow}+B_{\boldsymbol{k}\downarrow\uparrow}}{2}.
\end{equation}
Let us also define
\begin{equation}
    \tilde{V}_{\boldsymbol{k}\boldsymbol{k}'} = \frac{1}{2}(\bar{V}_{\boldsymbol{k}\boldsymbol{k}'}^{\downarrow\uparrow\uparrow\downarrow} + \bar{V}_{-\boldsymbol{k},-\boldsymbol{k}'}^{\uparrow\downarrow\downarrow\uparrow} -  \bar{V}_{-\boldsymbol{k},\boldsymbol{k}'}^{\downarrow\uparrow\downarrow\uparrow} - \bar{V}_{\boldsymbol{k},-\boldsymbol{k}'}^{\uparrow\downarrow\uparrow\downarrow}  ) = 2\bar{V}_{\boldsymbol{k}\boldsymbol{k}'}^{\downarrow\uparrow\uparrow\downarrow}.
\end{equation}
Writing out the sum over $\sigma_1\sigma_2$ in Eq.~\eqref{eq:gapeq} and recasting in terms of $ \Delta_{\boldsymbol{k}\uparrow\downarrow}^{O(s)}$ and $ \Delta_{\boldsymbol{k}\uparrow\downarrow}^{E(s)}$ gives the gap equation in Eq.~(3) of the main text. There,
\begin{equation}
\label{eq:V44}
    \mathcal{V}_{\boldsymbol{k}'\boldsymbol{k}} = \begin{pmatrix}
    \tilde{V}_{\boldsymbol{k}'\boldsymbol{k}}^{E(\boldsymbol{k}')E(\boldsymbol{k})}& \bar{V}_{\boldsymbol{k}'\boldsymbol{k}}^{\uparrow\downarrow\uparrow\uparrow E(\boldsymbol{k})} & \bar{V}_{\boldsymbol{k}'\boldsymbol{k}}^{\uparrow\downarrow\downarrow\downarrow E(\boldsymbol{k})} & \tilde{V}_{\boldsymbol{k}'\boldsymbol{k}}^{O(\boldsymbol{k}')E(\boldsymbol{k})} \\
    -2\bar{V}_{\boldsymbol{k}'\boldsymbol{k}}^{\uparrow\uparrow\uparrow\downarrow E(\boldsymbol{k}')} & \bar{V}_{\boldsymbol{k}'\boldsymbol{k}}^{\uparrow\uparrow\uparrow\uparrow} &   \bar{V}_{\boldsymbol{k}'\boldsymbol{k}}^{\uparrow\uparrow\downarrow\downarrow} & 2\bar{V}_{\boldsymbol{k}'\boldsymbol{k}}^{\uparrow\uparrow\uparrow\downarrow O(\boldsymbol{k}')} \\
    -2\bar{V}_{\boldsymbol{k}'\boldsymbol{k}}^{\downarrow\downarrow\uparrow\downarrow E(\boldsymbol{k}')} & \bar{V}_{\boldsymbol{k}'\boldsymbol{k}}^{\downarrow\downarrow\uparrow\uparrow} & \bar{V}_{\boldsymbol{k}'\boldsymbol{k}}^{\downarrow\downarrow\downarrow\downarrow} & 2\bar{V}_{\boldsymbol{k}'\boldsymbol{k}}^{\downarrow\downarrow\uparrow\downarrow O(\boldsymbol{k}')} \\
    \tilde{V}_{\boldsymbol{k}'\boldsymbol{k}}^{E(\boldsymbol{k}')O(\boldsymbol{k})}& \bar{V}_{\boldsymbol{k}'\boldsymbol{k}}^{\uparrow\downarrow\uparrow\uparrow O(\boldsymbol{k})} & \bar{V}_{\boldsymbol{k}'\boldsymbol{k}}^{\uparrow\downarrow\downarrow\downarrow O(\boldsymbol{k})} & \tilde{V}_{\boldsymbol{k}'\boldsymbol{k}}^{O(\boldsymbol{k}')O(\boldsymbol{k})}
    \end{pmatrix},
\end{equation}
with
\begin{align}
    \tilde{V}_{\boldsymbol{k}'\boldsymbol{k}}^{E(\boldsymbol{k}')E(\boldsymbol{k})}=& \frac{1}{4}(\tilde{V}_{-\boldsymbol{k}',-\boldsymbol{k}}+\tilde{V}_{-\boldsymbol{k}',\boldsymbol{k}}+\tilde{V}_{\boldsymbol{k}',\boldsymbol{k}}+\tilde{V}_{\boldsymbol{k}',-\boldsymbol{k}}), \\
    \tilde{V}_{\boldsymbol{k}'\boldsymbol{k}}^{E(\boldsymbol{k}')O(\boldsymbol{k})}=&  \frac{1}{4}(\tilde{V}_{-\boldsymbol{k}',-\boldsymbol{k}}-\tilde{V}_{-\boldsymbol{k}',\boldsymbol{k}}-\tilde{V}_{\boldsymbol{k}',\boldsymbol{k}}+\tilde{V}_{\boldsymbol{k}',-\boldsymbol{k}}), \\
    \tilde{V}_{\boldsymbol{k}'\boldsymbol{k}}^{O(\boldsymbol{k}')E(\boldsymbol{k})} =&  \frac{1}{4}(\tilde{V}_{-\boldsymbol{k}',-\boldsymbol{k}}+\tilde{V}_{-\boldsymbol{k}',\boldsymbol{k}}-\tilde{V}_{\boldsymbol{k}',\boldsymbol{k}}-\tilde{V}_{\boldsymbol{k}',-\boldsymbol{k}}), \\
    \tilde{V}_{\boldsymbol{k}'\boldsymbol{k}}^{O(\boldsymbol{k}')O(\boldsymbol{k})} =& \frac{1}{4}(\tilde{V}_{-\boldsymbol{k}',-\boldsymbol{k}}-\tilde{V}_{-\boldsymbol{k}',\boldsymbol{k}}+\tilde{V}_{\boldsymbol{k}',\boldsymbol{k}}-\tilde{V}_{\boldsymbol{k}',-\boldsymbol{k}}), \\
    \bar{V}_{\boldsymbol{k}'\boldsymbol{k}}^{\uparrow\downarrow\uparrow\uparrow E(\boldsymbol{k})} =& \frac{\bar{V}_{\boldsymbol{k}'\boldsymbol{k}}^{\uparrow\downarrow\uparrow\uparrow}+\bar{V}_{\boldsymbol{k}',-\boldsymbol{k}}^{\uparrow\downarrow\uparrow\uparrow}}{2}, \qquad \bar{V}_{\boldsymbol{k}'\boldsymbol{k}}^{\uparrow\downarrow\uparrow\uparrow O(\boldsymbol{k})} = \frac{\bar{V}_{\boldsymbol{k}'\boldsymbol{k}}^{\uparrow\downarrow\uparrow\uparrow}-\bar{V}_{\boldsymbol{k}',-\boldsymbol{k}}^{\uparrow\downarrow\uparrow\uparrow}}{2}, \\
    \bar{V}_{\boldsymbol{k}'\boldsymbol{k}}^{\uparrow\downarrow\downarrow\downarrow E(\boldsymbol{k})} =& \frac{\bar{V}_{\boldsymbol{k}'\boldsymbol{k}}^{\uparrow\downarrow\downarrow\downarrow} + \bar{V}_{\boldsymbol{k}',-\boldsymbol{k}}^{\uparrow\downarrow\downarrow\downarrow}}{2}, \qquad \bar{V}_{\boldsymbol{k}'\boldsymbol{k}}^{\uparrow\downarrow\downarrow\downarrow O(\boldsymbol{k})} = \frac{\bar{V}_{\boldsymbol{k}'\boldsymbol{k}}^{\uparrow\downarrow\downarrow\downarrow} - \bar{V}_{\boldsymbol{k}',-\boldsymbol{k}}^{\uparrow\downarrow\downarrow\downarrow}}{2}, \\
    2\bar{V}_{\boldsymbol{k}'\boldsymbol{k}}^{\uparrow\uparrow\uparrow\downarrow E(\boldsymbol{k}')} =& \bar{V}_{\boldsymbol{k}'\boldsymbol{k}}^{\uparrow\uparrow\uparrow\downarrow} + \bar{V}_{-\boldsymbol{k}',\boldsymbol{k}}^{\uparrow\uparrow\uparrow\downarrow}, \qquad 2\bar{V}_{\boldsymbol{k}'\boldsymbol{k}}^{\uparrow\uparrow\uparrow\downarrow O(\boldsymbol{k}')} = \bar{V}_{\boldsymbol{k}'\boldsymbol{k}}^{\uparrow\uparrow\uparrow\downarrow} - \bar{V}_{-\boldsymbol{k}',\boldsymbol{k}}^{\uparrow\uparrow\uparrow\downarrow}, \\
    2\bar{V}_{\boldsymbol{k}'\boldsymbol{k}}^{\downarrow\downarrow\uparrow\downarrow E(\boldsymbol{k}')} =& \bar{V}_{\boldsymbol{k}'\boldsymbol{k}}^{\downarrow\downarrow\uparrow\downarrow} + \bar{V}_{-\boldsymbol{k}',\boldsymbol{k}}^{\downarrow\downarrow\uparrow\downarrow}, \qquad 2\bar{V}_{\boldsymbol{k}'\boldsymbol{k}}^{\downarrow\downarrow\uparrow\downarrow O(\boldsymbol{k}')} = \bar{V}_{\boldsymbol{k}'\boldsymbol{k}}^{\downarrow\downarrow\uparrow\downarrow} - \bar{V}_{-\boldsymbol{k}',\boldsymbol{k}}^{\downarrow\downarrow\uparrow\downarrow}.
\end{align}

\subsubsection{Coupling functions}

\begin{figure}
    \begin{minipage}{.5\textwidth}
      \includegraphics[width=\linewidth]{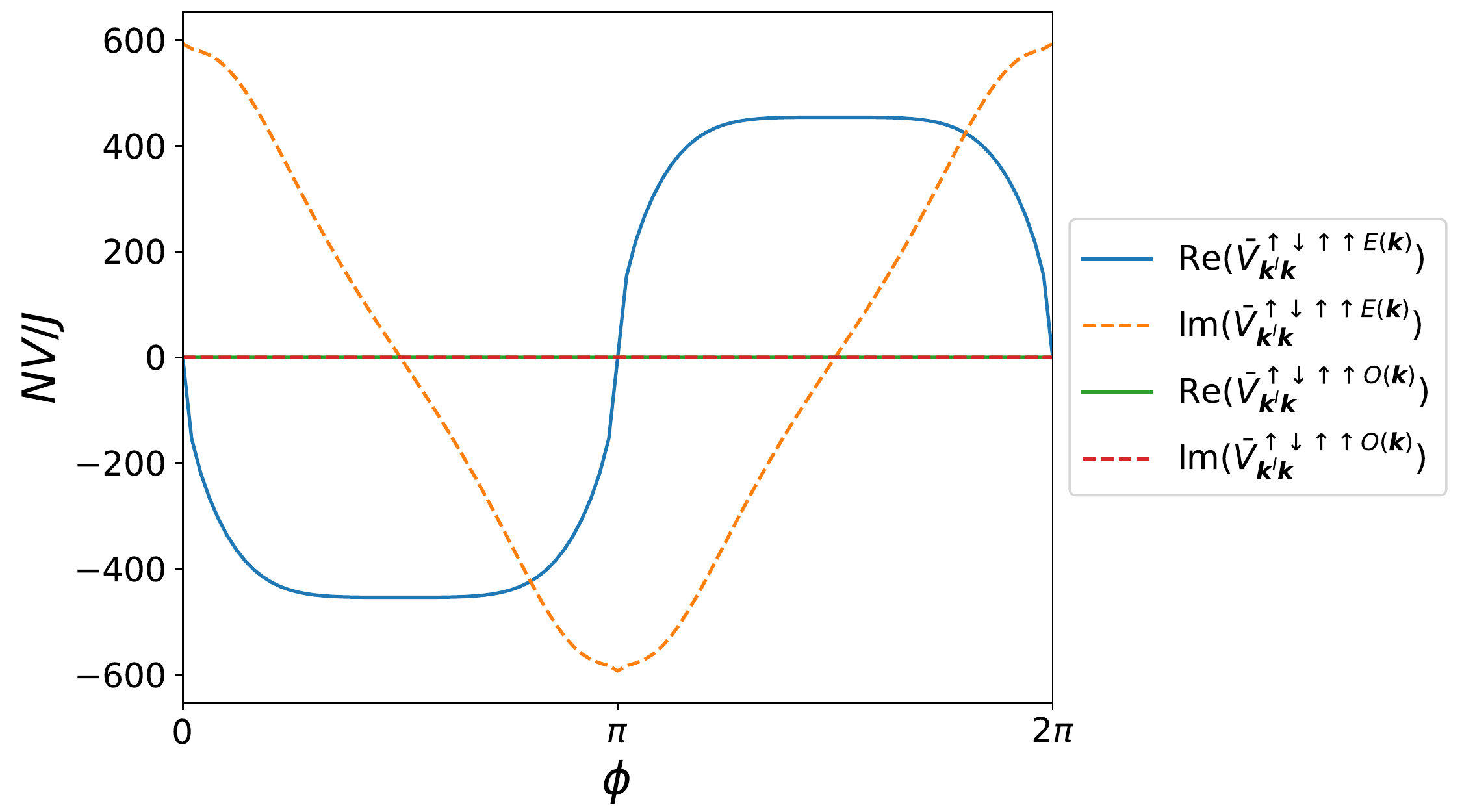}
    \end{minipage}%
    \begin{minipage}{.5\textwidth}
      \includegraphics[width=\linewidth]{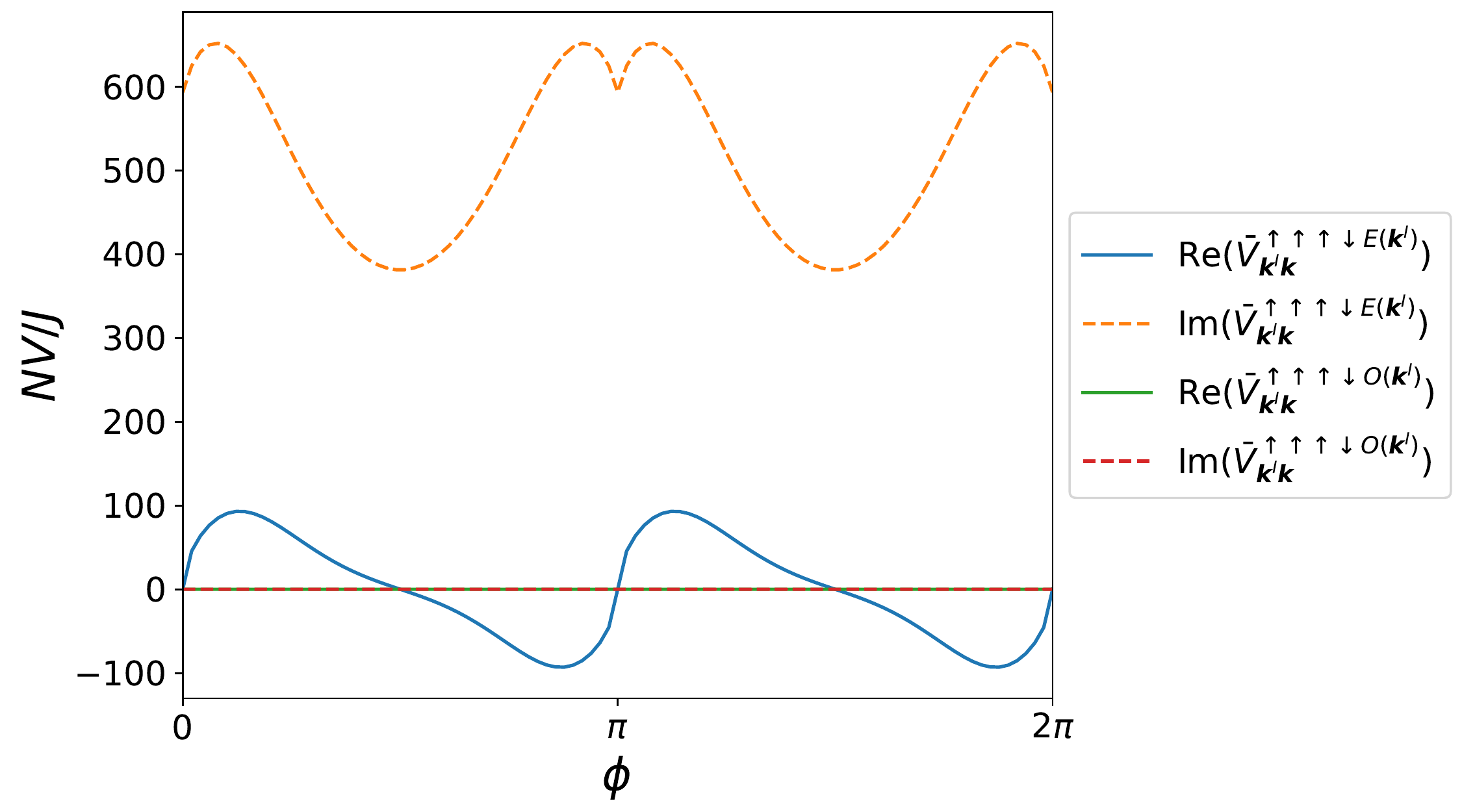}
    \end{minipage}%
    \\
    \begin{minipage}{.5\textwidth}
      \includegraphics[width=\linewidth]{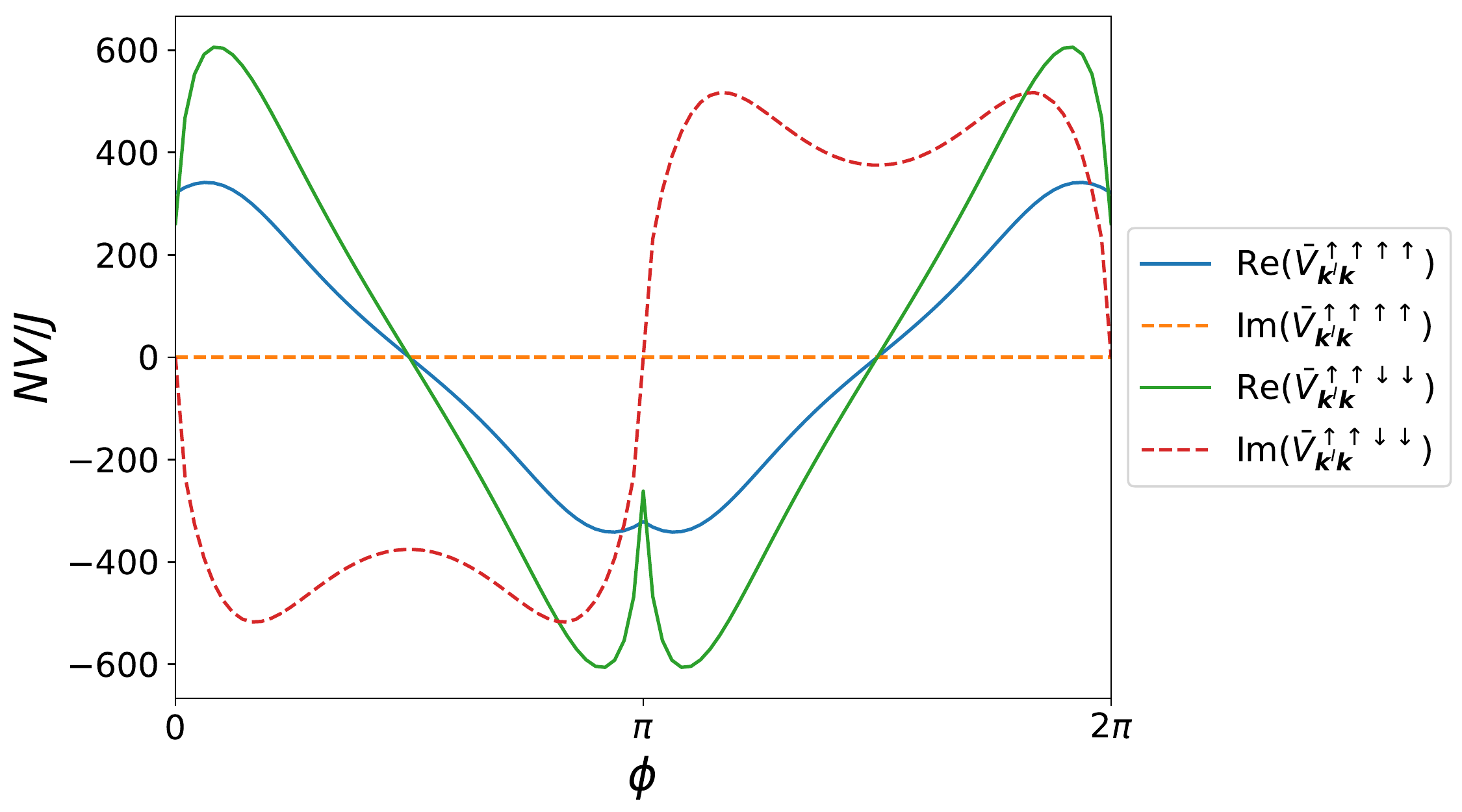}
    \end{minipage}%
    \begin{minipage}{.5\textwidth}
      \includegraphics[width=\linewidth]{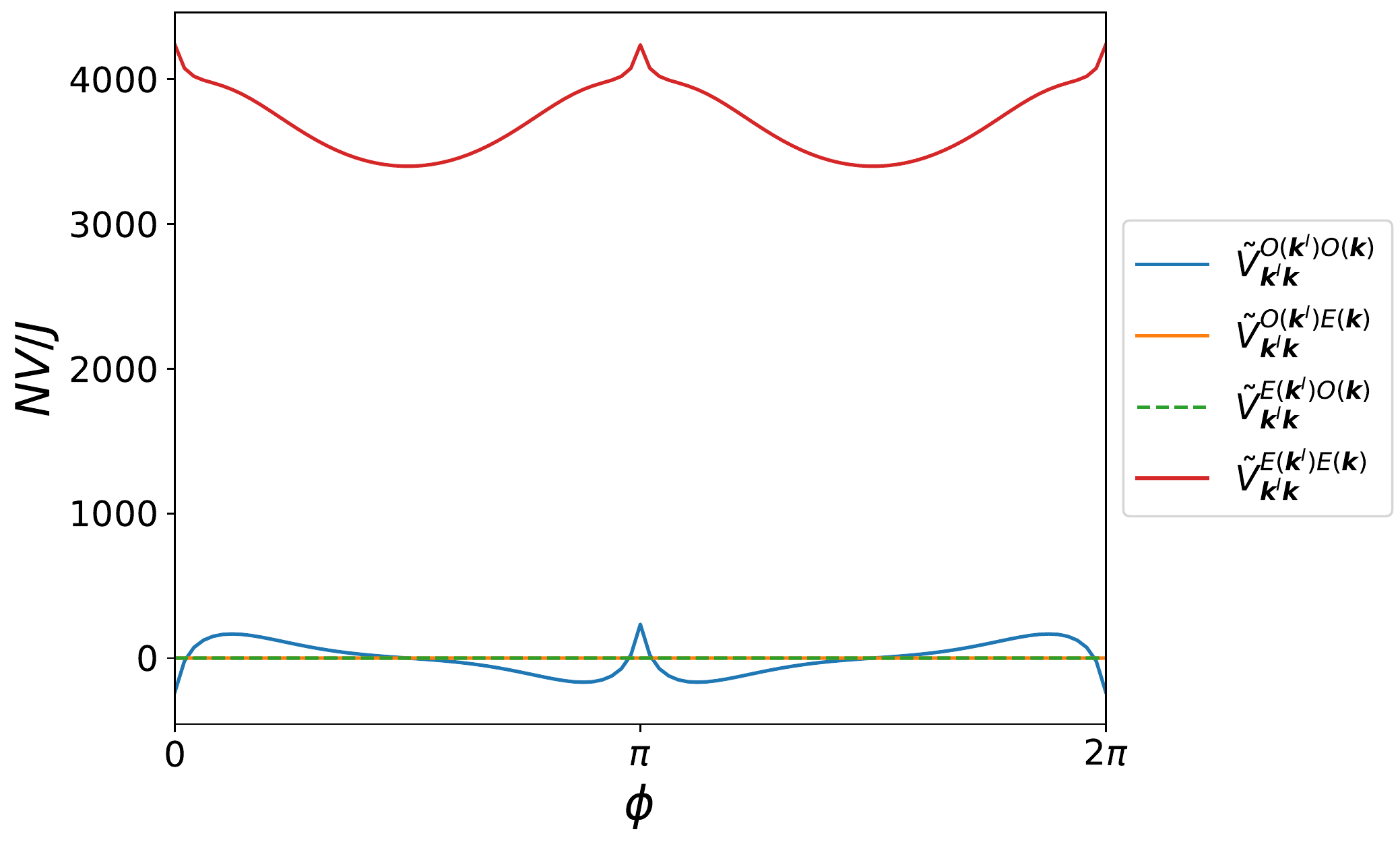}
    \end{minipage}%
    \caption{ Plots of some of the coupling functions that enter the gap equation. They are plotted as a function of the angle $\boldsymbol{k}'$ makes with the $k_x$-axis, with $\boldsymbol{k}'$ on the FS. The other parameters are $\boldsymbol{k} = (k_{\text{F}x},0)$, $t/J = 1000$, $K/J = 0.1$, $D/J = 2.16, U/J = 0.35$, $S=1$, $\mu/t = -5.5$, $\bar{J}/J = 50$. \label{fig:Vbar}}
\end{figure}

Some of the coupling functions that enter the rewritten gap equation are shown in Fig.~\ref{fig:Vbar}. We find that $\bar{V}_{\boldsymbol{k}'\boldsymbol{k}}^{\uparrow\downarrow\downarrow\downarrow}  = -\bar{V}_{\boldsymbol{k}'\boldsymbol{k}}^{\uparrow\downarrow\uparrow\uparrow*}$, $\bar{V}_{\boldsymbol{k}'\boldsymbol{k}}^{\downarrow\downarrow\uparrow\downarrow} = -\bar{V}_{\boldsymbol{k}'\boldsymbol{k}}^{\uparrow\uparrow\uparrow\downarrow*}$, $\bar{V}_{\boldsymbol{k}'\boldsymbol{k}}^{\downarrow\downarrow\downarrow\downarrow} = \bar{V}_{\boldsymbol{k}'\boldsymbol{k}}^{\uparrow\uparrow\uparrow\uparrow} \in \mathbb{R}$, $\bar{V}_{\boldsymbol{k}'\boldsymbol{k}}^{\downarrow\downarrow\uparrow\uparrow} = \bar{V}_{\boldsymbol{k}'\boldsymbol{k}}^{\uparrow\uparrow\downarrow\downarrow*}$, and $\tilde{V}_{\boldsymbol{k}\boldsymbol{k}'} \in \mathbb{R}$. Furthermore,
$
    \abs{\bar{V}_{\boldsymbol{k}'\boldsymbol{k}}^{\uparrow\downarrow\uparrow\uparrow E(\boldsymbol{k})}} \gg \abs{\bar{V}_{\boldsymbol{k}'\boldsymbol{k}}^{\uparrow\downarrow\uparrow\uparrow O(\boldsymbol{k})}}, \abs{\bar{V}_{\boldsymbol{k}'\boldsymbol{k}}^{\uparrow\uparrow\uparrow\downarrow E(\boldsymbol{k}')}} \gg \abs{\bar{V}_{\boldsymbol{k}'\boldsymbol{k}}^{\uparrow\uparrow\uparrow\downarrow O(\boldsymbol{k}')}},  \abs{\tilde{V}_{\boldsymbol{k}'\boldsymbol{k}}^{E(\boldsymbol{k}')E(\boldsymbol{k})}} \gg \abs{\tilde{V}_{\boldsymbol{k}'\boldsymbol{k}}^{O(\boldsymbol{k}')E(\boldsymbol{k})}} ,  \abs{\tilde{V}_{\boldsymbol{k}'\boldsymbol{k}}^{O(\boldsymbol{k}')O(\boldsymbol{k})}} \gg \abs{\tilde{V}_{\boldsymbol{k}'\boldsymbol{k}}^{E(\boldsymbol{k}')O(\boldsymbol{k})}} .
$
In SkX2 the small coupling functions are nonzero and so all four SC gaps couple. In SkX1, all the small coupling functions are in fact zero, and so $\Delta_{\boldsymbol{k}\uparrow\downarrow}^{E(s)}$ decouples from the other gaps.

\subsection{Linearized gap equation}
If we let $T$ approach $T_{\text{c}}$ from below, the gap is small and we can ignore it in $E_{\boldsymbol{k}\eta}$. Then $E_{\boldsymbol{k}\eta} \approx |\epsilon_{\boldsymbol{k}}|$ and $\chi_{\boldsymbol{k}\eta} = \chi_{\boldsymbol{k}} \approx \tanh(\beta |\epsilon_{\boldsymbol{k}}|)/2|\epsilon_{\boldsymbol{k}}|$. $B_{\boldsymbol{k}\sigma_4\sigma_3}$ drops out of the gap equation \eqref{eq:gapeq},
\begin{equation}
    \Delta_{\boldsymbol{k}\sigma_1\sigma_2} = -\sum_{\boldsymbol{k}'\sigma_3\sigma_4} \bar{V}_{\boldsymbol{k}'\boldsymbol{k}}^{\sigma_1\sigma_2\sigma_3\sigma_4}  \Delta_{\boldsymbol{k}'\sigma_4\sigma_3} \chi_{\boldsymbol{k}'}.
\end{equation}
In simplified notation,
\begin{equation}
    \Delta_{\boldsymbol{k}} = -\sum_{\boldsymbol{k}'} V_{\boldsymbol{k}'\boldsymbol{k}} \Delta_{\boldsymbol{k}'} \frac{1}{2|\epsilon_{\boldsymbol{k}'}|}\tanh\frac{\beta_{\text{c}}|\epsilon_{\boldsymbol{k}'}|}{2}.
\end{equation}
We neglect the radial dependence of $ V_{\boldsymbol{k}'\boldsymbol{k}}$ away from the FS and only keep its angular dependence. Following BCS theory \cite{SFsuperconductivity, Sigrist}, $V_{\boldsymbol{k}'\boldsymbol{k}}$ is approximated to be nonzero only within an energy $\omega_{\text{c}}$ from the FS, where $\omega_{\text{c}}$ is set to the maximum magnon energy, $\omega_{\text{c}} \sim 23J$,
\begin{equation}
    V_{\boldsymbol{k}'\boldsymbol{k}} \approx V(\phi', \phi)\Theta(\omega_{\text{c}} - |\epsilon_{\boldsymbol{k}}|)\Theta(\omega_{\text{c}}-|\epsilon_{\boldsymbol{k}'}|).    
\end{equation}
The angles are defined by $\phi = \operatorname{atan2}(k_y, k_x)$, i.e.~the angle $\boldsymbol{k}$ makes with the $k_x$ axis. Similarly, $\Delta_{\boldsymbol{k}} \approx \Delta(\phi)\Theta(\omega_{\text{c}} - |\epsilon_{\boldsymbol{k}}|)$ is assumed.
Assuming $\boldsymbol{k}$ and $\boldsymbol{k}'$ are close enough to the FS that $V_{\boldsymbol{k}'\boldsymbol{k}}$ and $\Delta_{\boldsymbol{k}}$ are nonzero we get
\begin{equation}
    \Delta(\phi) = -\frac{N}{A_{\text{eBZ}}} \int_0^{2\pi} d \phi' V(\phi', \phi) \Delta(\phi') \int dk' k' \frac{1}{2|\epsilon_{k'\phi'}|}\tanh\frac{\beta_{\text{c}}|\epsilon_{k'\phi'}|}{2}.
\end{equation}
Now we use $D(\epsilon)d\epsilon = D(k)dk$ to transform the momentum integral into an energy integral,
\begin{equation}
    \Delta(\phi) = -\frac{N}{A_{\text{eBZ}}} \int_0^{2\pi} d \phi' V(\phi', \phi) \Delta(\phi') \int_{-\omega_{\text{c}}}^{\omega_{\text{c}}}  d\epsilon \frac{k' D(\epsilon)}{D(k')} \frac{1}{2|\epsilon|} \tanh\frac{\beta_{\text{c}}|\epsilon|}{2}.
\end{equation}
As an approximation, we replace $D(\epsilon)$ by its value at the FS, $D(\mu) = D_0$, and the angular integral by an FS average,
\begin{equation}
    \int_0^{2\pi} d \phi' V(\phi', \phi) \Delta(\phi') = 2\pi \langle V(\phi', \phi) \Delta(\phi') \rangle_{\text{FS},\phi'} = \frac{2\pi}{N_{\phi}} \sum_i  V(\phi'_i, \phi) \Delta(\phi'_i),
\end{equation}
where $N_{\phi}$ is the number of points sampled on the FS. Then, with $N_0 = D_0/2$ the DOS per spin,
\begin{equation}
    \Delta(\phi) = -N_0 \langle V(\phi', \phi) \Delta(\phi') \rangle_{\text{FS},\phi'} \int_{-\omega_{\text{c}}}^{\omega_{\text{c}}}  d\epsilon \frac{1}{2|\epsilon|} \tanh\frac{\beta_{\text{c}}|\epsilon|}{2}.
\end{equation}
We used that
$
    \frac{N}{A_{\text{eBZ}}}2\pi \frac{k' D_0}{D(k')} = N_0,
$
since $D(k) = \sqrt{3}Nk/2\pi$.
Now, define the dimensionless coupling constant $\lambda$,
\begin{equation}
    \frac{1}{\lambda} = \int_{-\omega_{\text{c}}}^{\omega_{\text{c}}}  d\epsilon \frac{1}{2|\epsilon|} \tanh\frac{\beta_{\text{c}}|\epsilon|}{2}.
\end{equation}
In the weak-coupling limit, $\lambda \ll 1$,
\begin{equation}
    \frac{1}{\lambda} = \ln\pqty{\frac{2}{\pi}e^\gamma \beta_{\text{c}} \omega_{\text{c}}},
\end{equation}
where $\gamma = 0.5772156649\dots$ is the Euler–Mascheroni constant \cite{SFsuperconductivity}.
The critical temperature for superconductivity is
\begin{equation}
    k_{\text{B}} T_{\text{c}} = \frac{2}{\pi}e^\gamma \omega_{\text{c}} e^{-1/\lambda} \approx 1.134\omega_{\text{c}} e^{-1/\lambda}.
\end{equation}
$\lambda$ is found from the linearized gap equation,
\begin{equation}
    \lambda \Delta(\phi) = -N_0 \langle V(\phi', \phi) \Delta(\phi') \rangle_{\text{FS},\phi'}.
\end{equation}
This can be solved as an eigenvalue equation by considering $ V(\phi', \phi)$ as a matrix in the chosen values of $\phi, \phi'$ and $\Delta(\phi)$ as eigenvectors. $\lambda$ is the largest eigenvalue of this eigenvalue equation \cite{EirikNMAFM}, since only the largest $T_{\text{c}}$ complies to the assumption that the gaps are small. Its corresponding eigenvector contains information about the $\boldsymbol{k}$ dependence of the gap along the FS.
By reinstating the full four-component gap vector and the $4\times4$ coupling matrix, we get the linearized gap equation stated in Eq.~(4) of the main text.

\subsection{Solutions to linearized gap equation}
\begin{figure}
    \centering
    \includegraphics[width=0.9\linewidth]{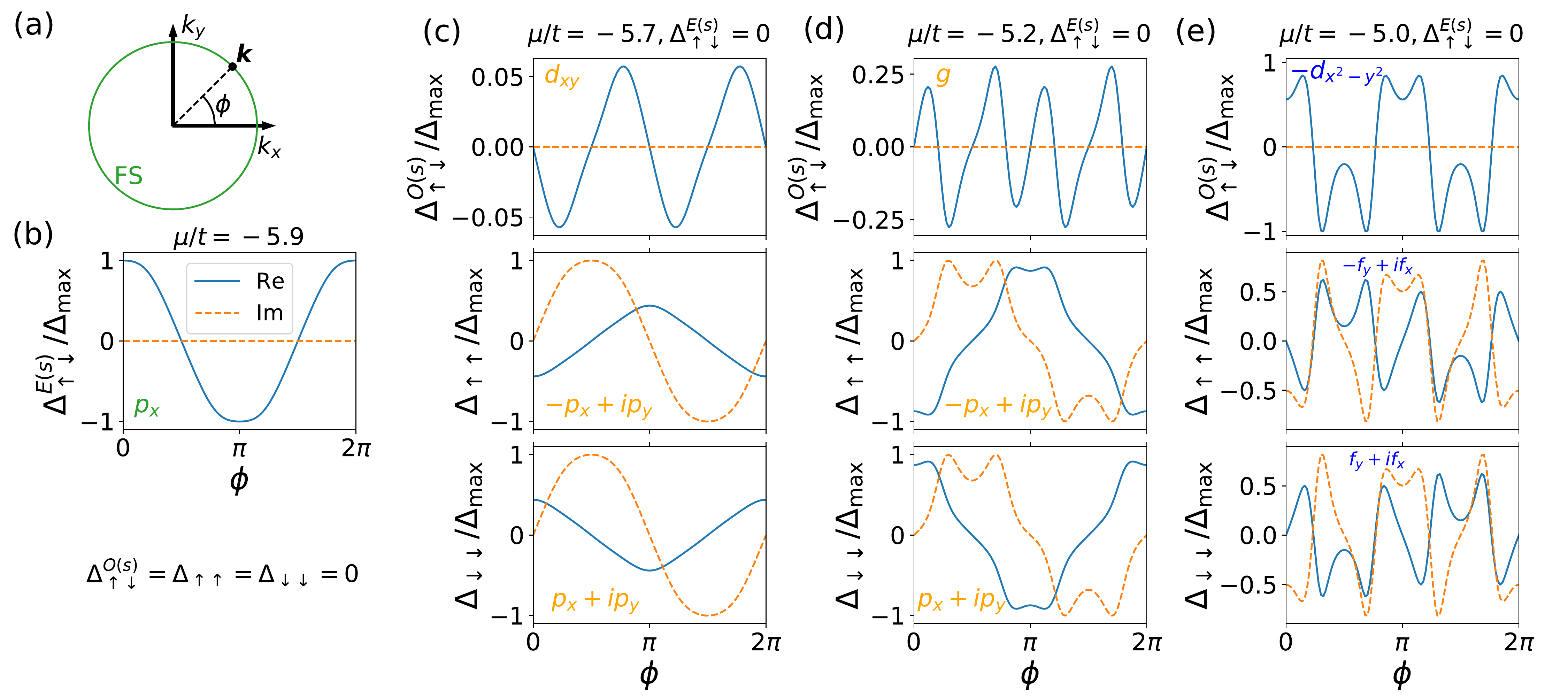}
    \caption{Solutions to the linearized gap equation for a selection of $\mu/t$ plotted as a function of the angle $\phi$ around the FS. The MML is in the SkX1 state. (a) shows a sketch of the FS in green, with a definition of the angle $\phi$. (b) For $\mu/t = -5.9$, only $\Delta_{\boldsymbol{k}\uparrow\downarrow}^{E(s)}$ is nonzero and displays $p_x$-wave symmetry. (c) For $\mu/t = -5.7$, $\Delta_{\boldsymbol{k}\uparrow\downarrow}^{O(s)}$ shows $d_{xy}$-wave symmetry, while $\Delta_{\boldsymbol{k}\downarrow\downarrow}$ has $p_x+ip_y$-wave symmetry. Note that $\Delta_{\boldsymbol{k}\uparrow\uparrow} = -\Delta_{\boldsymbol{k}\downarrow\downarrow}^*$. (d) For $\mu/t = -5.2$, $\Delta_{\boldsymbol{k}\uparrow\downarrow}^{O(s)}$ has changed to $g$-wave symmetry. (e) For $\mu/t = -5.0$, $\Delta_{\boldsymbol{k}\uparrow\downarrow}^{O(s)}$ shows $-d_{x^2-y^2}$-wave symmetry, while $\Delta_{\boldsymbol{k}\downarrow\downarrow}$ has $f_y+if_x$-wave symmetry. The parameters are $t/J = 1000, D/J = 2.16, U/J = 0.35, K/J = 0.1, S=1$, while the shown results do not depend on $\bar{J}$. $N_\phi = 100$ gives convergence of $\lambda$ to 5 significant digits.}
    \label{fig:gapslin}
\end{figure}

Let us define the amplitude of the gap as $\Delta_{\text{max}} = \text{max}_{\text{FS}}(|\Re \Delta|, |\Im \Delta|)$, i.e.~the largest amplitude of the real or imaginary part of the four gaps. The solution to the linearized gap equation is plotted relative to $\Delta_{\text{max}}$ for various $\mu$ in Fig.~\ref{fig:gapslin}. Note that the real and imaginary parts of $\Delta_{\boldsymbol{k}\downarrow\downarrow}$ in Fig.~\ref{fig:gapslin}(e) do not have their zeros exactly where sine and cosine functions would. Still, 2 of the 6 zeros are placed either at $\phi = 0, \pi$ or at $\phi = \pi/2, 3\pi/2$ giving the classification $f_y + if_x$. We also checked that no zeros occur for the same $\phi$ for the real and imaginary parts.  In SkX2 the results are similar, except that all four gaps couple and there is no part of the phase diagram where $\Delta_{\boldsymbol{k}\uparrow\downarrow}^{E(s)}$ dominates. The results in the $p_x+ip_y$ phase resemble Fig.~\ref{fig:gapslin}(c) and (d), and the results in the $f_y+if_x$ phase resemble Fig.~\ref{fig:gapslin}(e). The difference is that $\Delta_{\boldsymbol{k}\uparrow\downarrow}^{E(s)}$ is a comparatively small, nonzero $p_x$-wave gap.

In Ref.~\cite{QSkQTPT} we showed that the magnon bands of SkX1 and SkX2 are topologically nontrivial, and undergo quantum topological phase transitions (QTPTs) when the easy-axis anisotropy is tuned. For instance, in SkX1 there are 4 QTPTs between $K/J = 0.2$ and $K/J = 0.3$. In the phase diagrams in Figs.~2 and 4 of the main text we see no significant changes in the SC state that can be associated to these QTPTs. We attribute this to the fact that the Chern numbers of the magnon bands involve derivatives of the transformation matrix $T_{\boldsymbol{q}}$ \cite{QSkQTPT}, while the coupling functions for superconductivity depend directly on $T_{\boldsymbol{q}}$, see Eq.~\eqref{eq:couplingfunction}. Secondly, the coupling functions for superconductivity are defined as sums over all 15 bands, while the QTPTs only affect two bands at a time. Finally, the information about the QTPTs present in $T_{\boldsymbol{q}}$ appears at very small regions of $\boldsymbol{q}$ \cite{QSkQTPT}, while the calculations of superconductivity involve a great variation in $\boldsymbol{q}$. Hence, only a small part of the phase space is affected by the QTPTs. In total, this means that the topological nature of the bands have little influence on the SC state in this system. 
Ref.~\cite{TopoMagnonSC} considers a noncoplanar magnetic order in pyrochlore iridate thin films where the magnons are found to be topologically nontrivial. By doping the material, they find an unconventional TRS broken $d + id$ superconducting state. 

\subsection{Time-reversal symmetry}
From the solution to the linearized gap equation we find $\Delta_{\boldsymbol{k}\uparrow\downarrow} = \Delta_{\boldsymbol{k}\uparrow\downarrow}^{E(s)} + \Delta_{\boldsymbol{k}\uparrow\downarrow}^{O(s)} \in \mathbb{R}$  , $\Delta_{\boldsymbol{k}\downarrow\uparrow} = \Delta_{\boldsymbol{k}\uparrow\downarrow}^{E(s)} - \Delta_{\boldsymbol{k}\uparrow\downarrow}^{O(s)} \in \mathbb{R}$, $\Delta_{\boldsymbol{k}\uparrow\uparrow} = -\Delta_{\boldsymbol{k}\downarrow\downarrow}^*$.
Applying the time reversal operator, $\mathcal{T} = i\sigma_y K$, gives \cite{Sigrist}
\begin{equation}
    \mathcal{T} \hat{\Delta}_{\boldsymbol{k}} \mathcal{T}^{-1} = \sigma_y \hat{\Delta}_{-\boldsymbol{k}}^* \sigma_y = \begin{pmatrix}
    \Delta_{-\boldsymbol{k}\downarrow\downarrow}^* & -\Delta_{-\boldsymbol{k}\downarrow\uparrow}^*\\
    -\Delta_{-\boldsymbol{k}\uparrow\downarrow}^*  & \Delta_{-\boldsymbol{k}\uparrow\uparrow}^*
    \end{pmatrix} =  \begin{pmatrix}
    \Delta_{\boldsymbol{k}\uparrow\uparrow} & \Delta_{\boldsymbol{k}\uparrow\downarrow} \\
    \Delta_{\boldsymbol{k}\downarrow\uparrow} & \Delta_{\boldsymbol{k}\downarrow\downarrow}
    \end{pmatrix} =\hat{\Delta}_{\boldsymbol{k}},
\end{equation}
using
\begin{align}
    \Delta_{-\boldsymbol{k}\downarrow\uparrow}^* &=  \Delta_{-\boldsymbol{k}\uparrow\downarrow}^{E(s)} - \Delta_{-\boldsymbol{k}\uparrow\downarrow}^{O(s)} = -\Delta_{\boldsymbol{k}\uparrow\downarrow}^{E(s)} - \Delta_{\boldsymbol{k}\uparrow\downarrow}^{O(s)} = -\Delta_{\boldsymbol{k}\uparrow\downarrow}, \\
    \Delta_{-\boldsymbol{k}\uparrow\downarrow}^* &=  \Delta_{-\boldsymbol{k}\uparrow\downarrow}^{E(s)} + \Delta_{-\boldsymbol{k}\uparrow\downarrow}^{O(s)} = -\Delta_{\boldsymbol{k}\uparrow\downarrow}^{E(s)} + \Delta_{\boldsymbol{k}\uparrow\downarrow}^{O(s)} = -\Delta_{\boldsymbol{k}\downarrow\uparrow}, \\
    \Delta_{-\boldsymbol{k}\downarrow\downarrow}^* &= -\Delta_{\boldsymbol{k}\downarrow\downarrow}^* =  \Delta_{\boldsymbol{k}\uparrow\uparrow}, \qquad \Delta_{-\boldsymbol{k}\uparrow\uparrow}^* = -\Delta_{\boldsymbol{k}\uparrow\uparrow}^* =  \Delta_{\boldsymbol{k}\downarrow\downarrow}.
\end{align}
So the SC state is TRS.

\subsection{Zero temperature gap equation}
The amplitude of the SC gap is approximately constant up to a large fraction of $T_{\text{c}}$. Hence, finding the gap at zero temperature will give a good indication of much of the low temperature behavior in the SC.
Our starting point is the gap equation on matrix form
\begin{equation}
    \boldsymbol{\Delta_k} = -\sum_{\boldsymbol{k}'} \mathcal{V}_{\boldsymbol{k}'\boldsymbol{k}} \sum_\eta \pqty{\frac{1}{2}\boldsymbol{\Delta}_{\boldsymbol{k}'}+\eta \boldsymbol{B}_{\boldsymbol{k}'} }\chi_{\boldsymbol{k}'\eta},
\end{equation}
where
\begin{equation}
    \chi_{\boldsymbol{k}'\eta} = \frac{1}{2E_{\boldsymbol{k}'\eta}} \tanh\frac{\beta E_{\boldsymbol{k}'\eta}}{2} \underbrace{\to}_{\beta \to \infty} \frac{1}{2E_{\boldsymbol{k}'\eta}}.
\end{equation}
Employing FS avarages, and going to an integral over polar coordinates gives
\begin{align}
    \Delta(\phi) =& -\frac{N}{A_{\text{eBZ}}} \int_0^{2\pi} d\phi' V(\phi',\phi) \sum_\eta \pqty{\frac{1}{2}\Delta(\phi')+\eta B(\phi')}  \int dk' k' \frac{1}{2\sqrt{\epsilon_{k'\phi'}^2+\frac{1}{2}\Tr\hat{\Delta}(\phi') \hat{\Delta}^\dagger (\phi')+\eta \frac{1}{2}\sqrt{A(\phi')}}}.
\end{align}
Using $D(\epsilon)d\epsilon = D(k) dk$ and approximating $D(\epsilon) \approx D(\mu) = D_0$ gives
\begin{align}
    \Delta(\phi) =& -\frac{D_0}{4\pi}\int_0^{2\pi} d\phi' V(\phi',\phi)\sum_\eta \pqty{\frac{1}{2}\Delta(\phi')+\eta B(\phi')}  \int_{-\omega_{\text{c}}}^{\omega_{\text{c}}}  d\epsilon \frac{1}{2\sqrt{\epsilon^2+\frac{1}{2}\Tr\hat{\Delta}(\phi') \hat{\Delta}^\dagger (\phi') +\eta \frac{1}{2}\sqrt{A(\phi')}}}.
\end{align}
The energy integral can be performed analytically,
\begin{equation}
    \Delta(\phi) = -\frac{D_0}{4\pi}\int_0^{2\pi} d\phi' V(\phi',\phi)\sum_\eta \pqty{\frac{1}{2}\Delta(\phi')+\eta B(\phi')} \operatorname{arsinh} \pqty{\frac{\omega_{\text{c}}}{\sqrt{\frac{1}{2}\Tr\hat{\Delta}(\phi') \hat{\Delta}^\dagger (\phi')+\eta \frac{1}{2}\sqrt{A(\phi')}}}}.
\end{equation}
Rewriting in terms of a FS average gives
\begin{equation}
    \Delta(\phi) = -N_0 \left\langle V(\phi',\phi)\sum_\eta \pqty{\frac{1}{2}\Delta(\phi')+\eta B(\phi')} \operatorname{arsinh} \pqty{\frac{\omega_{\text{c}}}{\sqrt{\frac{1}{2}\Tr\hat{\Delta}(\phi') \hat{\Delta}^\dagger (\phi')+\eta \frac{1}{2}\sqrt{A(\phi')}}}} \right\rangle_{\text{FS}, \phi'}.
\end{equation}
Inserting the $4\cross4$ matrix $\mathcal{V}_{\boldsymbol{k}'\boldsymbol{k}}$ from Eq.~\eqref{eq:V44} and the 4-vectors $\boldsymbol{\Delta}_{\boldsymbol{k}} = (\Delta_{\boldsymbol{k}\uparrow\downarrow}^{O(s)}, \Delta_{\boldsymbol{k}\uparrow\uparrow}, \Delta_{\boldsymbol{k}\downarrow\downarrow}, \Delta_{\boldsymbol{k}\uparrow\downarrow}^{E(s)})^T$, and $\boldsymbol{B}_{\boldsymbol{k}} = (B_{\boldsymbol{k}\uparrow\downarrow}^{O(s)}, B_{\boldsymbol{k}\uparrow\uparrow}, B_{\boldsymbol{k}\downarrow\downarrow}, B_{\boldsymbol{k}\uparrow\downarrow}^{E(s)})^T$ for the gaps yields the zero temperature gap equation in Eq.~(5) of the main text.

A naive iteration is not sufficient, as the procedure tends to diverge away from a solution. Instead, we employ self-consistent iteration. We start from an initial guess $\boldsymbol{\Delta}_0(\phi)$. Then, we calculate derivatives in real and imaginary parts of each element in $\boldsymbol{\Delta}_0(\phi)$ looking at how small changes in the gap changes the function
\begin{equation}
    f[\boldsymbol{\Delta}(\phi)] = \boldsymbol{\Delta}(\phi)+N_0 \left\langle \mathcal{V}(\phi',\phi)\sum_\eta \Big(\frac{\boldsymbol{\Delta}(\phi')}{2}+\eta \boldsymbol{B}(\phi')\Big)\text{arsinh} \pqty{\frac{\omega_{\text{c}}}{\sqrt{\frac{1}{2}\Tr\hat{\Delta}(\phi') \hat{\Delta}^\dagger (\phi')+\eta \frac{1}{2}\sqrt{A(\phi')}}}} \right\rangle_{\text{FS}, \phi'},
\end{equation}
which should be zero. From these derivatives, a new $\boldsymbol{\Delta}_i(\phi)$ is set up from the previous $\boldsymbol{\Delta}_{i-1}(\phi)$ using
\begin{equation}
    \boldsymbol{\Delta}_i(\phi) = \boldsymbol{\Delta}_{i-1}(\phi)  -\alpha \boldsymbol{R}(\phi) -i\alpha \boldsymbol{I}(\phi),
\end{equation}
where $\boldsymbol{R}(\phi), \boldsymbol{I}(\phi)$ are vectors containing the derivatives. $\alpha$ is the mixing parameter, mostly set to the $\Delta_{\text{max}}/50$. We use 3 runs of maximum 100 iterations in hopes of finding convergence. For each run, $\alpha$ is reduced by 1/10. The derivatives are defined by
\begin{equation}
    R_{j}(\phi) = \frac{f[\boldsymbol{\Delta}(\phi)+\boldsymbol{h}_{j}]-f[\boldsymbol{\Delta}(\phi)-\boldsymbol{h}_{j}]}{2h},
\qquad
    I_{j}(\phi) = \frac{f[\boldsymbol{\Delta}(\phi)+i\boldsymbol{h}_{j}]-f[\boldsymbol{\Delta}(\phi)-i\boldsymbol{h}_{j}]}{2h},
\end{equation}
where $\boldsymbol{h}_{j}$ is a vector of the same length as $\boldsymbol{\Delta}(\phi)$ with all elements 0 except element number $j$ which is $h$. $h$ is some small number, often $\Delta_{\text{max}}/100$. This procedure is continued until $f[\boldsymbol{\Delta}(\phi)]$ is smaller than a chosen tolerance, set to $\Delta_{\text{max}}/10000$.

\begin{figure}
    \centering
    \includegraphics[width=0.9\linewidth]{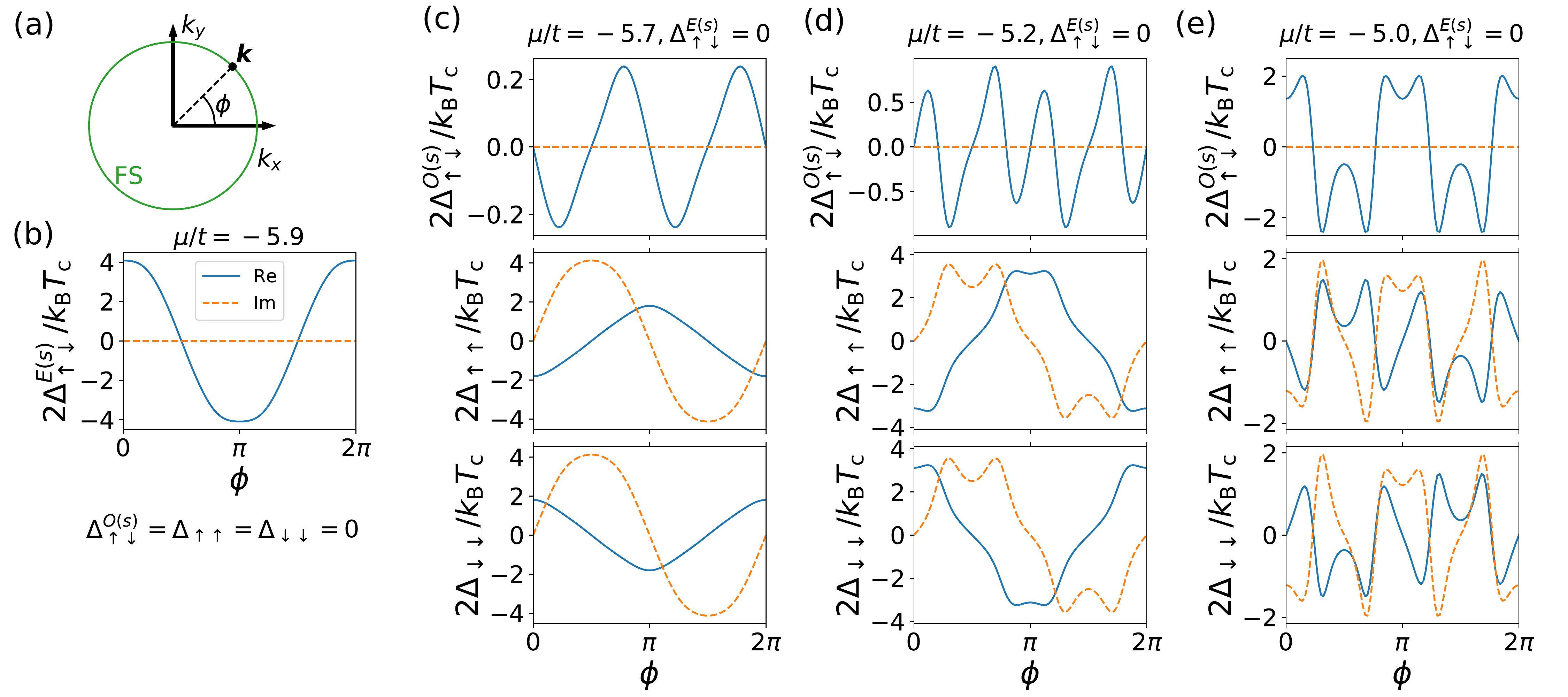}
    \caption{Solutions to the zero temperature gap equation for a selection of $\mu/t$ plotted as a function of the angle $\phi$ around the FS. The MML is in the SkX1 state. (a) shows a sketch of the FS, with a definition of the angle $\phi$. For these choices of parameters, the symmetries at zero temperature are the same as those close to $T_{\text{c}}$, see Fig.~\ref{fig:gapslin}. The parameters are $t/J = 1000, D/J = 2.16, U/J = 0.35, K/J = 0.1, S=1,$ and $N_\phi = 100$. The shown results have a very weak dependence on $\bar{J}$.}
    \label{fig:gapsT0}
\end{figure}

Solutions of the zero temperature gap equation are shown in Fig.~\ref{fig:gapsT0}. We see that well within the phases, the main change from the result at $T$ close to $T_{\text{c}}$ is the amplitude of the gaps. Note that the amplitude of the gaps in the $f_y+if_x$ are more comparable to the BCS result $2\Delta(0)/k_{\text{B}} T_{\text{c}} = 2\pi e^{-\gamma} \approx 3.528$ than the maximum of $E_{\boldsymbol{k},-}$ as mentioned in the main text. Similar results exist at other $K/J$ and for the MML in the SkX2 state. When the MML is in the SkX2 state there is a small, coexisting $p_x$-wave solution for $\Delta_{\boldsymbol{k}\uparrow\downarrow}^{E(s)}$.

From solutions of the zero temperature gap equation we can construct a phase diagram at zero temperature, see Fig.~4 in the main text. We limit ourselves to ansatze with the same symmetries as those found for $T$ close to $T_{\text{c}}$, i.e. chiral $f$-wave, chiral $p$-wave and $p_x$-wave. For $\mu$ close to a phase transition, the gap symmetries on either side both yield convergence. For the transition between $p_x+ip_y$ to $f_x+if_y$, we set a lower limit where $f_x+if_y$ does not converge given the tolerance of $\Delta_{\text{max}}/10000$. The upper limit is given when it turns out that a $p_x+ip_y$ ansatz converges to a $h+if_y$ ($h$: $l=5$, 10 sign changes, odd in $\boldsymbol{k}$) state with much lower amplitude than the $f_y+if_x$ gap. Then, we assume the $f_y+if_x$-wave gap dominates. For the transition between $p_x$ and $p_x+ip_y$ we set the lower limit where $p_x+ip_y$ ceases to converge with a tolerance of $\Delta_{\text{max}}/10000$. $p_x$ tends to yield convergence up to larger $\mu$, so we claim that when the gap amplitude of $p_x$ is 1/50 of the gap amplitude of $p_x+ip_y$ the $p_x+ip_y$-wave state will be preferred. Even when their amplitudes are comparable, the $p_x+ip_y$-symmetry is more effective in producing a large binding energy.

Note that a more accurate determination of the phase diagram in Fig.~4 of the main text would be to compute the free energy of competing phases which converge. In addition, the solutions of the gap equation are only guaranteed to extremize the free energy. With FS averaged results, we do not have a full momentum space knowledge of the gap, making a detailed study of the free energy beyond the scope of this study. It is well known that the creation of Cooper pairs create a binding energy which lowers the free energy \cite{SFsuperconductivity}. Hence, we can be confident that any SC instability will be preferred over the normal state for $T < T_{\text{c}}$. A consideration of other competing phases is left outside the scope of this study, though we will mention that even when the FS is similar to the mBZ in size, it is still approximately circular, see Fig.~1(a) in the main text and Fig.~\ref{fig:RUmpklapp}. Hence, there is a lack of nesting vectors meaning that a competing spin-density wave is unlikely \cite{ArneAFMNM_Umklapp}. 

\section{Topological superconductivity} \label{sec:TSC}
The bulk topological invariant in Eq.~(6) of the main text is only defined if $\Delta_{\uparrow\downarrow}^{O(s)} = \Delta_{\uparrow\downarrow}^{E(s)} = 0$. However, if removing these gaps in a continuous fashion does not close the bulk gap of the SC, the topological classification is the same \cite{Bernevig2013}.
The Bogoliubov-de Gennes (BdG) Hamiltonian is
\begin{equation}
    H_{\text{BdG}}(x) = \frac{1}{2} \sum_{\boldsymbol{k}} \Psi_{\boldsymbol{k}}^\dagger H_{\boldsymbol{k}}(x) \Psi_{\boldsymbol{k}},
\end{equation}
with $ \Psi_{\boldsymbol{k}} = (c_{\boldsymbol{k}\uparrow}, c_{\boldsymbol{k}\downarrow}, c_{-\boldsymbol{k}\downarrow}^\dagger, c_{-\boldsymbol{k}\uparrow}^\dagger)^T$, and
\begin{equation}
    H_{\boldsymbol{k}}(x) = 
    \begin{pmatrix}
    \epsilon_{\boldsymbol{k}} & 0 & \Delta_{\boldsymbol{k}\uparrow\downarrow}(1-x) & \Delta_{\boldsymbol{k}\uparrow\uparrow} \\
    0 & \epsilon_{\boldsymbol{k}} & \Delta_{\boldsymbol{k}\downarrow\downarrow} & \Delta_{\boldsymbol{k}\downarrow\uparrow}(1-x) \\
    \Delta_{\boldsymbol{k}\uparrow\downarrow}^\dagger(1-x) & \Delta_{\boldsymbol{k}\downarrow\downarrow}^\dagger & -\epsilon_{\boldsymbol{k}} & 0 \\
    \Delta_{\boldsymbol{k}\uparrow\uparrow}^\dagger & \Delta_{\boldsymbol{k}\downarrow\uparrow}^\dagger(1-x) & 0 & -\epsilon_{\boldsymbol{k}}  \\
    \end{pmatrix}.
\end{equation}
\begin{figure}
    \centering
    \begin{minipage}{0.5\textwidth}
    \includegraphics[width=0.6\linewidth]{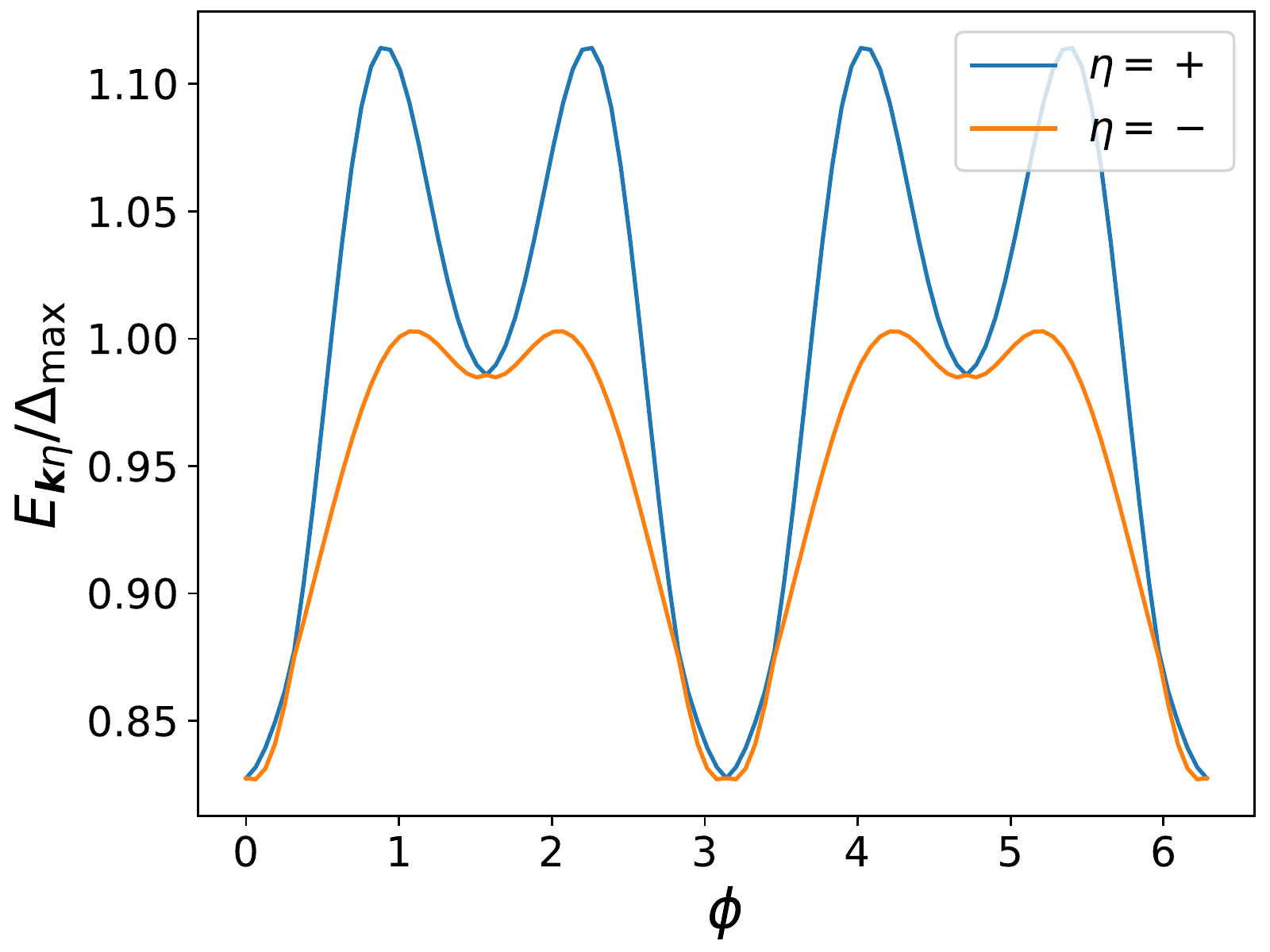}
    \end{minipage}%
    \begin{minipage}{0.5\textwidth}
    \includegraphics[width=0.6\linewidth]{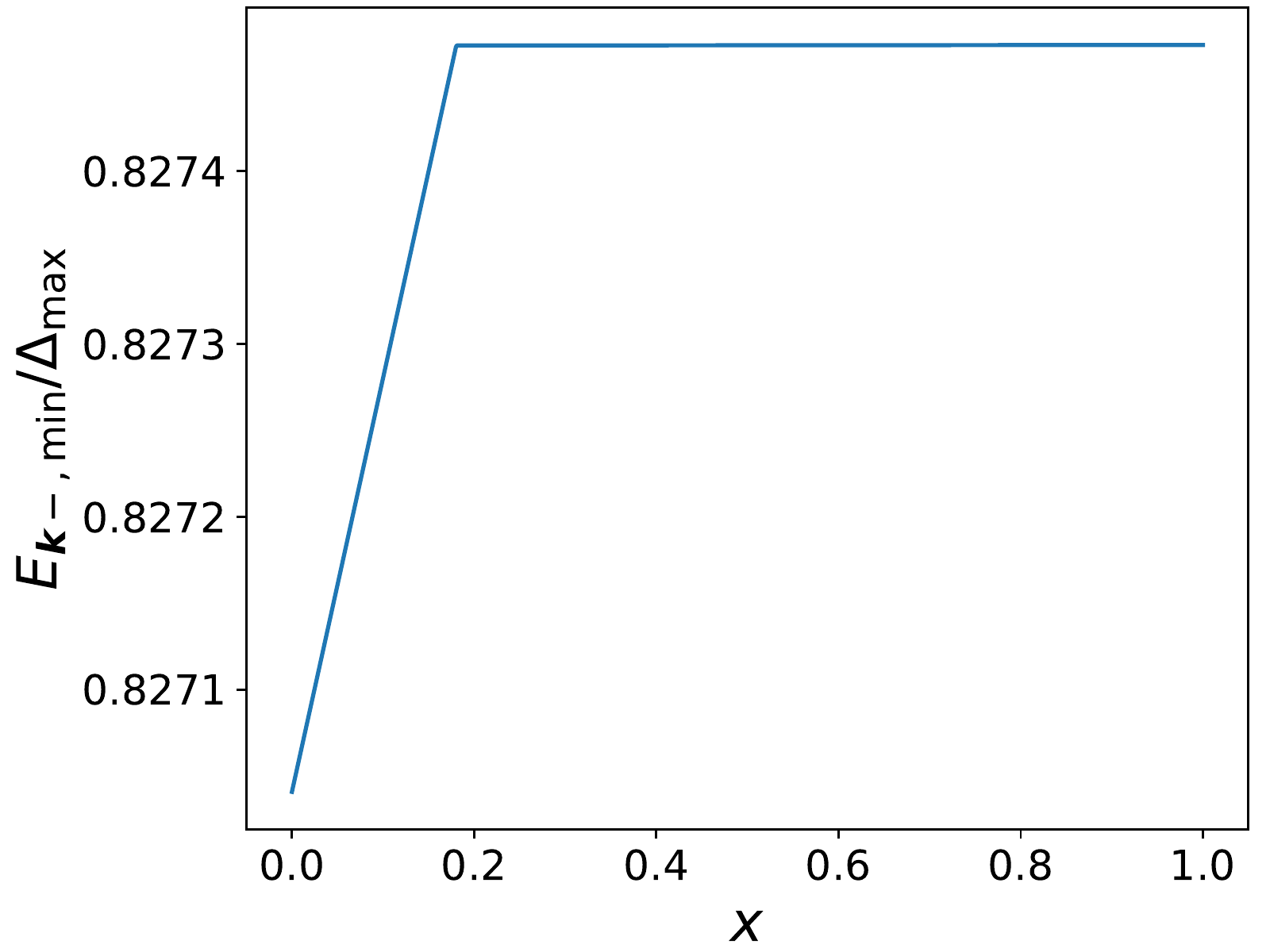}
    \end{minipage}%
    \caption{Left: $E_{\boldsymbol{k}\eta}$ shown on the FS for $x = 0$ and $p_x+ip_y$-wave gap, showing that there is a bulk gap. Right: Minimum of $E_{\boldsymbol{k}-}$ shown as a function of $x$, i.e. when continuously removing the SC gaps that couple the spin blocks. The gap does not close when they are removed. The strange form is understood from the fact that first, the minimum of $E_{\boldsymbol{k}-}$ moves until $E_{\boldsymbol{k}\eta}$ both have their minimum for the same $\phi$. Then a much slower increase sets in where that minimum increases in size. The parameters are $\mu/t = -5.5, t/J = 1000, D/J = 2.16, U/J = 0.35, K/J = 0.1, S=1,$ and $N_\phi = 100$.  \label{fig:Eketagap}}
\end{figure}
Since the bulk gap is retained by letting $x$ go from 0 to 1, see Fig.~\ref{fig:Eketagap}, our system, $ H_{\text{BdG}}(0)$, is topologically equivalent to a spin decoupled system, $ H_{\text{BdG}}(1)$. Then, we can write the BdG Hamiltonian in two separated spin blocks, $H_{\text{BdG}\sigma} = (1/2) \sum_{\boldsymbol{k}} \Psi_{\boldsymbol{k}\sigma}^\dagger H_{\boldsymbol{k}\sigma} \Psi_{\boldsymbol{k}\sigma}$,
with $ \Psi_{\boldsymbol{k}\sigma} = (c_{\boldsymbol{k}\sigma}, c_{-\boldsymbol{k}\sigma}^\dagger)^T$ and $H_{\boldsymbol{k}\sigma}$ specified in the main text.

We know $\Delta_{\uparrow\uparrow}(\phi)$ at $N_\phi$ points of $\phi$. Usually that is not enough to get accurate momentum space derivatives of the gap. However, once $N_\phi$ is large enough that the gap is a smooth function of $\phi$ we can extrapolate to arbitrary $\phi$ density by assuming linear $\phi$ dependence between two discrete values of $\phi$. For all $k_x, k_y$ such that $\boldsymbol{k} \in $ eBZ we find $\phi_{\text{here}} = \operatorname{atan2}(k_y, k_x)$. Then we find which discrete $\phi$ points it lies between, name them $\phi_i$ and $\phi_{i+1}$, to get
\begin{equation}
    \Delta(\phi_{\text{here}} ) = \Delta(\phi_i) + [\Delta(\phi_{i+1})-\Delta(\phi_i)]\frac{\phi_{\text{here}}-\phi_i}{\phi_{i+1}-\phi_i}.
\end{equation}
This equation assumes no radial dependence of the gap. This can be added by multiplying with a function of the length of $\boldsymbol{k}$.
We calculate the momentum derivatives in the winding number using central difference, with $\Delta k = 10^{-5}$.

The great need for points on the FS to get accurate integrals requires adaptive integration \cite{AdaptQuad, QSkQTPT}. With parameters $t/J = 1000, K/J = 0.1, \mu/t = -5.5$ we know that the form of the gap is essentially independent of temperature below $T_{\text{c}}$. We use the result from $T$ close to $T_{\text{c}}$ with an amplitude set to approximately $2.9J$. By rigorous topological arguments the amplitude of the gap, i.e. the temperature ($<T_{\text{c}}$) and the value of $\bar{J}> 0$ does not affect the topological invariant. Using a somewhat large amplitude is simply a numerical convenience as it spreads the FS concentrated behavior a little bit further out. Then, $\approx 7\cdot10^4$ points in adaptive quadrature gives $N_\uparrow \approx -0.93$, $\approx 1.4\cdot10^5$ points gives $N_\uparrow \approx -1.00014,$ and $\approx 2.3\cdot10^5$ points gives $N_\uparrow \approx -0.999989.$
Given that the winding number by definition is an integer, we conclude that $N_\uparrow = -1$. That means $N_\downarrow = 1$ and so
\begin{equation}
    \nu_{\mathbb{Z}_2} = \frac{1}{2} (N_\uparrow - N_\downarrow) \text{ mod } 2 = -1 \text{ mod } 2 = 1.
\end{equation}
We checked that the same holds for several other choices of $\mu, K$ within the SkX1 state. Similarly, the $p_x+ip_y$ phase from SkX2 is a topological SC, with $N_\downarrow = 1$ and $\nu_{\mathbb{Z}_2} = 1$.

As a double check, we tried a lower gap amplitude, $0.34J$. As expected, this merely required more points to get nice integer result. We also tried different assumptions on the radial dependence of the gap -- e.g.~no radial dependence or multiply by a Gaussian around the FS with width set by $|\epsilon_{\boldsymbol{k}}| < \omega_{\text{c}}$ -- and got the same result. In conclusion, $\bar{J}>0$, $T<T_{\text{c}}$ and the radial dependence of the gap do not influence the topological character of the SC. 


\begin{figure}
    \centering
    \begin{minipage}{0.5\textwidth}
    \includegraphics[width=0.6\linewidth]{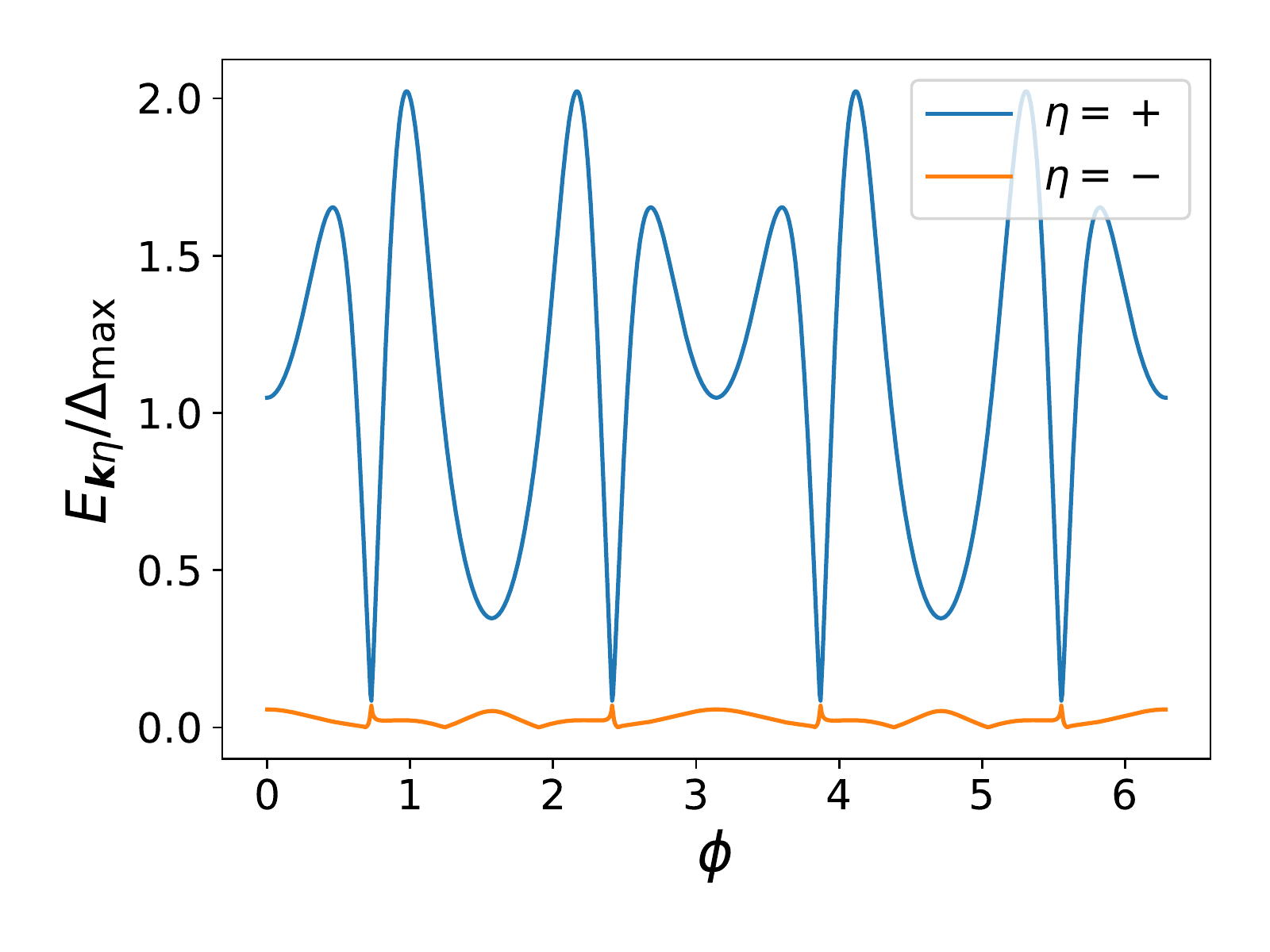}
    \end{minipage}%
    \begin{minipage}{0.5\textwidth}
    \includegraphics[width=0.6\linewidth]{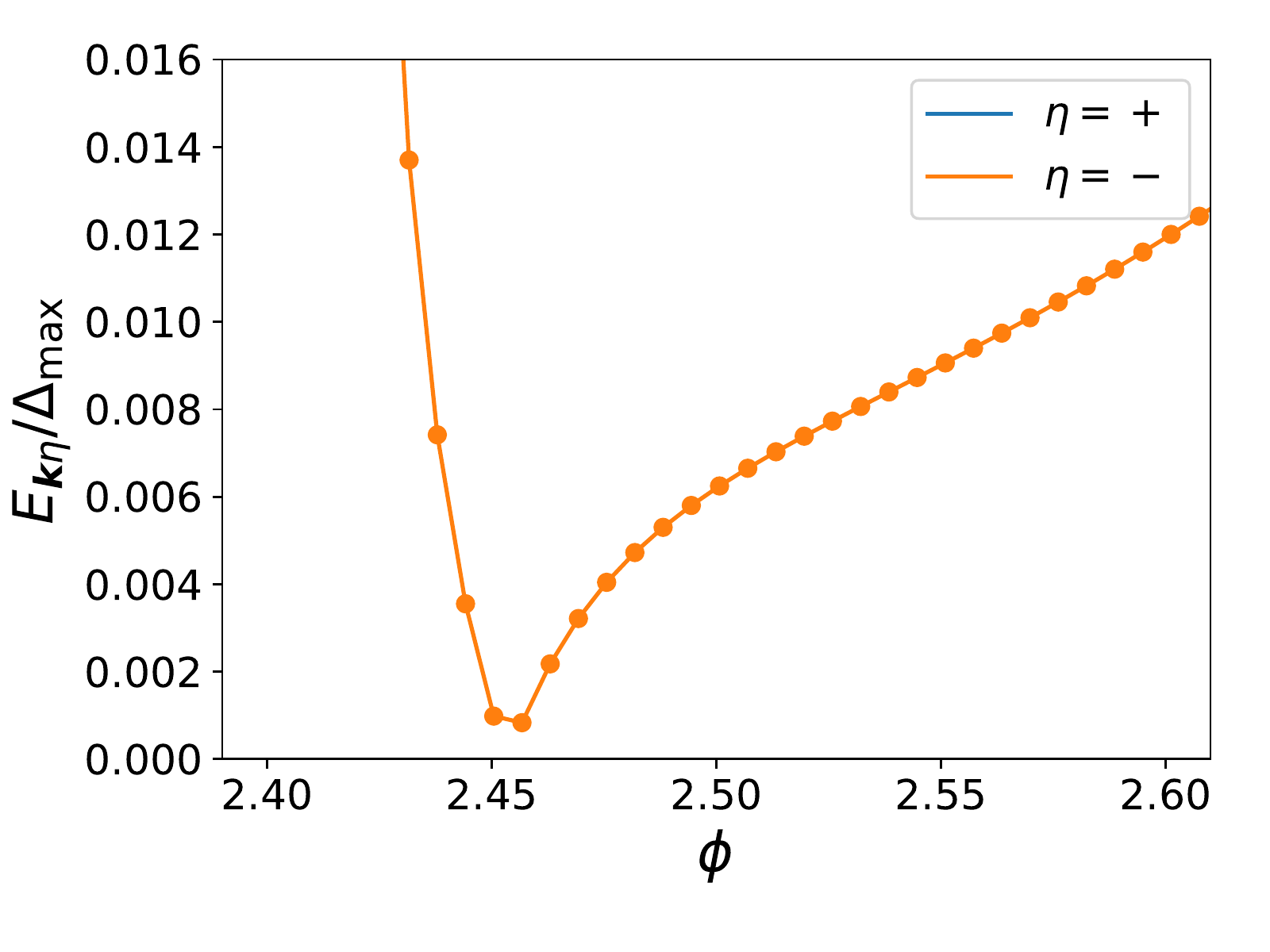}
    \end{minipage}%
    \caption{Left: $E_{\boldsymbol{k}\eta}$ shown on the FS for $x = 0$ and $f_y+if_x$-wave gap. As it turns out $E_{\boldsymbol{k}-}$ is small and has some very sharp valleys where it approaches zero. Right: Zoomed plot with markers at calculated points. Even with $N_\phi = 1000$ when solving the linearized gap equation, it is difficult to say for certain that the bulk is gapped. The parameters are $\mu/t = -5.0, t/J = 1000, D/J = 2.16, U/J = 0.35, K/J = 0.1,$ and $S=1$.  \label{fig:fwavegap}}
\end{figure}

Fig. \ref{fig:fwavegap} shows an attempt to determine whether the $f_y+if_x$ gap gives a gapped spectrum. We used $N_\phi = 1000$ to solve the linearized gap equation, which is a time-consuming calculation. This is not sufficient to unambiguously say that the bulk is gapped. With the solution to the linearized gap equation, it turns out that
$\frac{1}{2}\Tr\hat{\Delta}_{\boldsymbol{k}} \hat{\Delta}_{\boldsymbol{k}}^\dagger \gtrsim \frac{1}{2}\sqrt{A_{\boldsymbol{k}}}$. They are very similar in value, giving small $E_{\boldsymbol{k}-}$. Given numerical results with a discretized number of $\boldsymbol{k}$-values on the FS we cannot exclude the possibility that there is some $\boldsymbol{k}$ where $\frac{1}{2}\Tr\hat{\Delta}_{\boldsymbol{k}} \hat{\Delta}_{\boldsymbol{k}}^\dagger = \frac{1}{2}\sqrt{A_{\boldsymbol{k}}}$, and $E_{\boldsymbol{k}-} = 0$. Additionally, if we manage to show that $E_{\boldsymbol{k}-} > 0$ at some choice of $\mu, K$, we cannot rule out that the gap closes at another choice of $\mu, K$ since it is so small. Therefore, determining the bulk topological invariant would require very accurate solutions of the gap for all $\mu, K$ in the $f_y+if_x$ region.

\end{document}